\documentclass[prd,nofootinbib,showpacs,preprint,superscriptaddress]{revtex4-1}
\usepackage[T1]{fontenc}
\usepackage{amsmath,amssymb}
\usepackage{epsfig}
\usepackage{graphicx}
\usepackage[usenames,dvipsnames]{color}
\usepackage{slashed}
\usepackage[colorlinks,citecolor=blue]{hyperref}
\usepackage{color}

\newcommand{\beq}{\begin{equation}}
\newcommand{\eeq}{\end{equation}}
\newcommand{\beqa}{\begin{eqnarray}}
\newcommand{\eeqa}{\end{eqnarray}}

\newcommand{\tz}{{\theta_{23}}}

\newcommand{\dcp}{\delta_{\mathrm{cp}}}

\newcommand{\s}{\sigma}

\newcommand{\RN}[1]{%
  \textup{\uppercase\expandafter{\romannumeral#1}}%
}
\begin{document}
\title{Probing Majorana Neutrino Textures at DUNE }
\author{Kalpana Bora}
\email{kalpana.bora@gmail.com }
\affiliation{Department of Physics, Gauhati University, Guwahati, Assam 781014, India}
\author{Debasish Borah}
\email{dborah@iitg.ernet.in}
\affiliation{Department of Physics, Indian Institute of Technology Guwahati, Assam 781039, India}
\author{Debajyoti Dutta}
\email{debajyotidutta@hri.res.in}
\affiliation{Harish-Chandra Research Institute, Chhatnag Road, Jhunsi, Allahabad 211019, India}
\affiliation{Homi Bhabha National Institute, Training School Complex,
Anushaktinagar, Mumbai - 400094, India}
\begin{abstract}
We study the possibility of probing different texture zero neutrino mass matrices at long baseline neutrino experiment DUNE particularly focusing on its sensitivity to the octant of atmospheric mixing angle $\theta_{23}$ and leptonic Dirac CP phase $\delta_{\text{cp}}$. Assuming a diagonal charged lepton basis and Majorana nature of light neutrinos, we first classify the possible light neutrino mass matrices with one and two texture zeros and then numerically evaluate the parameter space which satisfies the texture zero conditions. Apart from using the latest global fit $3\sigma$ values of neutrino oscillation parameters, we also use the latest bound on the sum of absolute neutrino masses $(\sum_i \lvert m_i \rvert)$ from the Planck mission data and the updated bound on effective neutrino mass $M_{ee}$ from neutrinoless double beta decay $(0\nu \beta \beta)$ experiment to find the allowed Majorana texture zero mass matrices. For the allowed texture zero mass matrices from all these constraints, we then feed the corresponding light neutrino parameter values satisfying the texture zero conditions into the numerical analysis in order to study the capability of DUNE to allow/exclude them once it starts taking data. We find that the DUNE will be able to exclude some of these texture zero mass matrices which restrict the $(\theta_{23}-\delta_{\text{cp}})$ to a very specific range of values, depending on the values of the parameters that Nature has chosen.
\end{abstract}

\maketitle

\section{Introduction}
\label{sec:intro}
Since the discovery of the Higgs boson in 2012, the large hadron collider (LHC) experiment has been confirming the validity of the standard model (SM) of particle physics again and again without any convincing signature of new physics beyond the standard model (BSM) till date. In spite of such null results, there are plenty of reasons: both theoretical and observational, which undoubtedly suggest the presence of BSM physics. Observation of tiny but non-zero neutrino masses and large leptonic mixing by neutrino oscillation experiments \cite{PDG, kamland08, T2K, chooz, daya, reno, minos} is one such reason that has led to a large number of BSM proposals in the last few decades. The current status of neutrino oscillation experiments can be summarised in terms of $3\sigma$ global fit values of neutrino parameters shown in table \ref{tab:data1}.
\begin{center}
\begin{table}[htb]
\begin{tabular}{|c|c|c|}
\hline
Parameters & Normal Hierarchy (NH) & Inverted Hierarchy (IH) \\
\hline
$ \frac{\Delta m_{21}^2}{10^{-5} \text{eV}^2}$ & $7.03-8.09$ & $7.02-8.09 $  \\
$ \frac{|\Delta m_{3l}^2|}{10^{-3} \text{eV}^2}$ & $2.407-2.643$ & $2.399-2.635 $  \\
$ \sin^2\theta_{12} $ &  $0.271-0.345 $ & $0.271-0.345 $  \\
$ \sin^2\theta_{23} $ & $0.385-0.635$ &  $0.393-0.640 $  \\
$\sin^2\theta_{13} $ & $0.01934-0.02392$ & $0.01953-0.02408 $  \\
$ \delta $ & $0^{\circ}-360^{\circ}$ & $145^{\circ}-390^{\circ}$  \\
\hline
\end{tabular}
\caption{Global fit $3\sigma$ values of neutrino oscillation parameters \cite{schwetz16}. Here $\Delta m_{3l}^2 \equiv \Delta m_{31}^2$ for NH and $\Delta m_{3l}^2 \equiv \Delta m_{32}^2$ for IH.}
\label{tab:data1}
\end{table}
\end{center}
where NH and IH refer to normal and inverted hierarchical neutrino masses respectively. The above data also allow the possibility of quasi-degenerate neutrinos $(m_1 \approx m_2 \approx m_3)$. Although the question of the absolute mass scale of neutrinos remain open at oscillation experiments, cosmology and neutrinoless double beta decay $(0\nu \beta \beta)$ experiments put an upper bound on the lightest neutrino mass. For example, the latest Planck experiment data constrain the sum of absolute neutrino masses as $\sum_i \lvert m_i \rvert < 0.17$ eV \cite{Planck15}. Similarly, constraints from the ongoing $0\nu \beta \beta$ experiments \cite{kamland, GERDA, kamland2, GERDA2} rule out light neutrino masses around $0.1$ eV and above. On the other hand, the leptonic Dirac CP phase $\delta_{\text{cp}}$ remains undetermined as of now, though a recent hint towards $\delta_{\text{cp}} \approx -\pi/2$ \cite{diracphase} appeared recently from the T2K collaboration. If neutrinos are Majorana fermions, then two Majorana CP phases come into the picture which however, do not affect neutrino oscillation probabilities, but can be probed at $0\nu \beta \beta$ experiments.

In the three flavour neutrino oscillation scenario, the oscillations are driven by the two mass square differences $\Delta m^2_{21}$ and $\vert \Delta m^2_{31} \rvert$. Out of the three mixing angles, $\theta_{12}$, $\theta_{13}$ and $\theta_{23}$ governing the oscillation amplitudes, $\theta_{12}$ and $\theta_{13}$ are very precisely measured. Although the atmospheric mixing angle $\theta_{23}$ was measured in 1998 at Super-Kamiokande atmospheric neutrino experiment, yet it is not confirmed if $\theta_{23} = 45^{o}$ i.e. maximal or not. If not, then it may lie either in the higher octant (HO) ( $\theta_{23}$ $ >$ 45$^o$) or in the lower octant (LO) ($\theta_{23}$ $ <$ 45$^o$). The preliminary results from the T2K experiment \cite{T2K, diracphase} prefers near maximal mixing \cite{t2k} while MINOS \cite{minos, minos2} and Super-Kamiokande (SK) disfavour maximal mixing \cite{sk1}. All the global fit analyses are inconsistent with maximal $\theta_{23}$ at less than 1$\sigma$ \cite{schwetz16, global}. Recent NO$\nu$A data exclude the maximal mixing scenario at 2.5$\sigma$ C.L. \cite{novaNEW} favouring the case of two degenerate solutions. So in the present scenario, we have to wait for the long baseline (LBL) accelerator experiments to resolve this ambiguity as well as to measure $\theta_{23}$ precisely. The capability of DUNE (Deep Underground Neutrino Experiment) \cite{dune1, dune11, dune111, dune2}, NO$\nu$A \cite{nova1, nova2}, T2K etc to resolve octant ambiguity has been studied in \cite{suprabh1, raj1, monoj}. Octant sensitivity at DUNE in conjunction with reactor experiments is studied in \cite{dev1, dev11}. These long baseline accelerator experiments are sensitive to both $\nu_{\mu} \rightarrow \nu_e$ (appearance) and $\nu_{\mu} \rightarrow \nu_{\mu}$ (disappearance) and synergy between them can enhance the octant sensitivity \cite{fogli1, hira, sam}. On the other hand, any terrestrial baseline of several hundred kilometres is sensitive to matter effects. So these LBL experiments, being sensitive to matter effects can discriminate between the two hierarchies \cite{pomita1, dune1}.

 After the precise measurement of $\theta_{13}$, now the whole neutrino community is focusing on the leptonic Dirac CP phase $\delta_{\text{cp}}$, which is the only unknown parameter in the three flavour oscillation framework. Although in the quark sector, CP violation is observed and can be explained due to complex Yukawa couplings or complex Higgs field vacuum expectation values \cite{aa, bb}, it is yet to be discovered in the leptonic sector with some significant degree of precision. The LBL accelerator experiments which are sensitive to $\nu_{\mu} \rightarrow \nu_e$ channel can probe CP violation in the leptonic sector \cite{dune1}. From this point of view, the DUNE at Fermilab, the U.S. is the best candidate which will start taking data from next decade. In the literature, recent studies related to CP violation discovery potential of DUNE as well as other super-beam experiments can be found \cite{raj2, suprabh2,suprabh3, dev2, kb}.

Before the neutrino oscillation experiments completely remove the existing ambiguities in neutrino parameters like the octant of $\theta_{23}$, Dirac CP phase as well as mass hierarchy at ongoing and near future facilities, it is important to study the predictions for these neutrino observables within different BSM frameworks. Depending on the predictions for such neutrino observables, one can discriminate between different BSM scenarios at the oscillation experiments. However, generic BSM frameworks come with several free parameters, making it difficult to have a robust prediction for a particular neutrino parameter. Reducing the number of free parameters can therefore lead to robust testable predictions of neutrino observables. Such reduction of free parameters can be possible within the framework of flavour symmetry models where the additional discrete or continuous symmetries either relate two or more terms or forbid some terms in the mass matrix. One such possibility arises within the context of texture zeros in leptonic mass matrices. For a complete survey of such texture zeros in lepton mass matrices, please refer to \cite{Ludl2014}. Different possible flavour symmetric interpretations for the origin of such texture zero mass matrices can be found in\cite{texturesym,texturesym1,texturesym2,texturesym3,texturesym4,texturesym5,texturesym6,texturesym7,texturesym8,texturesym9} within the framework of different seesaw models. Such texture zero mass matrices in general, have specific predictions for neutrino oscillation parameters, the consequences of which can be studied from cosmology as well as oscillation experiments point of view. In the light of neutrino oscillation experiment data, one-zero and two-zero texture mass matrices were studied in \cite{onezero,onezero1,onezero2,onezero3} and \cite{texturesym9, alltex,twozero,twozero1,twozero2,twozero3,twozero4,twozero5,twozero6,twozero7,twozero8,twozero9,twozero10,twozero11} respectively. These texture zero mass matrices were also studied from the perspective of baryogenesis through leptogenesis in \cite{maniprd,leptotext,leptotext1,leptotext2,leptotext3,leptotext4, leptoext5}.

 In this work, we do a phenomenological study of one-zero and two-zero texture mass matrices assuming the neutrinos to be of Majorana nature and the charged lepton mass matrix to be diagonal. The main motive of this work is two fold. Firstly, we intend to find out the allowed texture zero mass matrices from the requirement of satisfying the latest neutrino oscillation data as well as the cosmological bound on the sum of absolute neutrino masses from the latest Planck mission data \cite{Planck15} and the bound on effective neutrino mass $M_{ee}$ from the latest experimental lower bound on neutrinoless double beta decay half-life \cite{kamland2, GERDA2}. The second objective is to find if any of these textures which are allowed from all experimental constraints can be probed, focusing on the sensitivity of neutrino oscillation experiments like DUNE to the values of $\theta_{23}, \delta_{\text{cp}}$ predicted by a particular texture zero mass matrix. We first numerically determine the values of $\theta_{23}, \delta_{\text{cp}}$ along with other neutrino parameters which satisfy a particular texture zero condition and then use them in GLoBES software \cite{Huber, Kopp} to determine the corresponding experimental sensitivity. We show that these textures can be excluded for some combinations of $\tz$ and $\dcp$ at DUNE. If Nature chooses some values of $\tz$ and $\dcp$, then how DUNE can allow/exclude a particular mass matrix texture is the goal of this present study. We have found that the two-zero textures are very interesting as DUNE can constrain the $\tz-\dcp$ parameter space more tightly compared to the one-zero textures. We have observed that if Nature prefers normal (inverted) mass ordering, then DUNE can constrain a particular type of texture mass matrix denoted as B1 (B1 and B2) more tightly than any other textures.

This article is organised as follows. In section \ref{sec:texture}, we discuss about the different possible texture zero mass matrices. A brief discussion on DUNE is presented in section \ref{sec:dune}. Section \ref{sec:numeric} contains the details of our numerical analysis while in section \ref{sec:results} we present our results and discussions. Concluding remarks are given in section \ref{sec:conclude}. 
\section{Texture Zeros in Majorana Neutrino Mass Matrix}
\label{sec:texture}
If light neutrinos are of Majorana nature, the $3\times 3$ mass matrix $M_{\nu}$ is complex symmetric having six independent complex parameters. In such a scenario, the total number of structurally different Majorana neutrino mass matrices with $k$ texture zeros is given by 
\begin{equation}
\label{prmtn}
^6C_k=\frac{6!}{k!(6-k)!}
\end{equation}
A symmetric mass matrix with more than three texture zeros is not compatible with the observed leptonic mixing and masses. The Pontecorvo-Maki-Nakagawa-Sakata (PMNS) leptonic mixing matrix is related to the diagonalising matrices of charged lepton and neutrino mass matrices as 
\begin{equation}
U_{\text{PMNS}} = U^{\dagger}_l U_{\nu}
\label{pmns0}
\end{equation}
In the diagonal charged lepton basis, $U_{\text{PMNS}}$ is same as the diagonalising matrix $U_{\nu}$ of the neutrino mass matrix $M_{\nu}$. In this basis, it was shown \cite{Xing:2004ik} that a symmetric Majorana neutrino mass matrix with 3 texture zeros is not compatible with neutrino oscillation data. This leaves us with the possible one-zero and two-zero texture mass matrices. Going by the counting formula mentioned above \eqref{prmtn}, we can have six possible one-zero texture and fifteen possible two-zero texture mass matrices. The one-zero texture mass matrices can be written as 
\begin{center}

$ G_1 :\left(\begin{array}{ccc}
0& \times&\times\\
\times& \times&\times \\
\times& \times&\times 
\end{array}\right) , G_2 :\left(\begin{array}{ccc}
\times& 0&\times\\
0& \times&\times \\
\times& \times&\times 
\end{array}\right) , G_3 :\left(\begin{array}{ccc}
\times& \times&0\\
\times& \times&\times \\
0 & \times&\times 
\end{array}\right) ,   G_4 :\left(\begin{array}{ccc}
\times& \times&\times\\
\times & 0 &\times \\
\times& \times&\times 
\end{array}\right)  ,$
 
\end{center}
\begin{equation}
G_5 :\left(\begin{array}{ccc}
\times& \times&\times\\
\times& \times & 0 \\
\times& 0 &\times 
\end{array}\right) , 
 G_6 :\left(\begin{array}{ccc}
\times& \times&\times\\
\times& \times& \times \\
\times& \times & 0
\end{array}\right) 
\end{equation}

Where the crosses ``$\times$'' denote non-zero arbitrary elements of $M_{\nu}$. Similarly, the two-zero texture matrices, classified into six categories (following the notations of \cite{twozero7}) are given by
\begin{equation}
A_1 :\left(\begin{array}{ccc}
0& 0&\times\\
0& \times&\times \\
\times& \times&\times 
\end{array}\right) , 
 A_2 :\left(\begin{array}{ccc}
0& \times&0\\
\times& \times&\times \\
0& \times&\times 
\end{array}\right);
\end{equation}

\begin{equation}
B_1 :\left(\begin{array}{ccc}
\times& \times&0\\
\times& 0&\times \\
0& \times&\times 
\end{array}\right) , 
 B_2 :\left(\begin{array}{ccc}
\times& 0 &\times\\
0& \times&\times \\
\times& \times&0 
\end{array}\right),
 B_3 :\left(\begin{array}{ccc}
\times& 0&\times\\
0& 0&\times \\
\times& \times&\times 
\end{array}\right), 
B_4 :\left(\begin{array}{ccc}
\times& \times&0\\
\times& \times&\times \\
0& \times&0 
\end{array}\right);
\end{equation}

\begin{equation}
C :\left(\begin{array}{ccc}
\times& \times&\times\\
\times& 0&\times \\
\times& \times&0
\end{array}\right);
\end{equation}

\begin{equation}
D_1 :\left(\begin{array}{ccc}
\times& \times&\times\\
\times& 0&0 \\
\times& 0&\times 
\end{array}\right) , 
 D_2 :\left(\begin{array}{ccc}
\times& \times&\times\\
\times& \times&0 \\
\times& 0&0 
\end{array}\right);
\end{equation}

\begin{equation}
E_1 :\left(\begin{array}{ccc}
0& \times&\times\\
\times& 0&\times \\
\times& \times&\times 
\end{array}\right) , 
 E_2 :\left(\begin{array}{ccc}
0& \times &\times\\
\times& \times&\times \\
\times& \times&0 
\end{array}\right),
 E_3 :\left(\begin{array}{ccc}
0& \times&\times\\
\times& \times&0 \\
\times& 0&\times 
\end{array}\right);
\end{equation}

\begin{equation}
F_1 :\left(\begin{array}{ccc}
\times& 0&0\\
0& \times&\times \\
0& \times&\times 
\end{array}\right) , 
 F_2 :\left(\begin{array}{ccc}
\times& 0 &\times\\
0& \times&0 \\
\times& 0&\times 
\end{array}\right),
 F_3 :\left(\begin{array}{ccc}
\times& \times&0\\
\times& \times&0 \\
0& 0&\times 
\end{array}\right), 
\end{equation}
Where the crosses ``$\times$'' imply non-zero arbitrary elements of $M_{\nu}$. The latest neutrino oscillation data on mixing angles, mass squared differences and cosmology data on sum of absolute neutrino masses allow only six different two-zero textures $A_{1,2}$ and $B_{1,2,3,4}$ as shown by \cite{twozero7,twozero9}.

\section{DUNE at a glance}
\label{sec:dune}
The Deep Underground Neutrino Experiment (DUNE) is a future accelerator based experiment capable of answering all the three major issues in the neutrino sector mentioned earlier: a) determination of neutrino mass hierarchy, b) resolution of the octant of $\tz$ and c) determination of leptonic CP violation. The first test with the megawatt neutrino-
beam facility will begin in $\sim2026$ and the experiment itself will be fully operational in $\sim2027$ \cite{john, duneNEW}. The DUNE is designed in such a way that it can answer all the three most important questions in the neutrino sector mentioned above. The ${\nu_{\mu}(\bar{\nu}_{\mu}})$ super-beam from the Fermilab will be detected by a 35-40 kton Liquid Argon (LAr) far detector at a distance of 1300 km in the Homestake mine, South Dakota. The experiment plans to run for 10 years both in anti-neutrino (5 years) and neutrino (5 years) modes. A 1.2 MW - 120 GeV proton beam will deliver $10^{21}$ protons-on-target (POT) per year which corresponds to a total exposure of $35\times10^{22}$ kton-POT-yr.

All other details,
such as signal and background channels, signal efficiencies are taken from \cite{dune111, dune2} and tabulated in table \ref{Table2} and table \ref{Table3}.

 \begin{table}[!h]
\begin{center}
\begin{tabular}{|c|c|c|c|c|}
\hline 
Experiment & Signal & Signal  & Background   &   Calibration error   \tabularnewline
&   &    norm error & norm error   &   Signal \qquad	Background  \tabularnewline
\hline 
DUNE & $\nu_e$ & 2\%  & 10\% & 5$\%$ \qquad 5\% \tabularnewline
 
 & $\nu_{\mu}$ & 5\%   & 10\%  & 5$\%$ \qquad 5\% \tabularnewline
\hline 
\end{tabular}
\caption{Systematics uncertainties for DUNE}
\label{Table2}

\par\end{center}
\end{table}

 \begin{table}[!h]
\begin{center}
\begin{tabular}{|c|c|c|c|c|c|}
\hline 
Experiment& Signal & Signal &  Energy & Runtime (yrs)    &   Detector Mass (Type)   \tabularnewline
&	& Efficiencies & Resolutions & $\nu + \bar{\nu}$   &   \tabularnewline
\hline 

DUNE & $\nu_{e}^{CC}$&80\% & $0.15/\sqrt{E}$  & 5 + 5 & 35 kton (LArTPC) \tabularnewline
 
 & $\nu_{\mu}^{CC}$& 85\%& $0.20/\sqrt{E}$   &  & \tabularnewline
\hline 
\end{tabular}
\caption{Simulation details like signal efficiencies, energy resolutions, total exposures and detector mass for DUNE}
\label{Table3}

\par\end{center}
\end{table}

\section{Numerical Analysis}
\label{sec:numeric}
In the diagonal charged lepton basis, $U_{\text{PMNS}}$ is same as the diagonalising matrix $U_{\nu}$ which can be parametrised as
\begin{equation}
U_{\text{PMNS}}=\left(\begin{array}{ccc}
c_{12}c_{13}& s_{12}c_{13}& s_{13}e^{-i\delta_{\text{cp}}}\\
-s_{12}c_{23}-c_{12}s_{23}s_{13}e^{i\delta_{\text{cp}}}& c_{12}c_{23}-s_{12}s_{23}s_{13}e^{i\delta_{\text{cp}}} & s_{23}c_{13} \\
s_{12}s_{23}-c_{12}c_{23}s_{13}e^{i\delta_{\text{cp}}} & -c_{12}s_{23}-s_{12}c_{23}s_{13}e^{i\delta_{\text{cp}}}& c_{23}c_{13}
\end{array}\right) \text{diag}(1, e^{i\alpha}, e^{i(\beta+\delta_{\text{cp}} )})
\label{matrixPMNS}
\end{equation}
where $c_{ij} = \cos{\theta_{ij}}, \; s_{ij} = \sin{\theta_{ij}}$. $\delta_{\text{cp}}$ is the Dirac CP phase and  $\alpha, \beta$ are the Majorana CP phases. Using the parametric form of PMNS matrix shown in \eqref{matrixPMNS}, the Majorana neutrino mass matrix $M_{\nu}$ can be found as
\begin{equation}
 M_{\nu}=U_{\text{PMNS}} M^{\text{diag}}_{\nu}U^T_{\text{PMNS}}
 \label{numatrix}
\end{equation}
where
\begin{equation}
M^{\text{diag}}_{\nu}=\left(\begin{array}{ccc}
m_1& 0&0\\
0& m_2& 0  \\
0& 0 &m_3
\end{array}\right),
\end{equation}
with $m_1, m_2$ and $m_3$ being the three neutrino mass eigenvalues. For the case of normal hierarchy, the three neutrino mass eigenvalues can be written as 
$$M^{\text{diag}}_{\nu}
= \text{diag}(m_1, \sqrt{m^2_1+\Delta m_{21}^2}, \sqrt{m_1^2+\Delta m_{31}^2})$$ while for the case of inverted hierarchy, they can be written as 
$$M^{\text{diag}}_{\nu} = \text{diag}(\sqrt{m_3^2+\Delta m_{23}^2-\Delta m_{21}^2}, \sqrt{m_3^2+\Delta m_{23}^2}, m_3)$$ 
The analytical expressions of the elements of this mass matrix are given in Appendix \ref{appen1}. 
\begin{figure*}
\begin{center}
\includegraphics[width=0.47\textwidth]{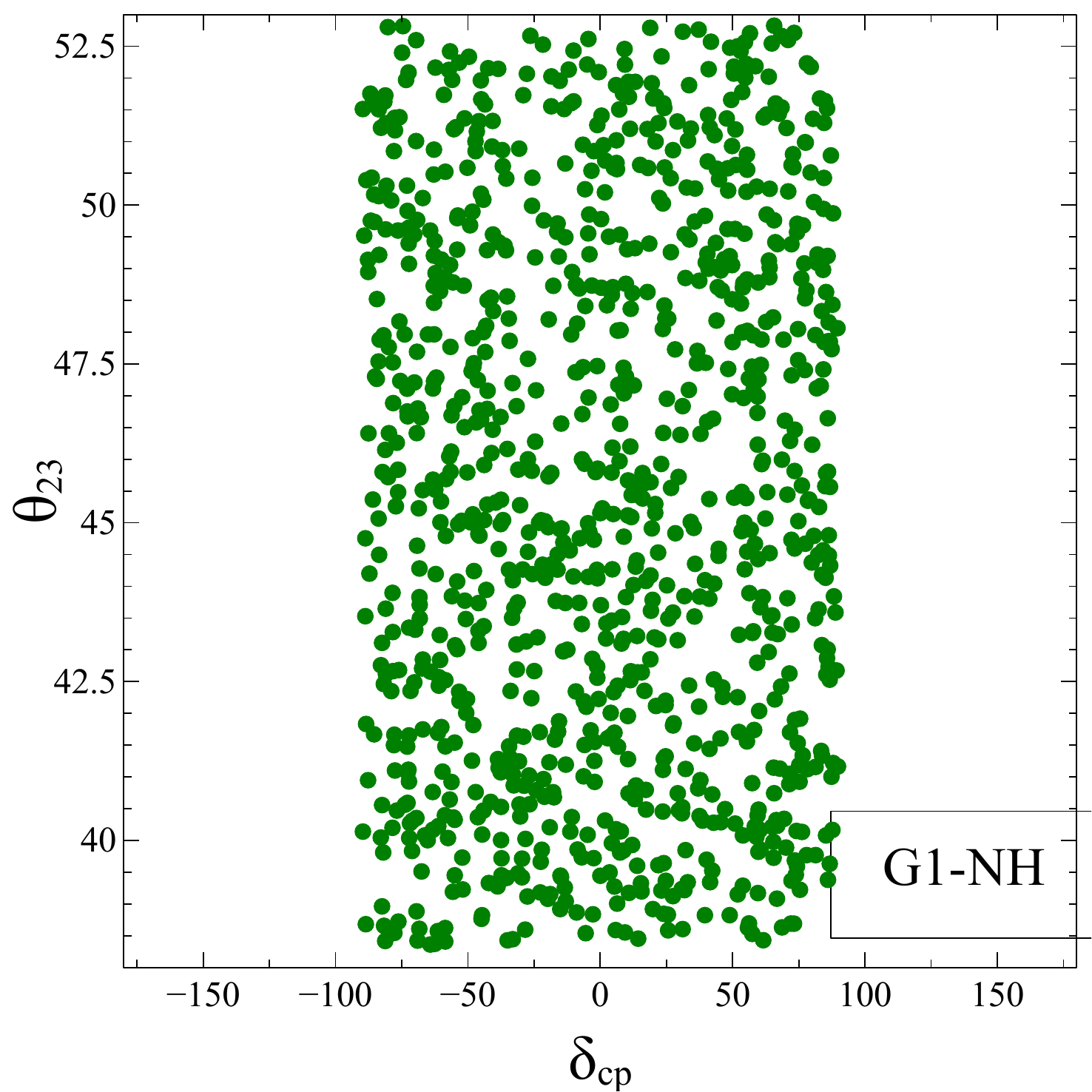}
\includegraphics[width=0.47\textwidth]{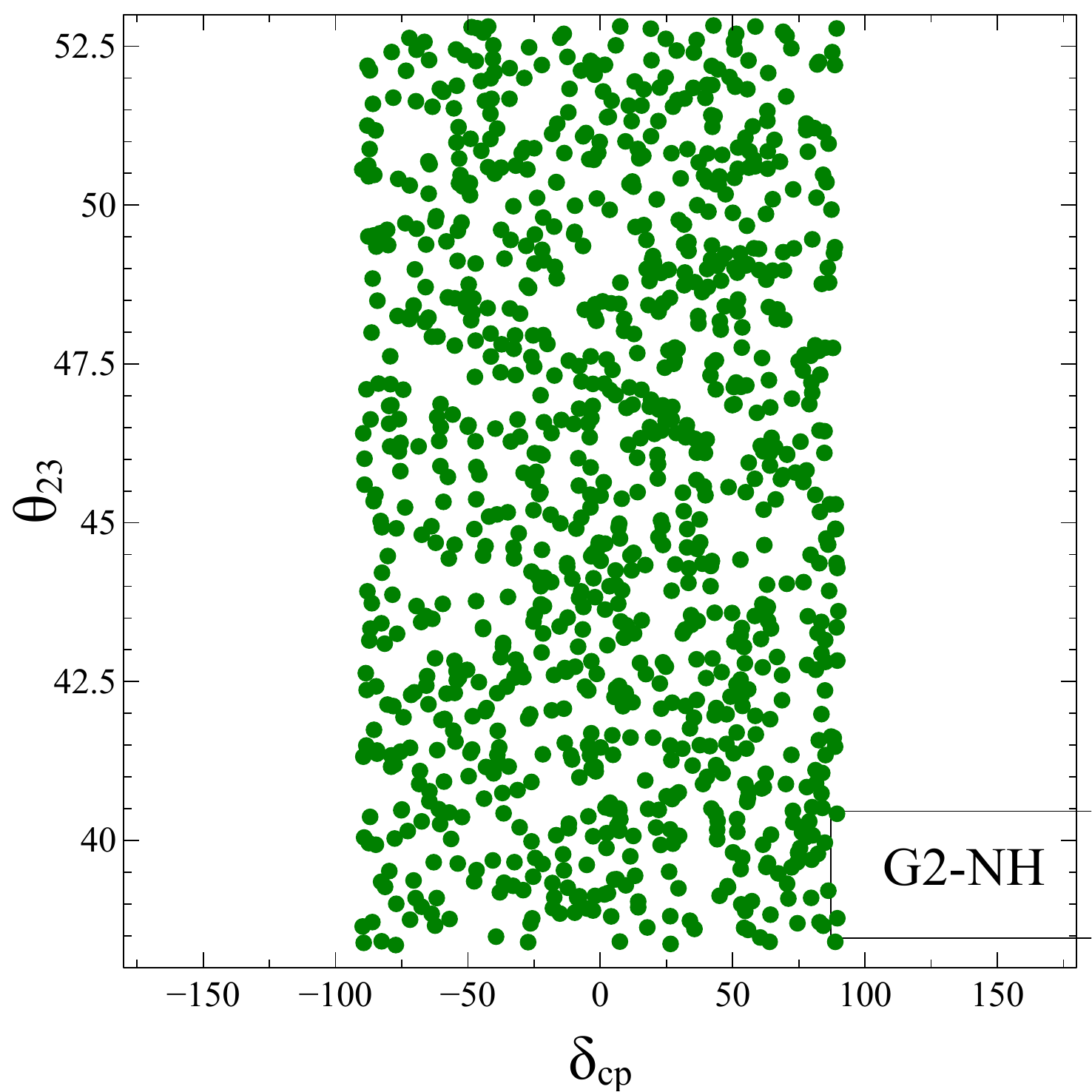}\\
\includegraphics[width=0.47\textwidth]{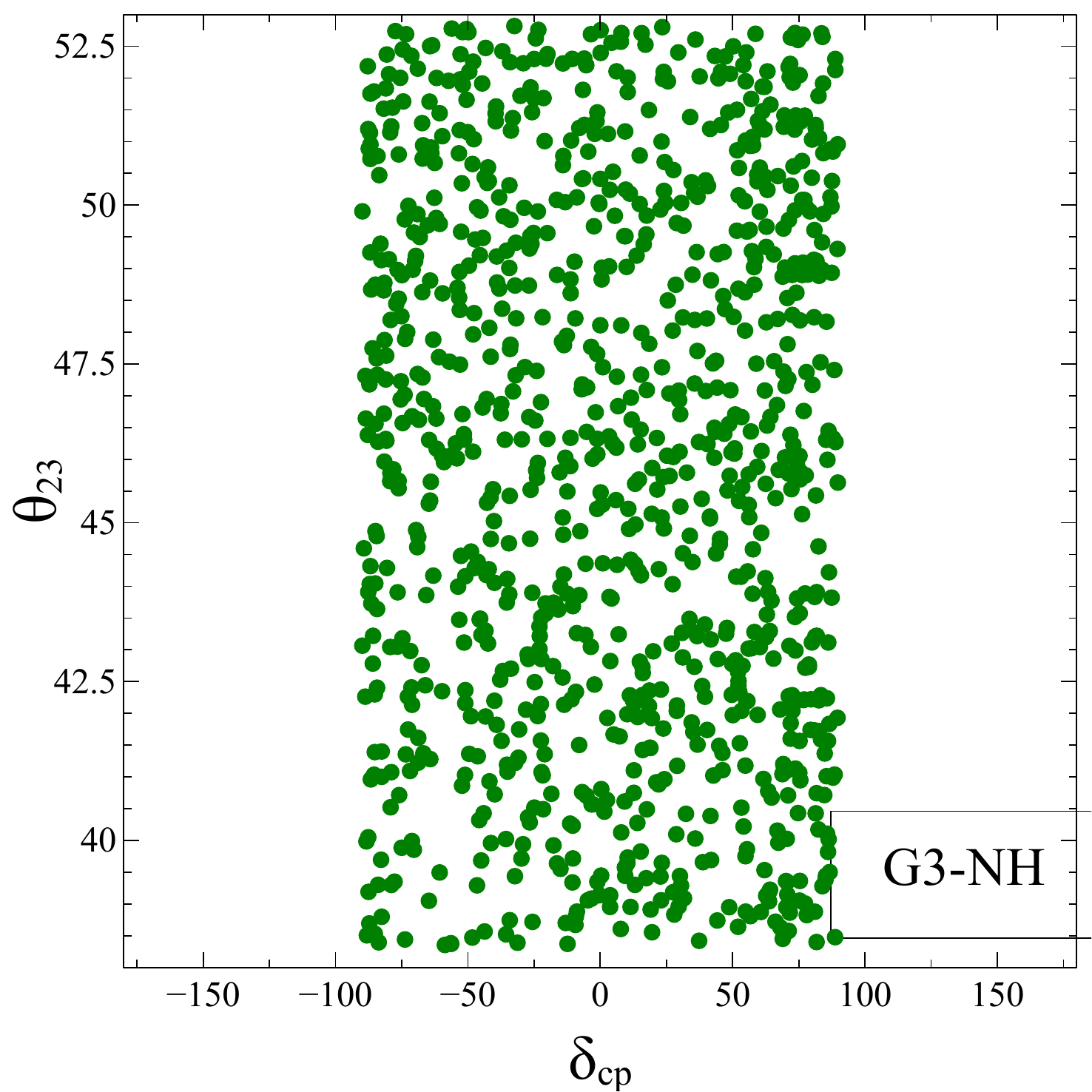}
\includegraphics[width=0.47\textwidth]{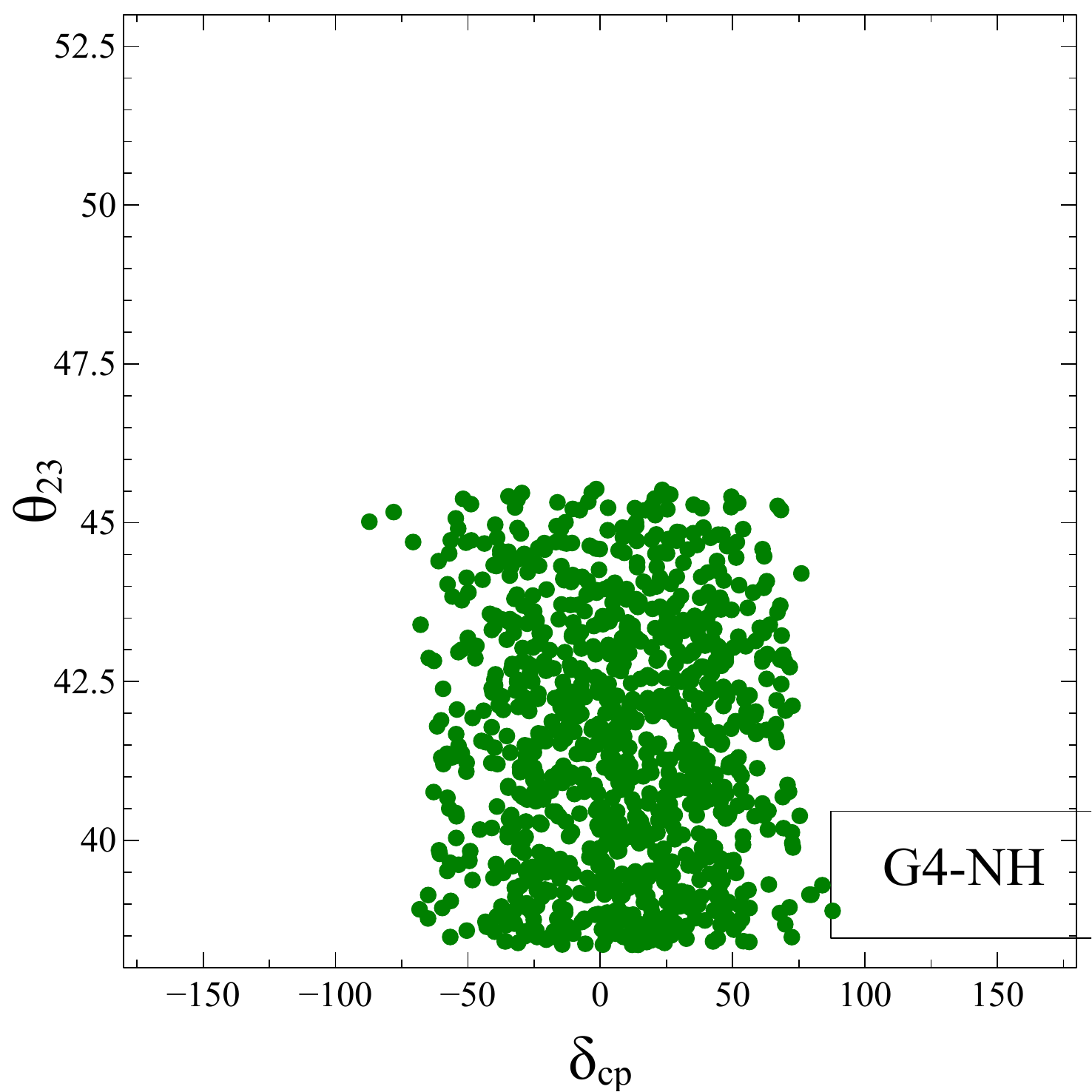} \\
\end{center}
\begin{center}
\caption{Neutrino parameters corresponding to one-zero textures for NH}
\label{fig3}
\end{center}
\end{figure*}
\begin{figure*}
\begin{center}
\includegraphics[width=0.47\textwidth]{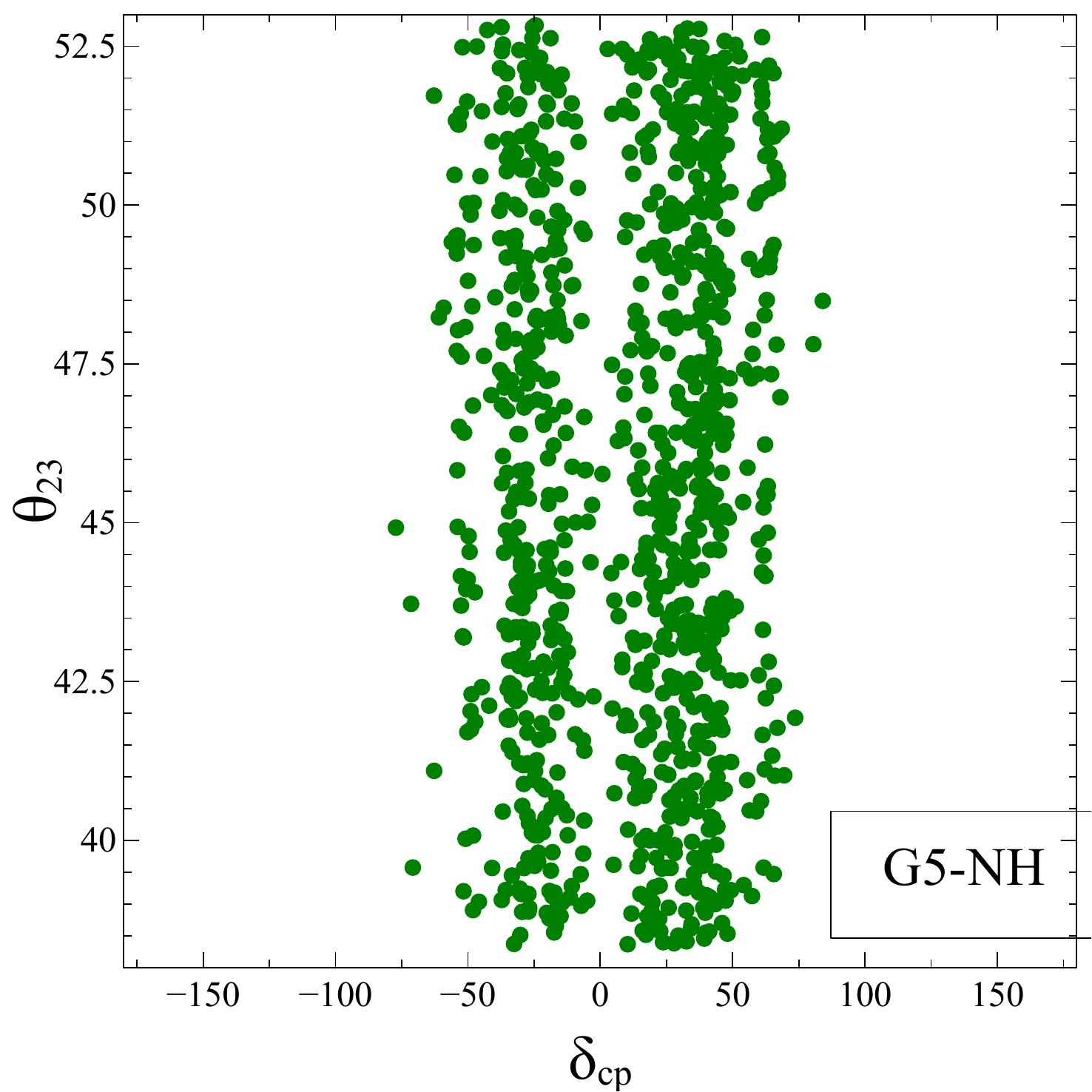}
\includegraphics[width=0.47\textwidth]{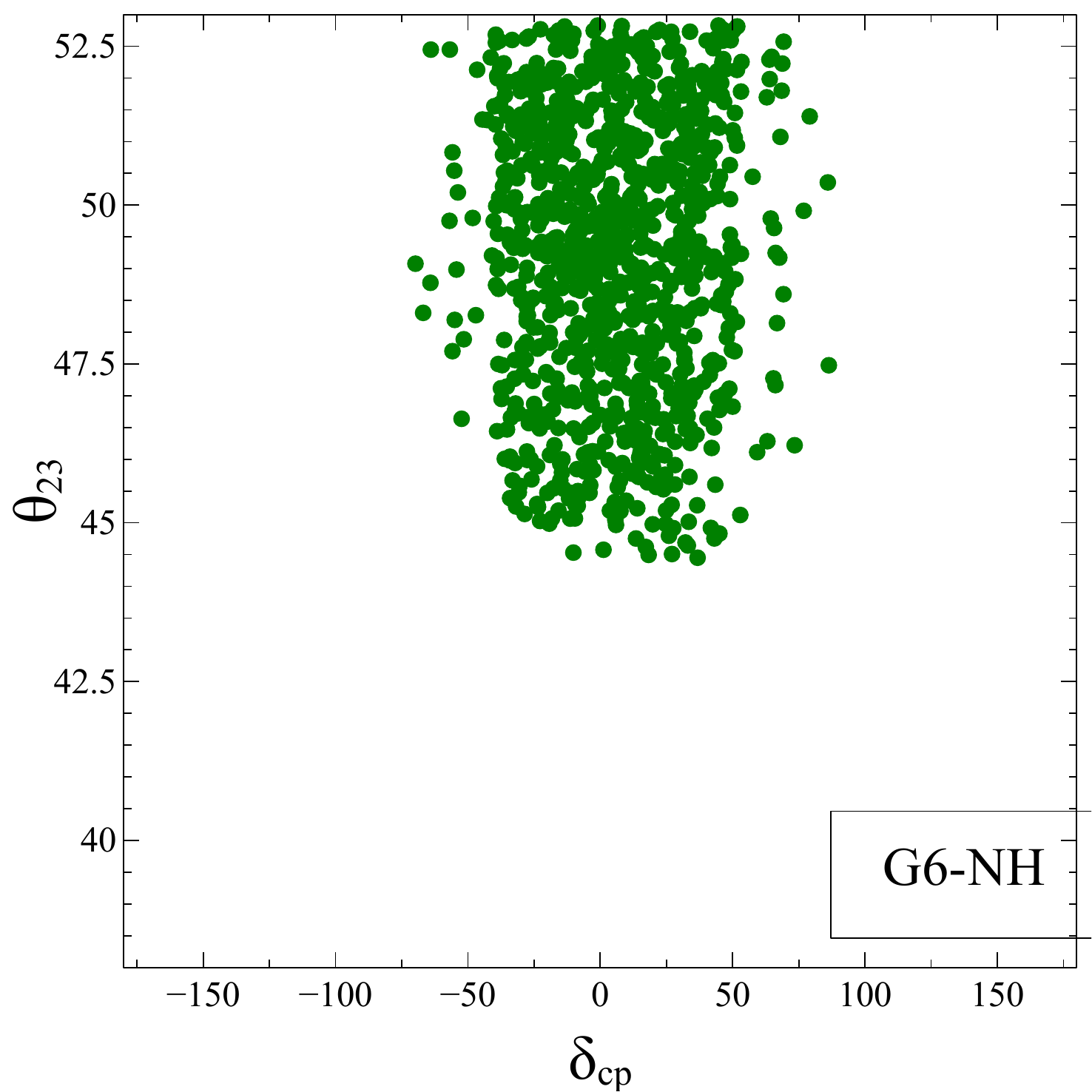}
\end{center}
\begin{center}
\caption{Neutrino parameters corresponding to one-zero textures for NH}
\label{fig3a}
\end{center}
\end{figure*}

\begin{figure*}
\begin{center}
\includegraphics[width=0.47\textwidth]{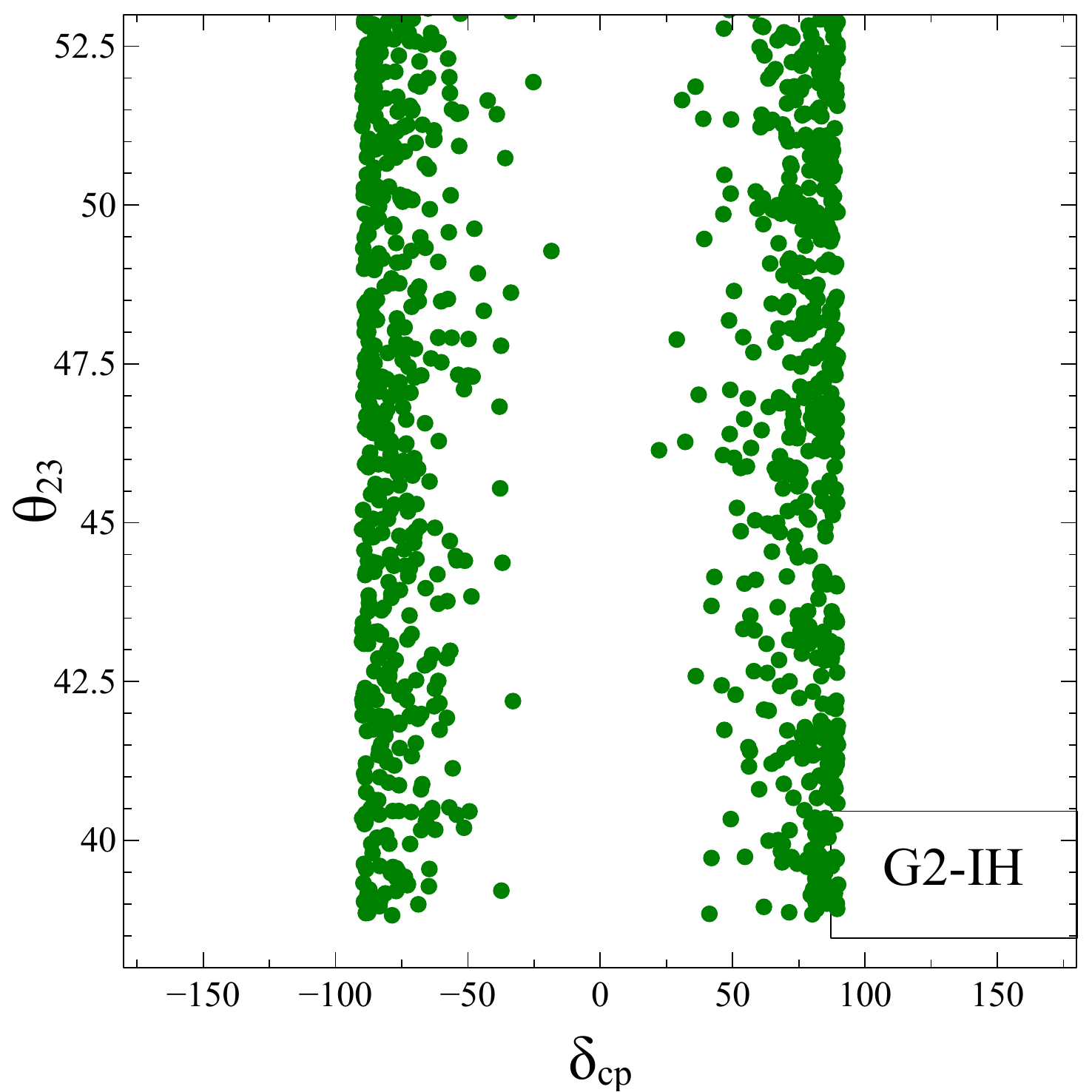}
\includegraphics[width=0.47\textwidth]{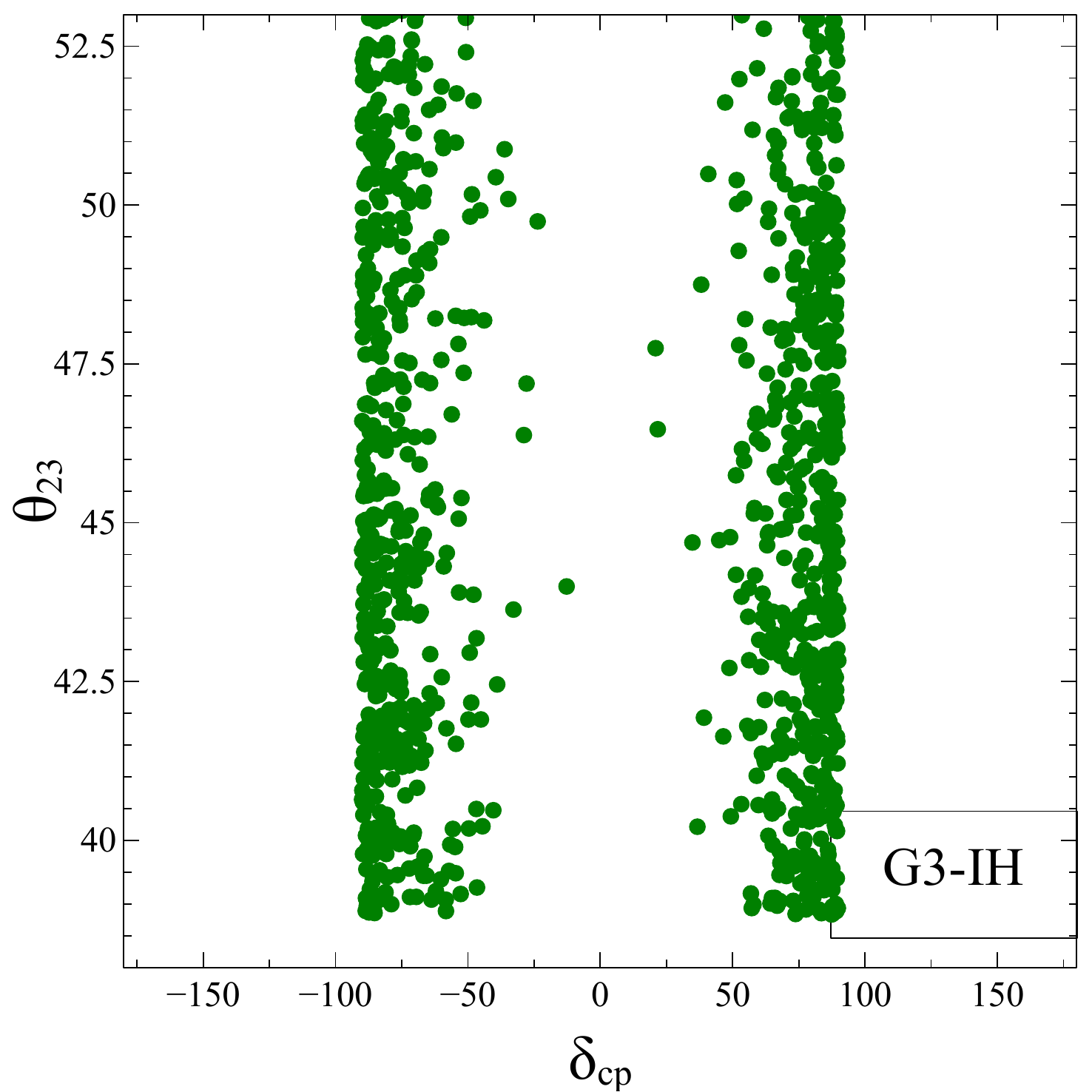} \\
\includegraphics[width=0.47\textwidth]{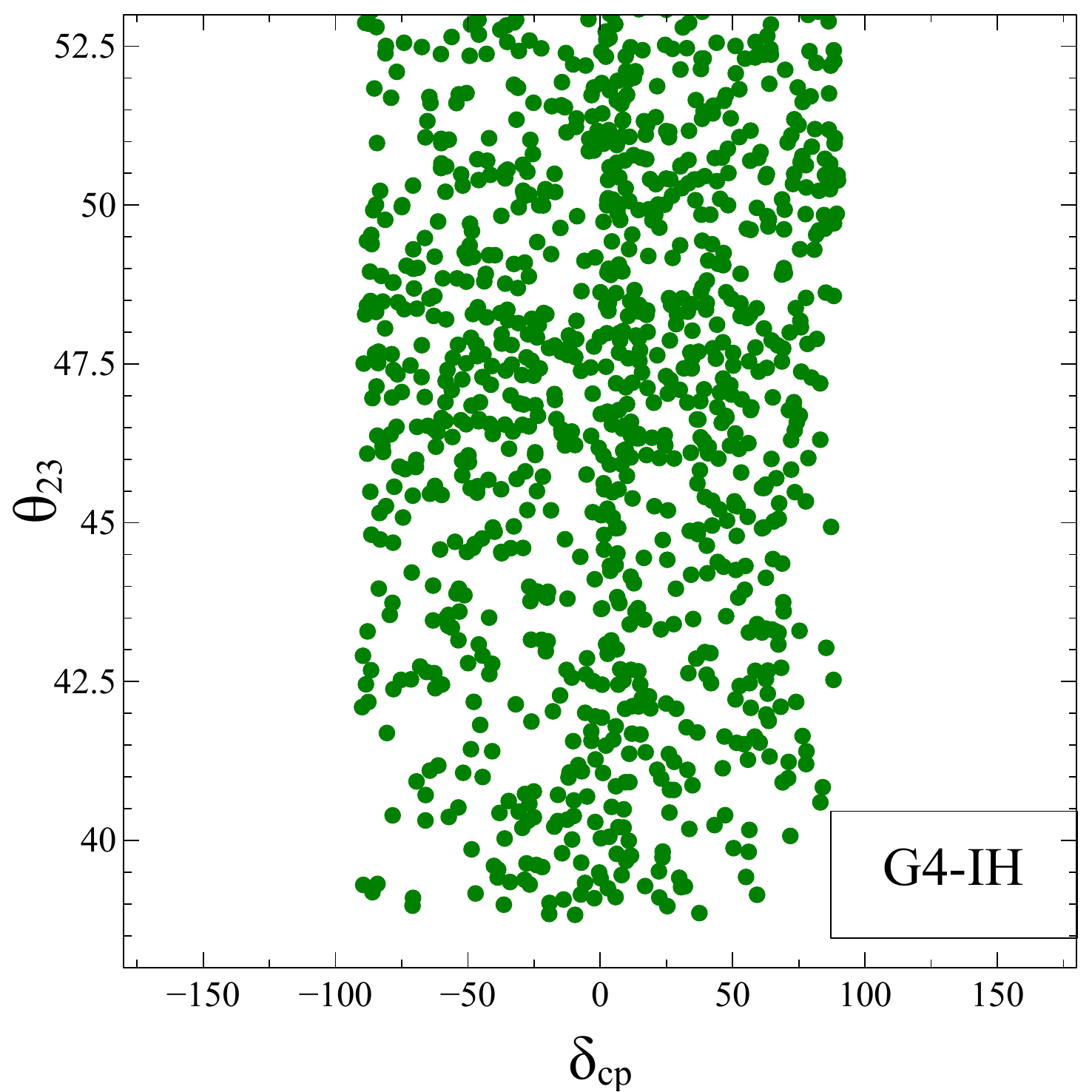}
\includegraphics[width=0.47\textwidth]{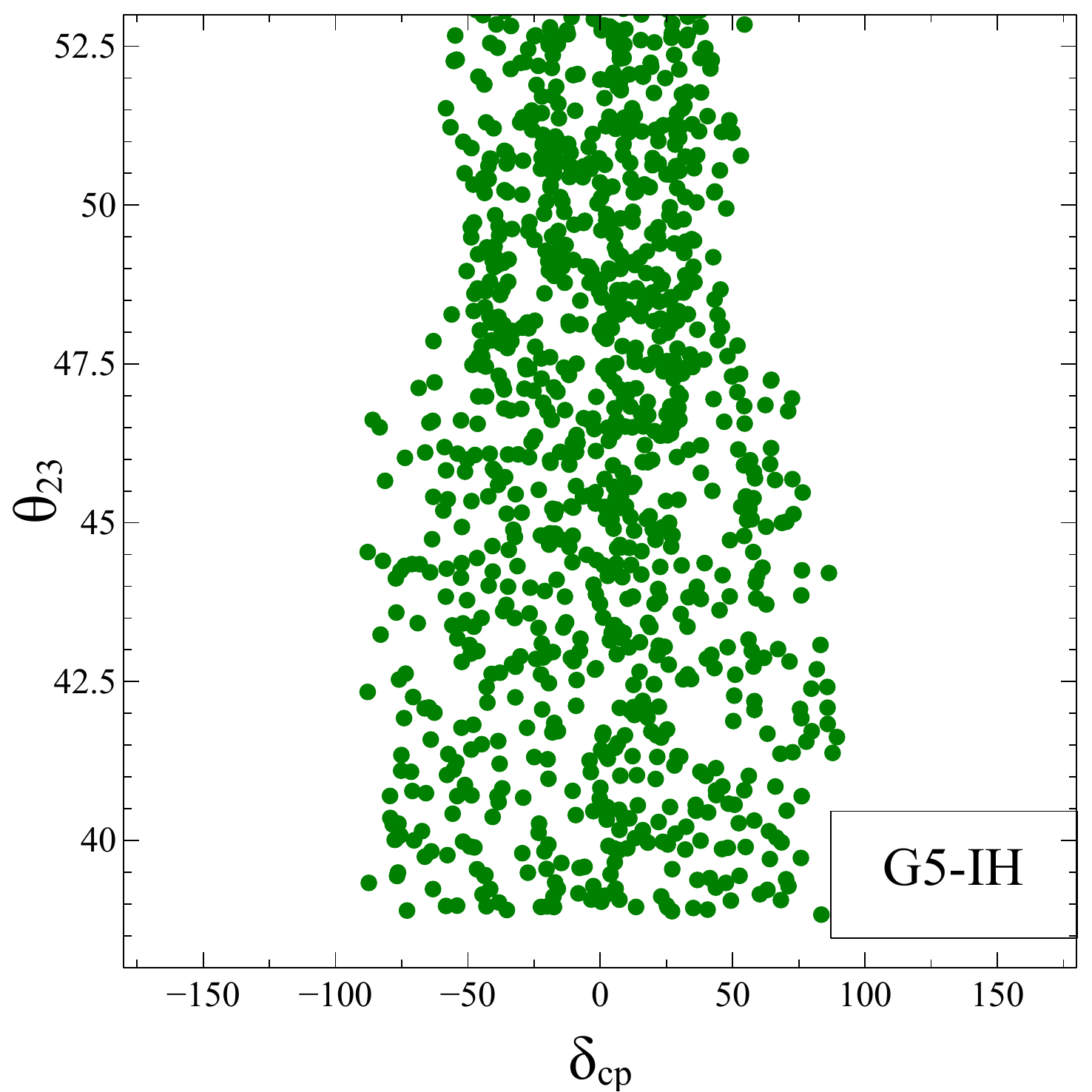}
\end{center}
\begin{center}
\caption{Neutrino parameters corresponding to one-zero textures for IH}
\label{fig4}
\end{center}
\end{figure*}
\begin{figure*}
\begin{center}
\includegraphics[width=0.5\textwidth]{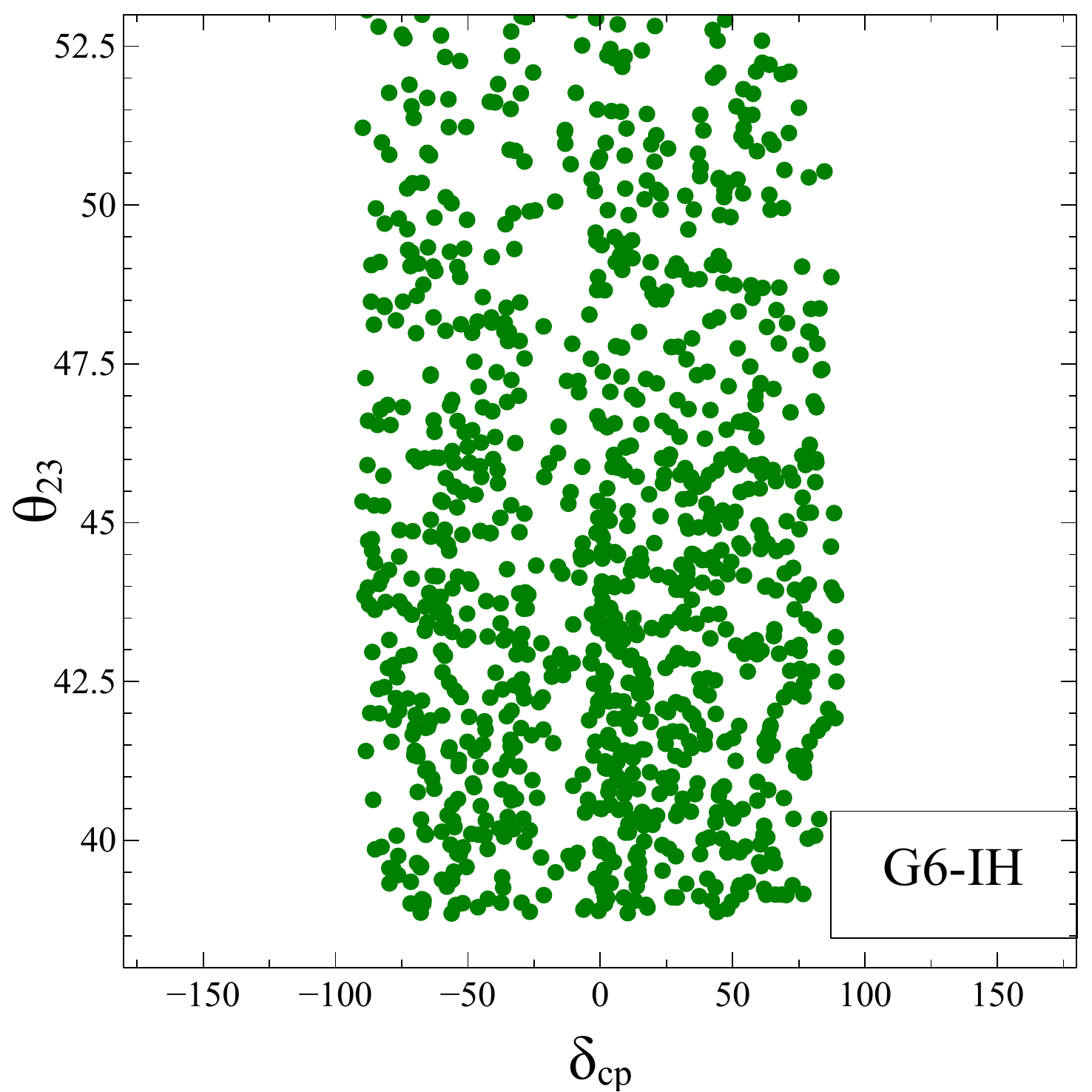}
\end{center}
\begin{center}
\caption{Neutrino parameters corresponding to one-zero texture $G_6$ for IH.}
\label{fig4a}
\end{center}
\end{figure*}
\begin{figure*}
\begin{center}
\includegraphics[width=0.5\textwidth]{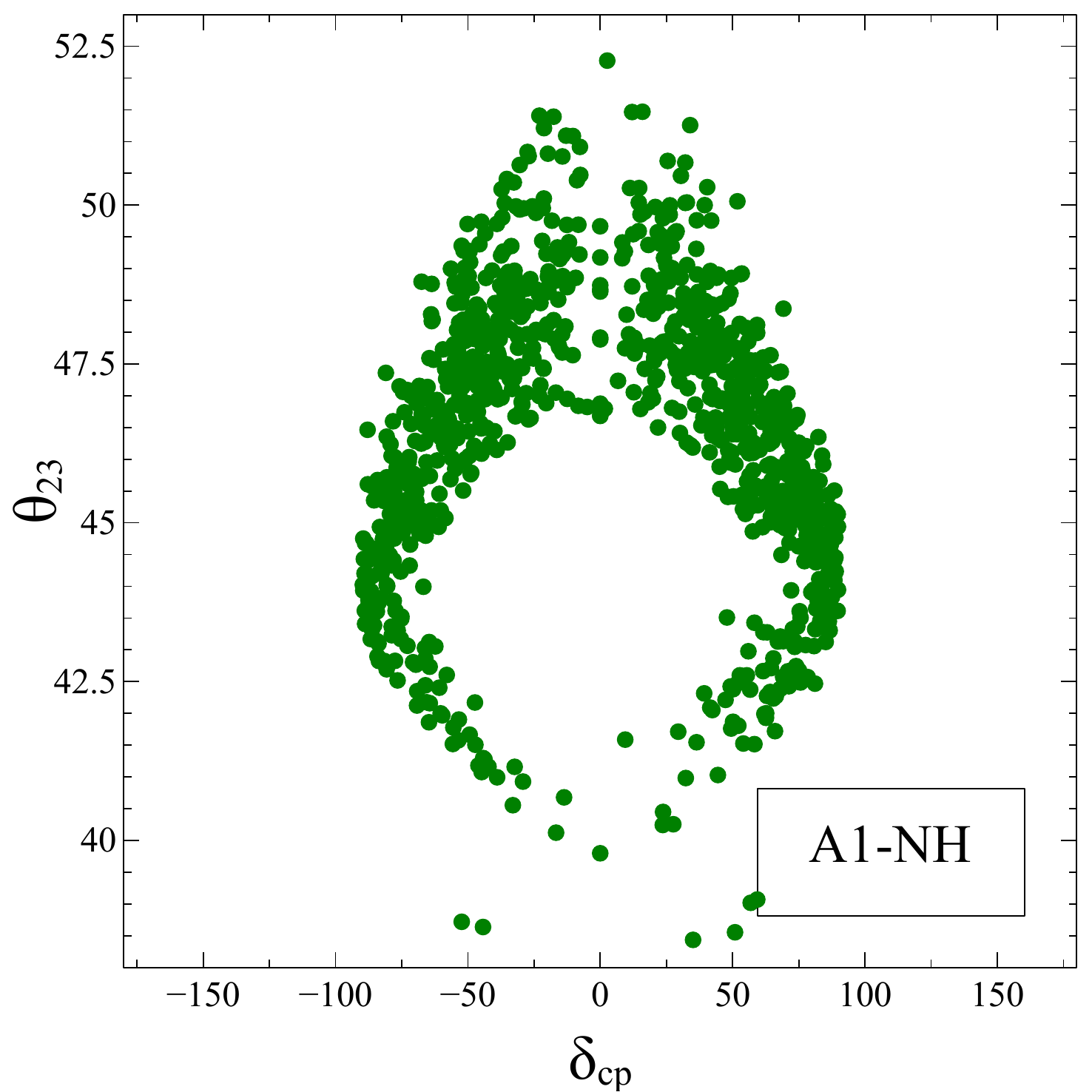}
\end{center}
\begin{center}
\caption{Neutrino parameters corresponding to two-zero texture $A_1$ for NH.}
\label{fig4b}
\end{center}
\end{figure*}

\begin{figure}[h]
\begin{center}
\includegraphics[width=0.47\textwidth]{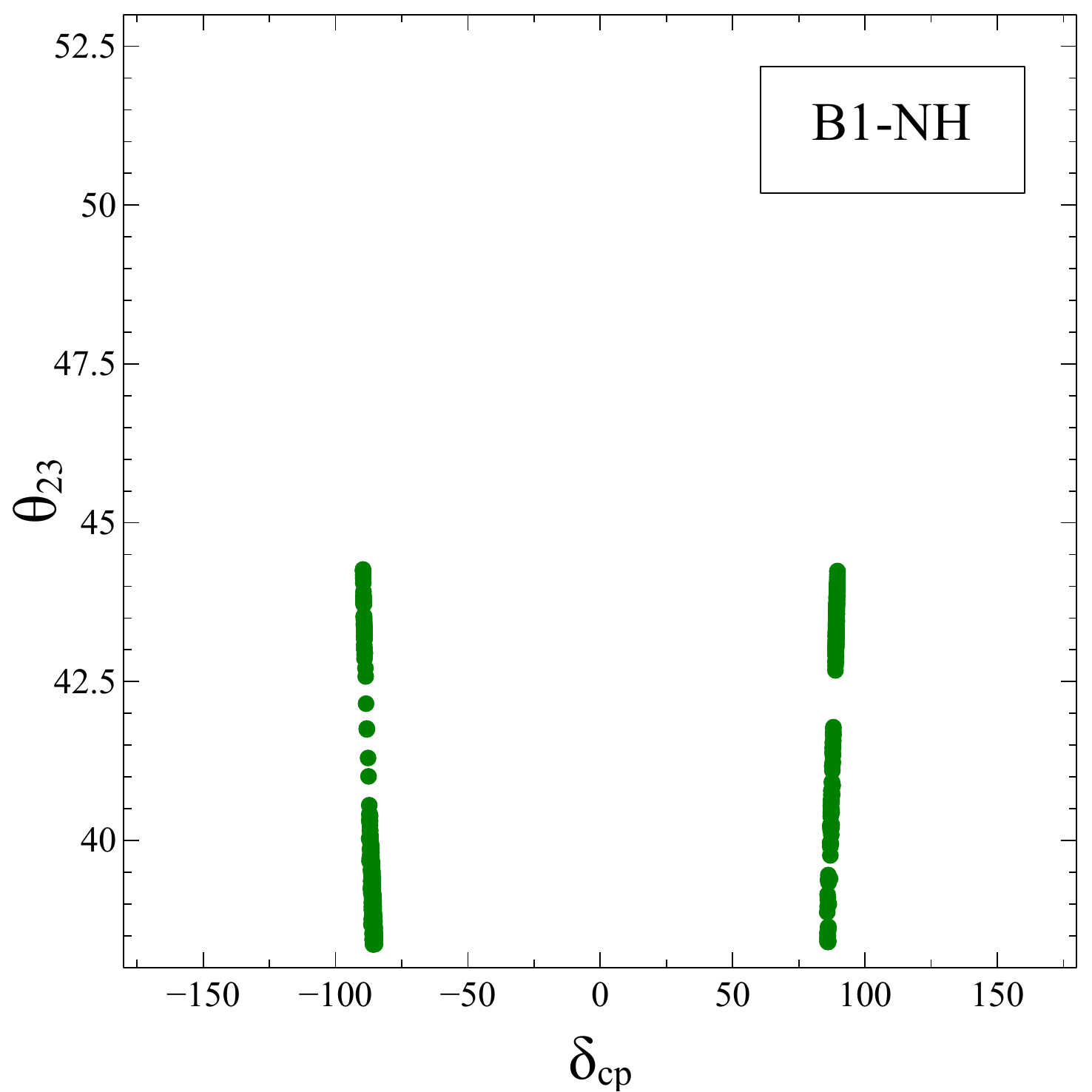}
\includegraphics[width=0.47\textwidth]{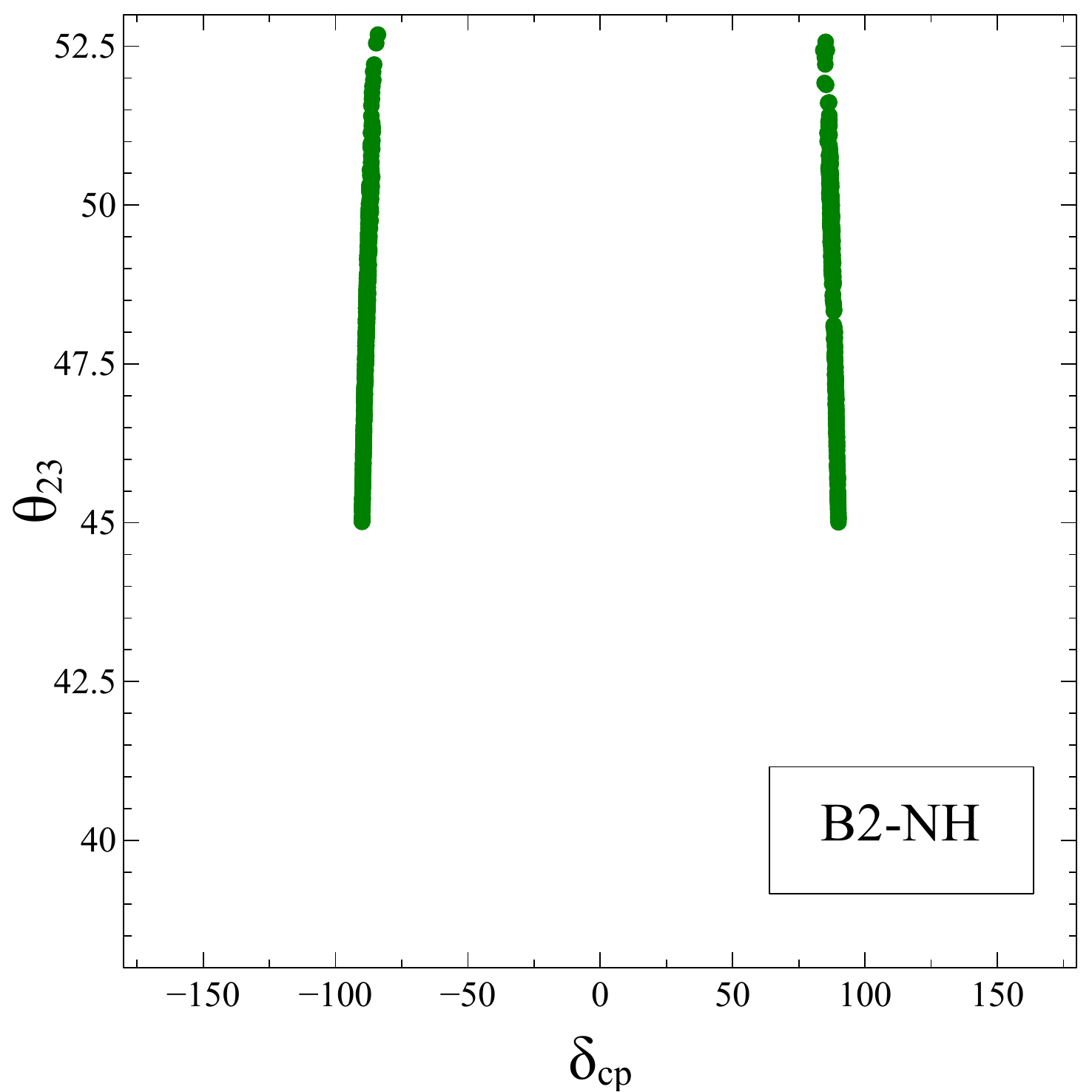} \\
\includegraphics[width=0.47\textwidth]{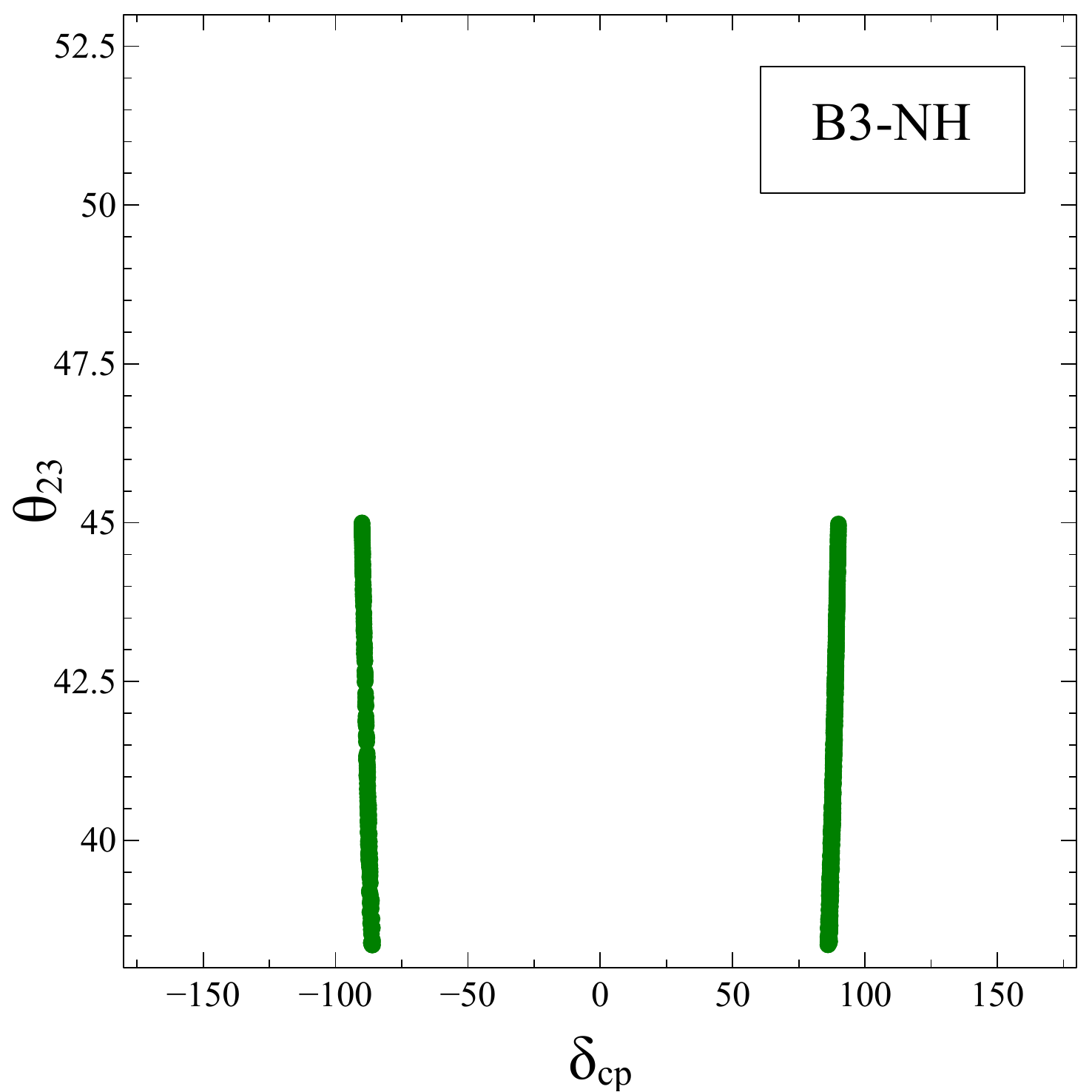}
\includegraphics[width=0.47\textwidth]{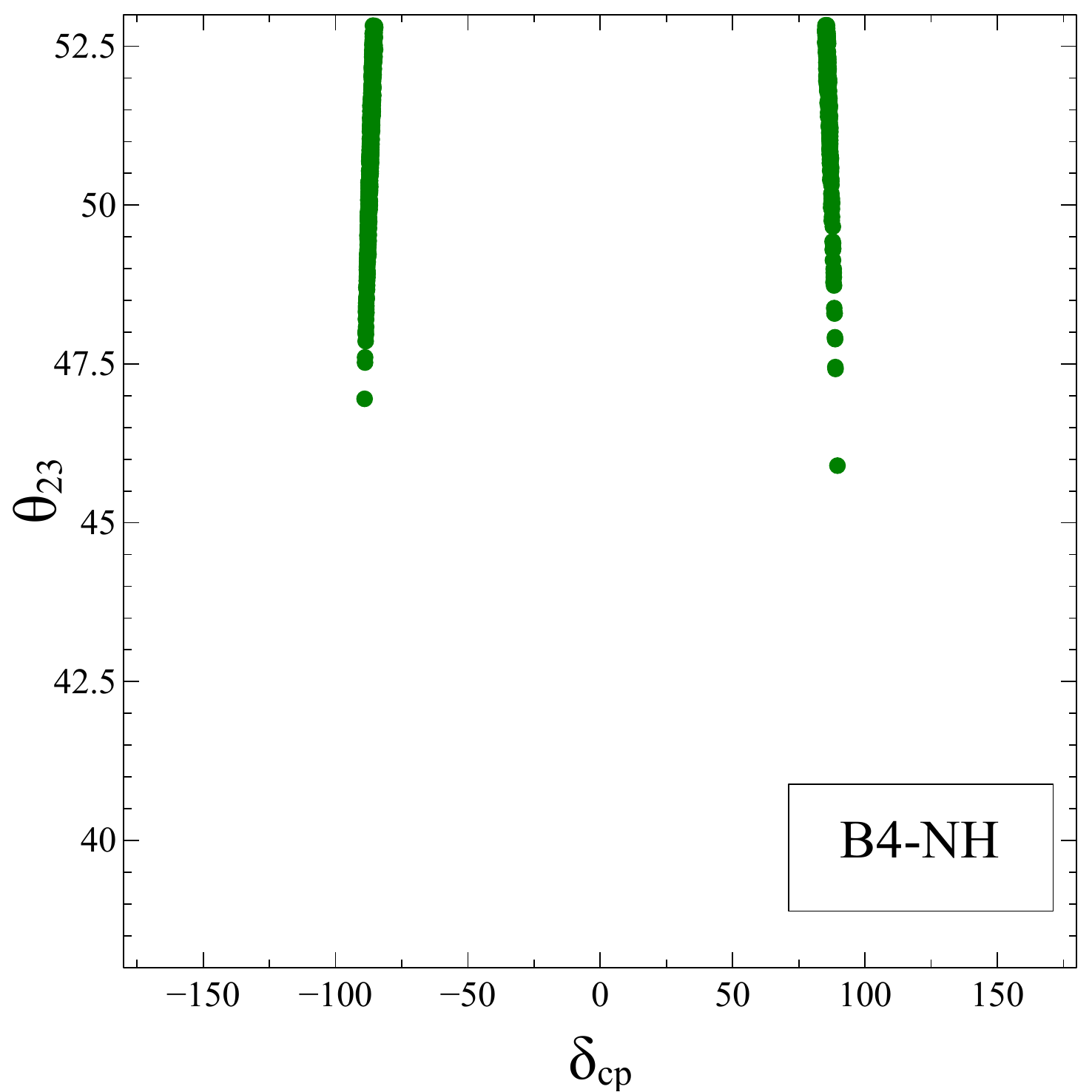}
\end{center}
\begin{center}
\caption{Neutrino parameters corresponding to two-zero textures for NH}
\label{fig1}
\end{center}
\end{figure}

\begin{figure*}
\begin{center}
\includegraphics[width=0.45\textwidth]{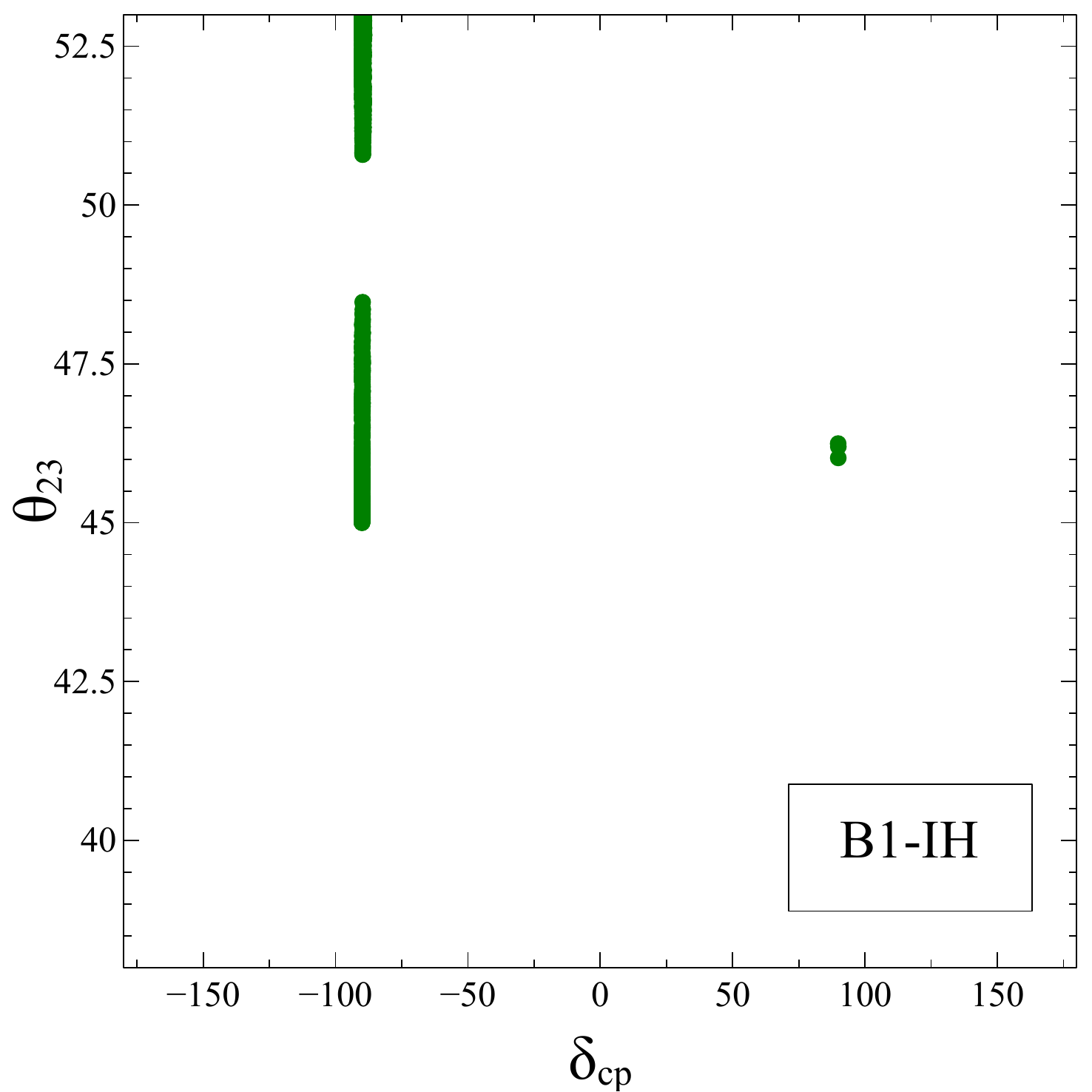}
\includegraphics[width=0.45\textwidth]{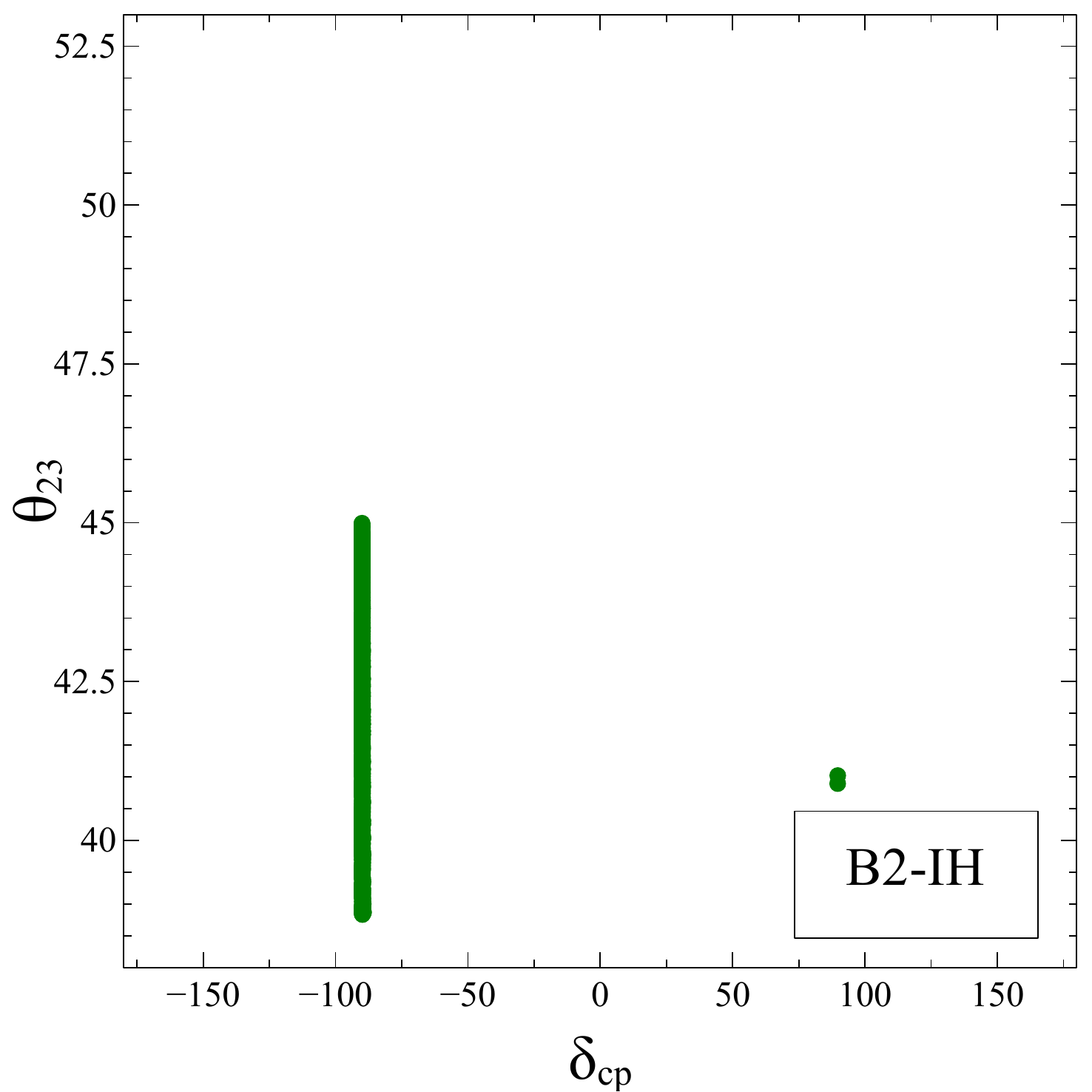} \\
\includegraphics[width=0.45\textwidth]{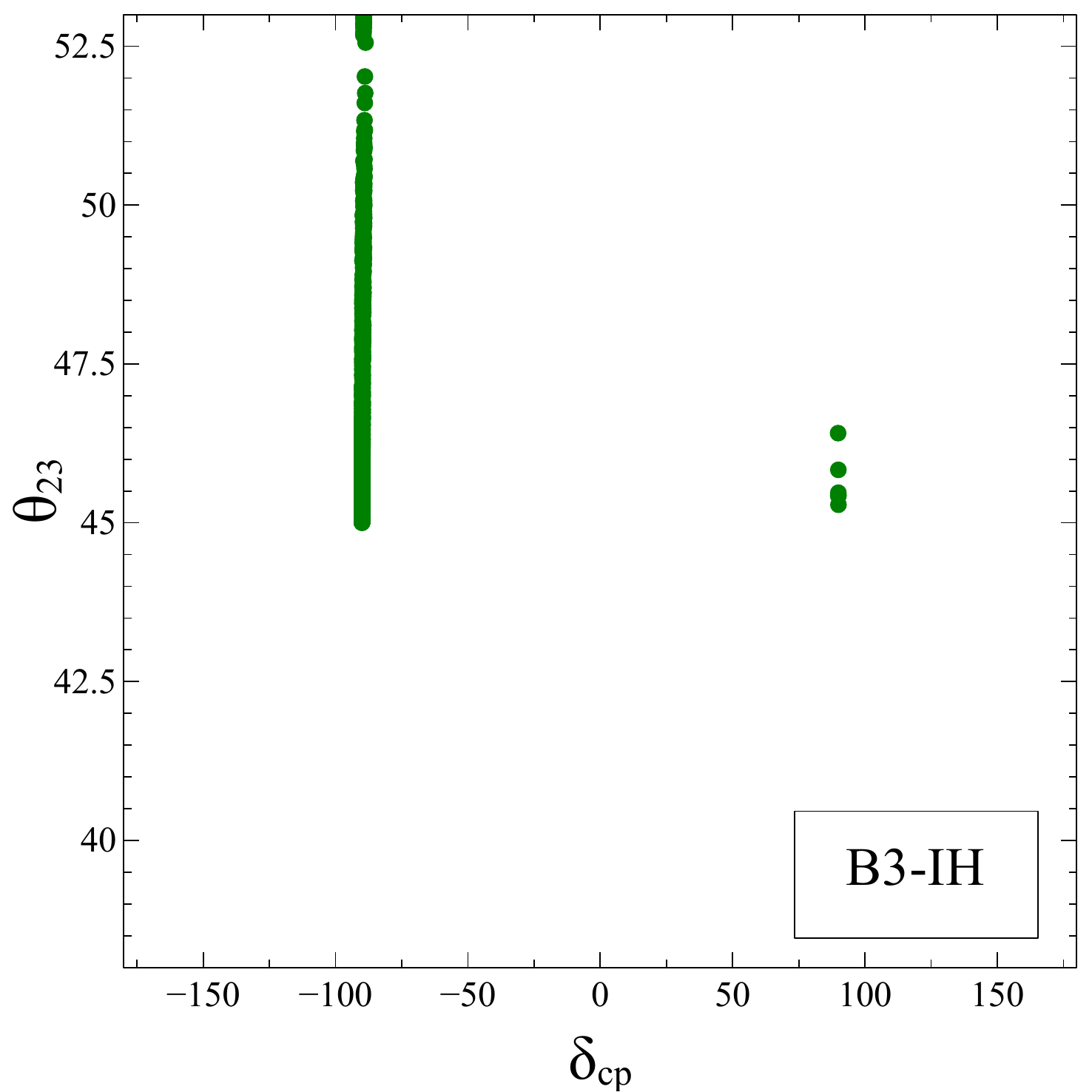}
\includegraphics[width=0.45\textwidth]{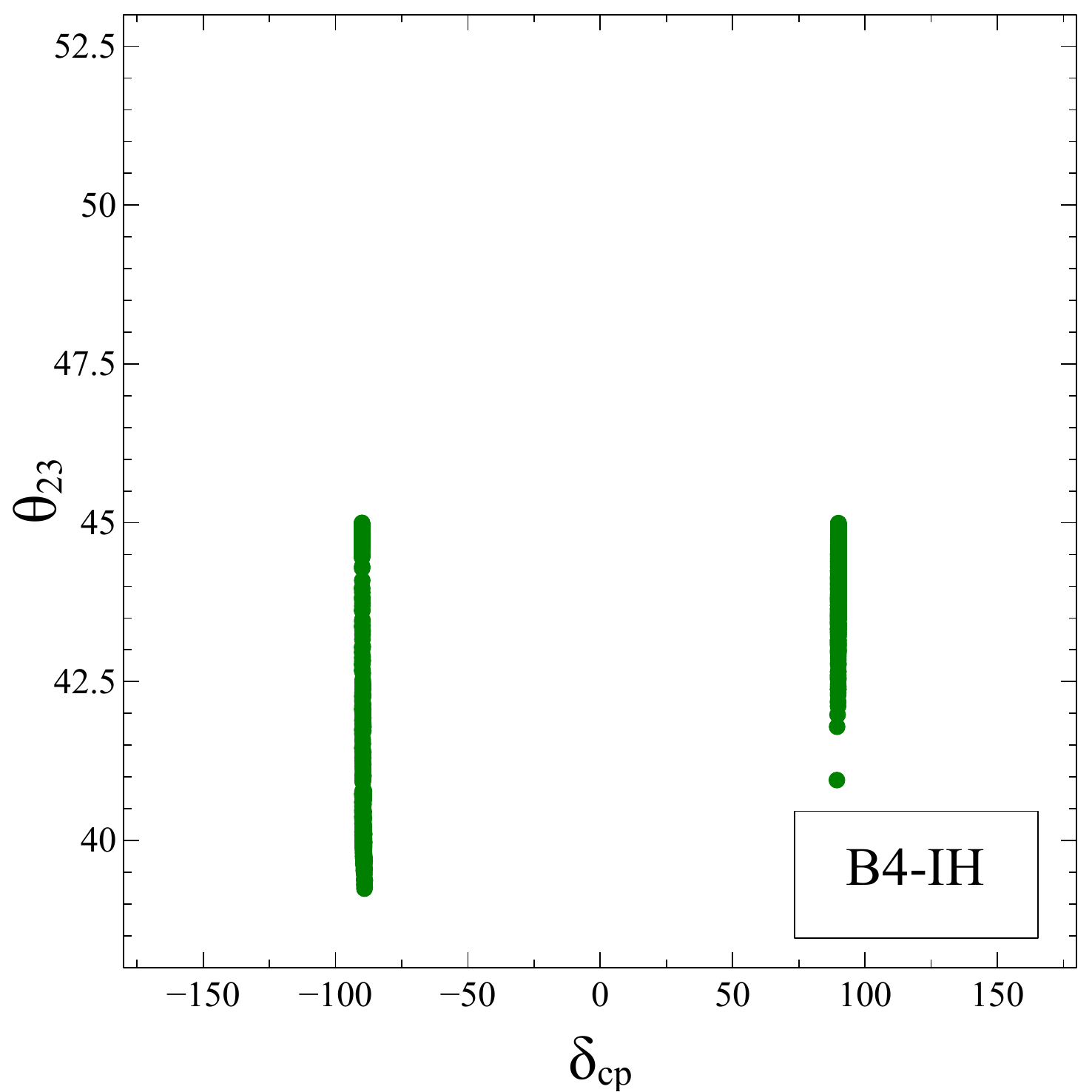}
\end{center}
\begin{center}
\caption{Neutrino parameters corresponding to two-zero textures for IH}
\label{fig2}
\end{center}
\end{figure*}
\begin{figure}
\begin{center}
\includegraphics[width=0.45\textwidth]{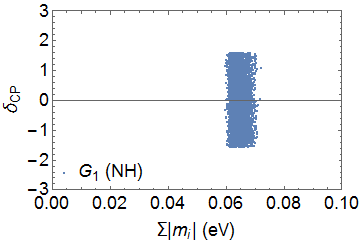}
\includegraphics[width=0.45\textwidth]{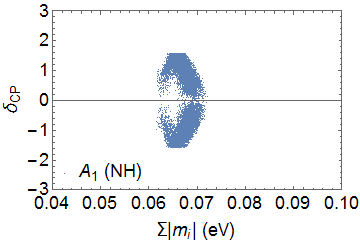}
\end{center}
\begin{center}
\caption{$\dcp$ with respect to the sum of absolute neutrino masses for the allowed one and two zero textures $G_1, A_1$ for which $M_{ee}=0$.}
\label{fig34new1}
\end{center}
\end{figure}
\begin{figure}
\begin{center}
\includegraphics[width=0.47\textwidth]{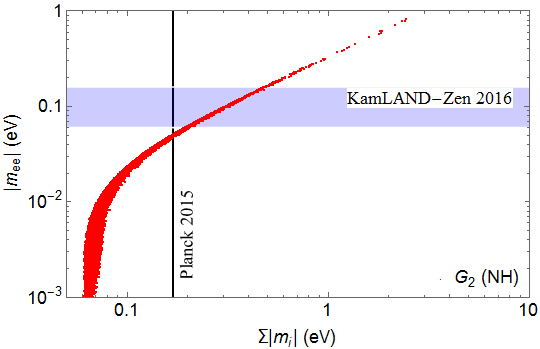}
\includegraphics[width=0.47\textwidth]{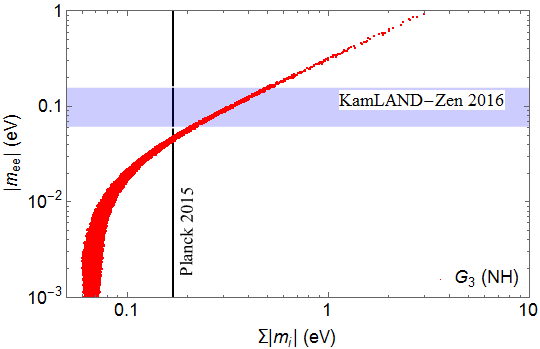}\\
\includegraphics[width=0.47\textwidth]{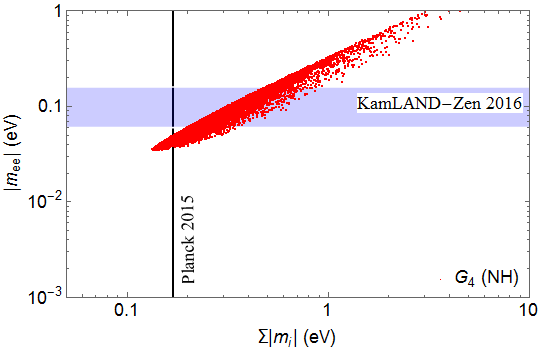}
\includegraphics[width=0.47\textwidth]{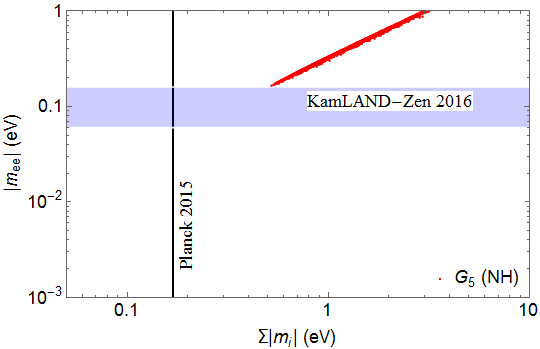}\\
\includegraphics[width=0.47\textwidth]{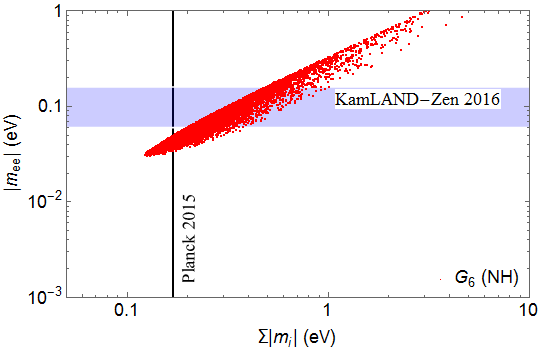}
\includegraphics[width=0.47\textwidth]{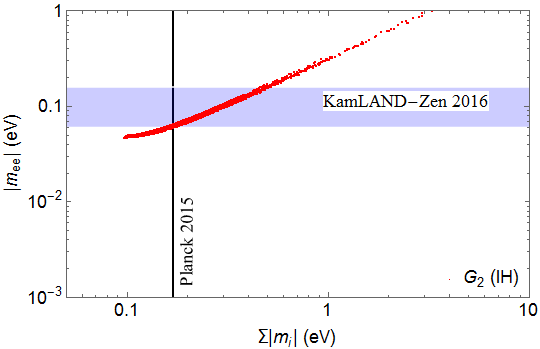} 
\end{center}
\begin{center}
\caption{The contribution to $0\nu\beta\beta$ for one zero texture mass matrices shown with respect to the sum of absolute neutrino masses. The blue horizontal band corresponds to the 2016 KamLAND-Zen bound $\lvert M_{ee} \rvert < 0.061-0.165$ eV. The vertical solid black line corresponds to the Planck 2015 bound $\sum_i \lvert m_i \rvert < 0.17$ eV.}
\label{fig34new2}
\end{center}
\end{figure}
\begin{figure}
\begin{center}
\includegraphics[width=0.47\textwidth]{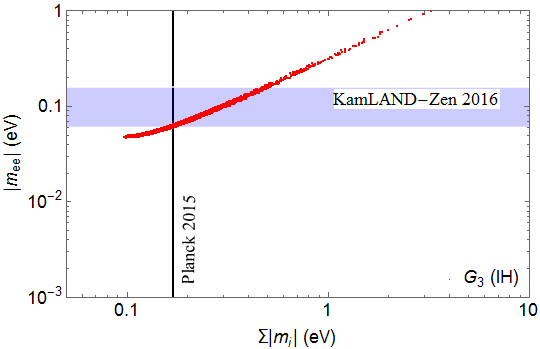} 
\includegraphics[width=0.47\textwidth]{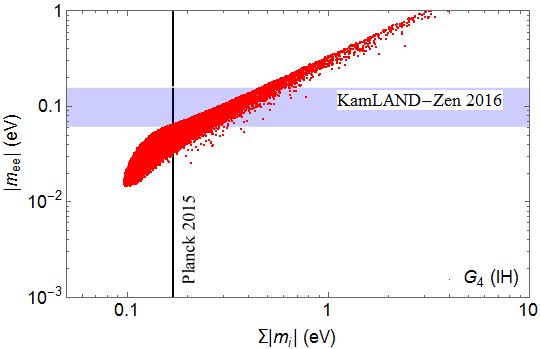}\\
\includegraphics[width=0.47\textwidth]{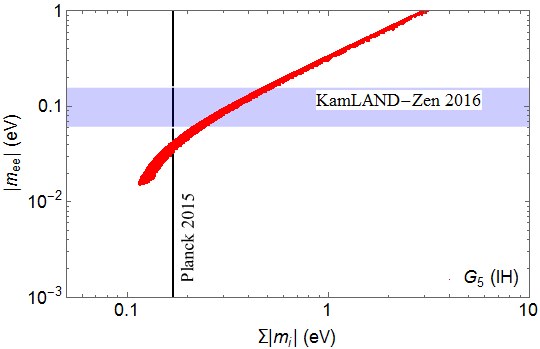}
\includegraphics[width=0.47\textwidth]{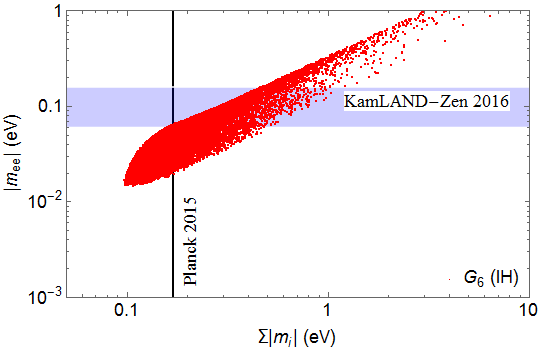}
\end{center}
\begin{center}
\caption{The contribution to $0\nu\beta\beta$ for one zero texture mass matrices shown with respect to the sum of absolute neutrino masses. The blue horizontal band corresponds to the 2016 KamLAND-Zen bound $\lvert M_{ee} \rvert < 0.061-0.165$ eV. The vertical solid black line corresponds to the Planck 2015 bound $\sum_i \lvert m_i \rvert < 0.17$ eV.}
\label{fig34new21}
\end{center}
\end{figure}

\begin{figure}
\begin{center}
\includegraphics[width=0.47\textwidth]{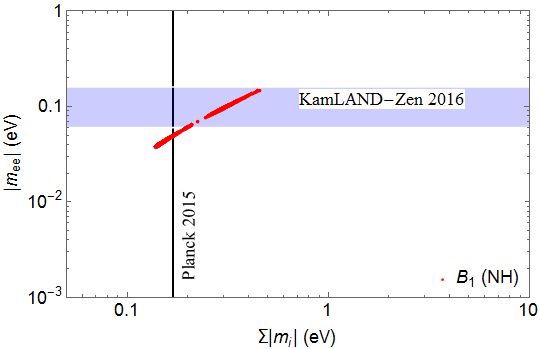}
\includegraphics[width=0.47\textwidth]{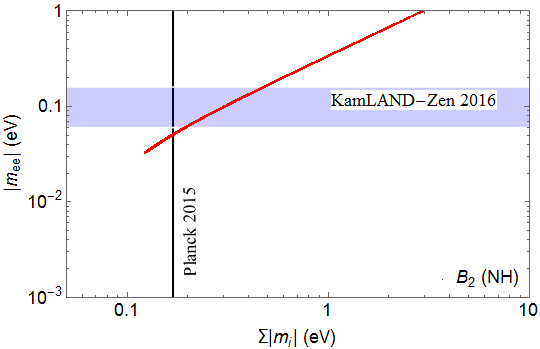}\\
\includegraphics[width=0.47\textwidth]{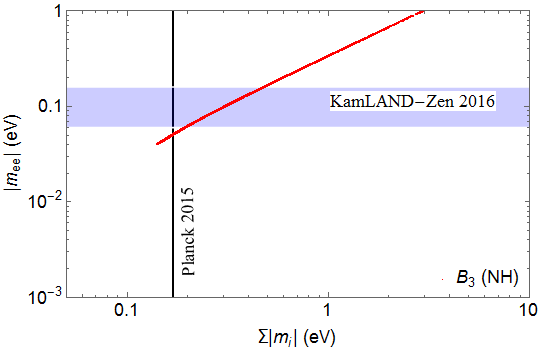}
\includegraphics[width=0.47\textwidth]{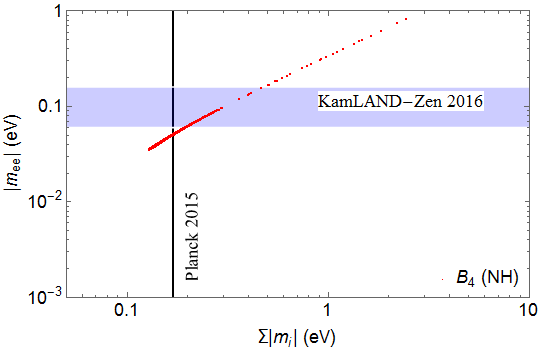}\\
\includegraphics[width=0.47\textwidth]{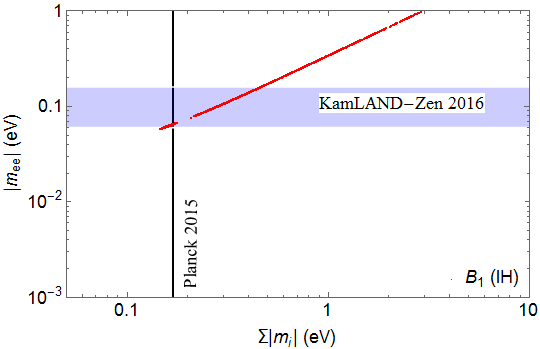} 
\includegraphics[width=0.47\textwidth]{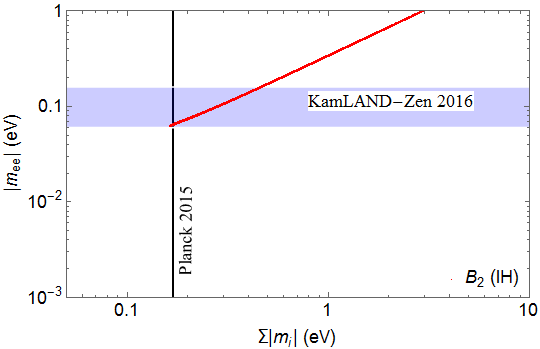}\\
\includegraphics[width=0.47\textwidth]{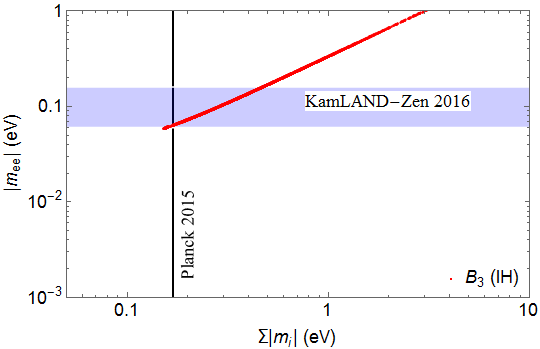}
\includegraphics[width=0.47\textwidth]{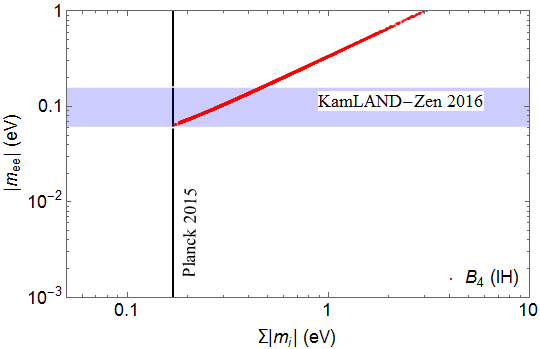}
\end{center}
\begin{center}
\caption{The contribution to $0\nu\beta\beta$ for two zero texture mass matrices shown with respect to the sum of absolute neutrino masses. The blue horizontal band corresponds to the 2016 KamLAND-Zen bound $\lvert M_{ee} \rvert < 0.061-0.165$ eV. The vertical solid black line corresponds to the Planck 2015 bound $\sum_i \lvert m_i \rvert < 0.17$ eV.}
\label{fig34new4}
\end{center}
\end{figure}
From the parametrisation of light neutrino mass matrix, it can be seen that there are nine parameters: three masses, three angles and three phases. Out of these five parameters namely, two mass squared differences and three mixing angles are measured in several experiments and their best fit values along with $3\sigma$ ranges are shown in table \ref{tab:data1}. It is worth noting that there still remains some ambiguity in determining the octant of $\theta_{23}$. Apart from this, the three CP phases and the lightest neutrino mass and hence the type of mass hierarchy are yet to be determined experimentally. Due to the predictive nature of texture zero mass matrices and upcoming neutrino experiment like DUNE having the potential to settle the issue of $\theta_{23}$ octant, Dirac CP phase $\delta_{\text{cp}}$ as well as mass hierarchy we first numerically evaluate the neutrino parameters for a particular texture zero mass matrix using the $3\sigma$ values of two angles $\theta_{12}, \theta_{13}$, two mass squared differences from the global fit reference \cite{schwetz16}. For each possible one-zero texture mass matrix, we have one complex or two real equations which can be solved numerically to evaluate $\theta_{23}, \delta_{\text{cp}}$ along with restricting other parameters. For the one-zero texture mass matrices, we solve the two real equations corresponding to the texture zero condition and determine the parameter space in terms neutrino parameters. While solving these equations, we vary the lightest neutrino mass in the range $10^{-6}-0.1$ eV and the Majorana CP phases in the range $-\pi < \alpha, \beta <\pi$. 

To use the bound on neutrino parameters from $0\nu \beta \beta$ experiments, we use the standard light neutrino contribution to this rare decay process. This rare decay is a lepton number violating process where a heavier nucleus decays into a lighter one and two electrons $(A, Z) \rightarrow (A, Z+2) + 2e^- $ without any (anti) neutrinos in the final state thereby violating lepton number by two units. For a review on $0\nu\beta \beta$, please refer to \cite{NDBDrev}. The amplitude of the light neutrino contribution to $0\nu \beta \beta$ can be written as
\begin{equation}
A_{\nu L L} \propto G^2_F \sum_i \frac{m_i U^2_{ei}}{p^2} 
\end{equation}
with $p$ being the average momentum exchange for the process. In the above expression, $m_i$ are the masses of light neutrinos for $i=1,2,3$. $G_F = 1.17 \times 10^{-5} \; \text{GeV}^{-2}$ is the Fermi coupling constant and $U=U_{\text{PMNS}}$ is the light neutrino mixing matrix. The corresponding half-life can be written as 
\begin{align}
\frac{1}{T^{0\nu}_{1/2}} = G^{0\nu}_{01} \lvert \mathcal{M}^{0\nu}_\nu  (\eta^L_{\nu}) \rvert^2
\end{align}
where $ \eta^L_{\nu} = \sum_i \frac{m_i U^2_{ei}}{m_e}=\frac{M_{ee}}{m_e}$. Here Here $m_e$ is the mass of electron, $\mathcal{M}^{0\nu}_\nu$ is the nuclear matrix element (NME) and $G^{0\nu}_{01}$ is the phase space factor. Since the ongoing experiments have not observed this rare decay process, they give a lower bound on its half-life or equivalently an upper bound on the amplitude. The latest data from the KamLAND-Zen experiment in 2016 have given a lower limit on $0\nu \beta \beta$ half-life of $T^{0\nu}_{1/2} > 1.07 \times 10^{26}$ yr at $90\%$ confidence level \cite{kamland2}. Using the commonly adopted values of NME, this lower bound on half-life corresponds to an upper bound on the effective Majorana neutrino mass in the range $\lvert M_{ee} \rvert < 0.061-0.165$ eV. More recently in 2017 , the GERDA collaboration has given a similar but slightly weaker lower bound on half-life as $T^{0\nu}_{1/2} > 5.3 \times 10^{25}$ yr at $90\%$ confidence level \cite{GERDA2} which corresponds to an upper bound $\lvert M_{ee} \rvert < 0.15-0.33$ eV. We use the most conservative bound from the KamLAND-Zen collaboration to constrain the texture zero mass matrices in our analysis. Along with this, we also use the cosmology bound on the sum of absolute neutrino masses $\sum_i \lvert m_i \rvert < 0.17$ eV from the Planck collaboration results \cite{Planck15} that appeared in 2015.

After numerically evaluating the neutrino oscillation parameters for the texture zero mass matrices, we move on to studying the capabilities of DUNE to exclude these textures once it starts taking data. For this, we have used the present best fit values of the standard 3$\nu$ oscillation parameters as the `true values' or `data'. The best fit values of the oscillation parameters are taken from \cite{schwetz16}. To test a particular texture, the values of the neutrino oscillation parameters from the texture zero conditions are used as `fit' values and then $\chi^2$ is calculated between `data' and `fit'. We define $\chi^2$ (statistical) as: 
\begin{equation}
\chi^2 = \sum_{i}\sum_{j=1}^{2}\frac{[N^{true}_{i,j} - N^{test}_{i,j}]^2}{N^{true}_{i,j}},
\end{equation}
where $N^{true}_{i,j}$ and $N^{test}_{i,j}$ are the event rates that correspond to `data' and `fit'. Here, index $i$ corresponds to the number of bins and $j$ corresponds to the type of neutrinos i.e. $ j=1$ for neutrinos and $ j=2$ for antineutrinos. We have considered 39 bins for DUNE in the energy range $0.5-10$ $\rm GeV$ having bin width of $250$ $\rm MeV$. The systematics uncertainties like signal and background related uncertainties, calibration errors are introduced in the $\chi^2$ by pull method and they are allowed to deviate from their
standard values in the calculation of the expected rate in the $\{i,j\}$-th bins $N^{test}_{i,j}$. If
the $\rm k^{th}$ input gets deviated from its standard value by $\sigma_{k}\xi_{k}$, then our final modified $\chi^2$ will be:
$$\chi^2(\xi_{k})=\sum_{i}\sum_{j=1}^{2}\frac{[N^{test}_{i,j}(std)+\sum_{k=1}^{\rm npull} c_{i,j}^{k}\xi_{k}-N^{true}_{i,j}]}{N^{true}_{i,j}}+\sum_{k=1}^{\rm npull} \xi^2_{k}$$ and hence,
$$\chi^2_{\rm pull} = \rm min_{\xi_{k}}[\chi^2(\xi_{k})]$$
Here, $N^{test}_{i,j}(std)$ is the events corresponding to `fit' in $\{i,j\}$-th bin computed with the standard values of the inputs, npull is the no of uncertainties, $\sigma_{k}$ is the uncertainties associated with the $\rm k^{th}$ inputs. $\xi_{k}$ are the pull variables or nuisance parameters and they are measure of how $\rm k^{th}$ inputs deviates from its standard value. They describe how the event rates
depends on the various sources of systematical errors. In the expression, $c^k_{i,j}$ is the change in the expected event spectra when the $k^{th}$ input changes by $\sigma_{k}$. Nuisance systematics parameters are given in table \ref{Table2}. Since, we are testing each textures at DUNE and hence all the `fit values' of the oscillation parameters are as predicted by the texture zero conditions. We have not considered any priors on them in this analysis. Variation of all the six neutrino oscillation parameters are taken into account in `fit' according to texture zero conditions.

\section{Results and Discussions}
\label{sec:results}
\subsection{Allowed Texture Zero Mass Matrices}
In this subsection, we discuss the results we obtain for the first part of our analysis that is, constraining the texture zero mass matrices from the latest experimental bounds on neutrino oscillation parameters, sum of absolute neutrino masses and $0\nu\beta\beta$ half-life. 

For one zero texture mass matrices, we numerically solve the two real texture zero equations or one complex texture zero condition. We solve it for two parameters while randomly varying other parameters in their allowed $3\sigma$ range. The lightest neutrino mass is varied in the range $10^{-6}-0.1$ eV. For the texture zero mass matrices allowed by neutrino oscillation data alone, we show the corresponding regions of parameter space in $\theta_{23}-\dcp$ plane in figure \ref{fig3}, \ref{fig3a}, \ref{fig4}, \ref{fig4a} for both NH and IH. It should be noted that due to only two real constraint equations and more unknown neutrino parameters, the one-zero textures are not very predictive. This is obvious from the broad ranges of values allowed in the $(\theta_{23}, \dcp)$ plane. We find that the latest neutrino oscillation data allow all the one zero texture mass matrices except $G_1$ with IH.

Similarly, we numerically solve the two complex or four real equations for the two-zero texture mass matrices $A_{1,2}, B_{1,2,3,4}$ and extract the resulting parameter space. In this case also, we vary the lightest neutrino mass in the range $10^{-6}-0.1$ eV and solve the four real equations numerically in order to evaluate $\theta_{23}, \delta_{\text{cp}}$ along with two Majorana phases. They are shown in figure \ref{fig4b}, \ref{fig1}, \ref{fig2} for both NH and IH. Since there are four real constraint equations, two zero texture mass matrices are much more predictive as can be seen from the narrow range of parameters allowed from the texture zero conditions. We find that the use of the latest $3\sigma$ global fit values of solar and reactor mixing angles as well as the two mass squared differences allow only five two-zero textures with NH and four two-zero textures with IH. This is in contrast with the allowed two-zero texture matrices shown in earlier works \cite{twozero7} as we are using more precise values of neutrino oscillation parameters from the latest global fit data. The allowed textures we have got from fitting neutrino oscillation data are same as the ones obtained in the more recent work \cite{twozero9}. This relatively recent work \cite{twozero9} however did not include the textures $A_{1,2}$ in their analysis as they correspond to vanishing amplitude for $0\nu \beta \beta$. Here we include them and found the $A_1$ texture to be allowed in case of normal hierarchy. It should be noted that for all these plots in figure \ref{fig3}, \ref{fig3a}, \ref{fig4}, \ref{fig4a}, \ref{fig4b}, \ref{fig1}, \ref{fig2} we have not imposed the constraints from the Planck and KamLAND-Zen experiments on sum of absolute neutrino masses and effective light neutrino mass respectively. At this stage the viability of the Majorana texture zero mass matrices is studied only in the light of latest neutrino oscillation data.

After finding the allowed texture zero mass matrices from the requirement of satisfying the latest neutrino oscillation data, we further check their viabilities in the light of cosmology and $0\nu \beta \beta$ bounds mentioned before. Since the textures $G_1, A_1$ predict vanishing $0\nu \beta \beta$ amplitude, we show $\dcp$ versus sum of absolute neutrino masses for these two textures in figure \ref{fig34new1}. We find that all the points allowed from neutrino data also obey the Planck 2015 upper bound on $\sum_i \lvert m_i \rvert$. The contributions to $0\nu \beta \beta$ are shown in figure \ref{fig34new2}, \ref{fig34new21}, \ref{fig34new4}. It is interesting to see from figure \ref{fig34new2}, \ref{fig34new21} and \ref{fig34new4} that most of the texture zero mass matrices can saturate the experimental bounds on $0\nu \beta \beta$ as well as the sum of absolute neutrino masses. The two zero texture mass matrix $A_1$ however predicts $\sum_i \lvert m_i \rvert $ far below the Planck 2015 bound. We find that one texture zero mass matrix belonging to previously allowed one zero texture mass matrices become disallowed after incorporating these additional constraints. This is namely, the $G_5$ texture with normal hierarchy as seen from figure \ref{fig34new2}. If we consider the most conservative KamLAND-Zen bound corresponding to the lower most value of $\lvert M_{ee} \rvert$ in the blue band shown in figure \ref{fig34new2}, \ref{fig34new21} and \ref{fig34new4} then two texture zero mass matrices belonging to previously allowed two zero texture mass matrices become disallowed. They are namely, $B_2$ and $B_4$ with inverted hierarchy as seen from figure \ref{fig34new4}. In fact $B_4$ texture with inverted hierarchy is disallowed from the Planck 2015 bound as seen from the same figure \ref{fig34new4}. Thus, all the two zero texture mass matrices with NH studied in this work are allowed except $A_2$ and three such texture zero matrices with IH namely $B_{1,2,3}$ are allowed. In our subsequent analysis from the DUNE point of view, we will consider these allowed texture zero mass matrices only.

We finally summarise our results from first part of our analysis in the table \ref{tablepart1} and table  \ref{tablepart2} for one zero and two zero texture mass matrices respectively.
\begin{table}[!h]
\begin{tabular}{|c|c|c|c|}
 \hline
Patterns     &Neutrino Data IH (NH) & $0\nu \beta \beta$ Bound IH (NH)& Planck Bound IH (NH)\\
        \hline \hline
\mbox{$G_1$}     & $\times$($\checkmark$)    & $\times$($\checkmark$)        &$\times$($\checkmark$) \\
\mbox{$G_2$}     & $\checkmark$($\checkmark$)    & $\checkmark$($\checkmark$)        & $\checkmark$($\checkmark$)\\ 
\mbox{$G_3$}     & $\checkmark$($\checkmark$)        & $\checkmark$($\checkmark$)        &$\checkmark$($\checkmark$)\\
\mbox{$G_4$}     & $\checkmark$($\checkmark$)    & $\checkmark$($\checkmark$)    &$\checkmark$($\checkmark$)\\
\mbox{$G_5$}     & $\checkmark$($\checkmark$)            & $\checkmark$($\times$)    &$\checkmark$($\times$)\\
\mbox{$G_6$}     & $\checkmark$($\checkmark$)       & $\checkmark$($\checkmark$)    &$\checkmark$($\checkmark$)\\
        \hline
\end{tabular}
 
\caption{Summary of results for one zero texture with inverted and normal hierarchy. 
The symbol $\checkmark$ ($\times$) is used when the particular texture zero mass matrix is (not) consistent with the respective experimental bound.}
\label{tablepart1}
\end{table}

\begin{table}[!h]
\begin{tabular}{|c|c|c|c|}
 \hline
Patterns     &Neutrino Data IH (NH) & $0\nu \beta \beta$ Bound IH (NH)& Planck Bound IH (NH)\\
        \hline \hline
\mbox{$A_1$}     & $\times$($\checkmark$)    & $\times$($\checkmark$)         &$\times$($\checkmark$) \\
\mbox{$A_2$}     & $\times$($\times$)    & $\times$($\times$)        & $\times$($\times$)\\ 
\mbox{$B_1$}     & $\checkmark$($\checkmark$)        & $\checkmark$($\checkmark$)        &$\checkmark$($\checkmark$)\\
\mbox{$B_2$}     & $\checkmark$($\checkmark$)    & $\textcolor{red}{\checkmark}$($\checkmark$)    &$\checkmark$($\checkmark$)\\
\mbox{$B_3$}     & $\checkmark$($\checkmark$)            & $\checkmark$($\checkmark$)    &$\checkmark$($\checkmark$)\\
\mbox{$B_4$}     & $\checkmark$($\checkmark$)       & $\textcolor{red}{\checkmark}$($\checkmark$)    &$\times$($\checkmark$)\\
        \hline
\end{tabular}
 
\caption{Summary of results for two zero texture with inverted and normal hierarchy. The symbol $\checkmark$ ($\times$) is used when the particular texture zero mass matrix is (not) consistent with the respective experimental bound. The $\checkmark$ in red colour means that the particular texture is only marginally allowed due to the broad band of upper bound on $M_{ee}$ from $0\nu \beta \beta$ experiment.}
\label{tablepart2}
\end{table}

\subsection{Probe of Allowed Textures at DUNE}
In this section, we present our results. We have tested two possible texture zero mass matrices at DUNE namely, one-zero and two-zero. Each set of textures are tested for both the hierarchies.

\begin{figure*}
\begin{center}
\includegraphics[width=0.47\textwidth]{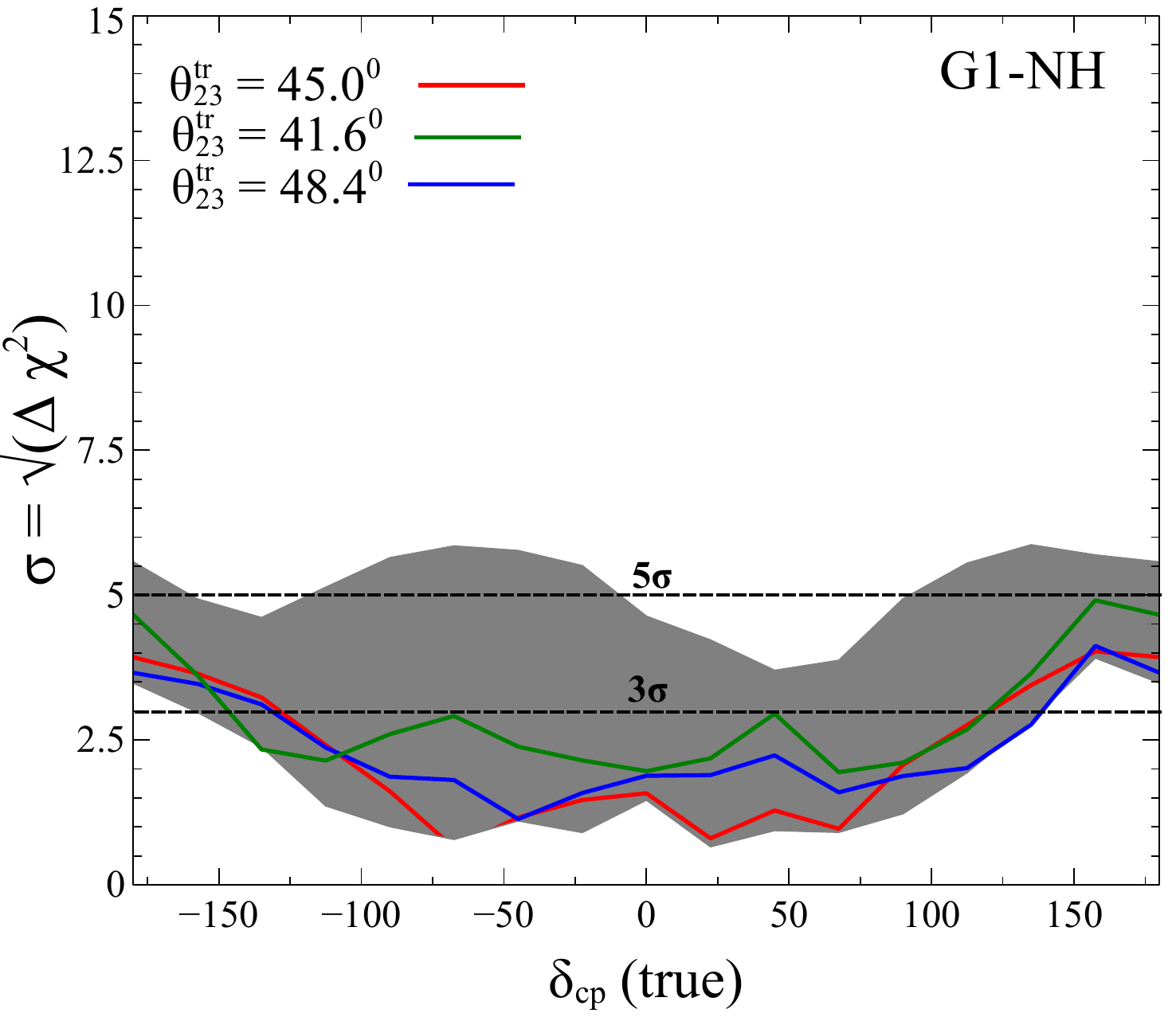}
\includegraphics[width=0.47\textwidth]{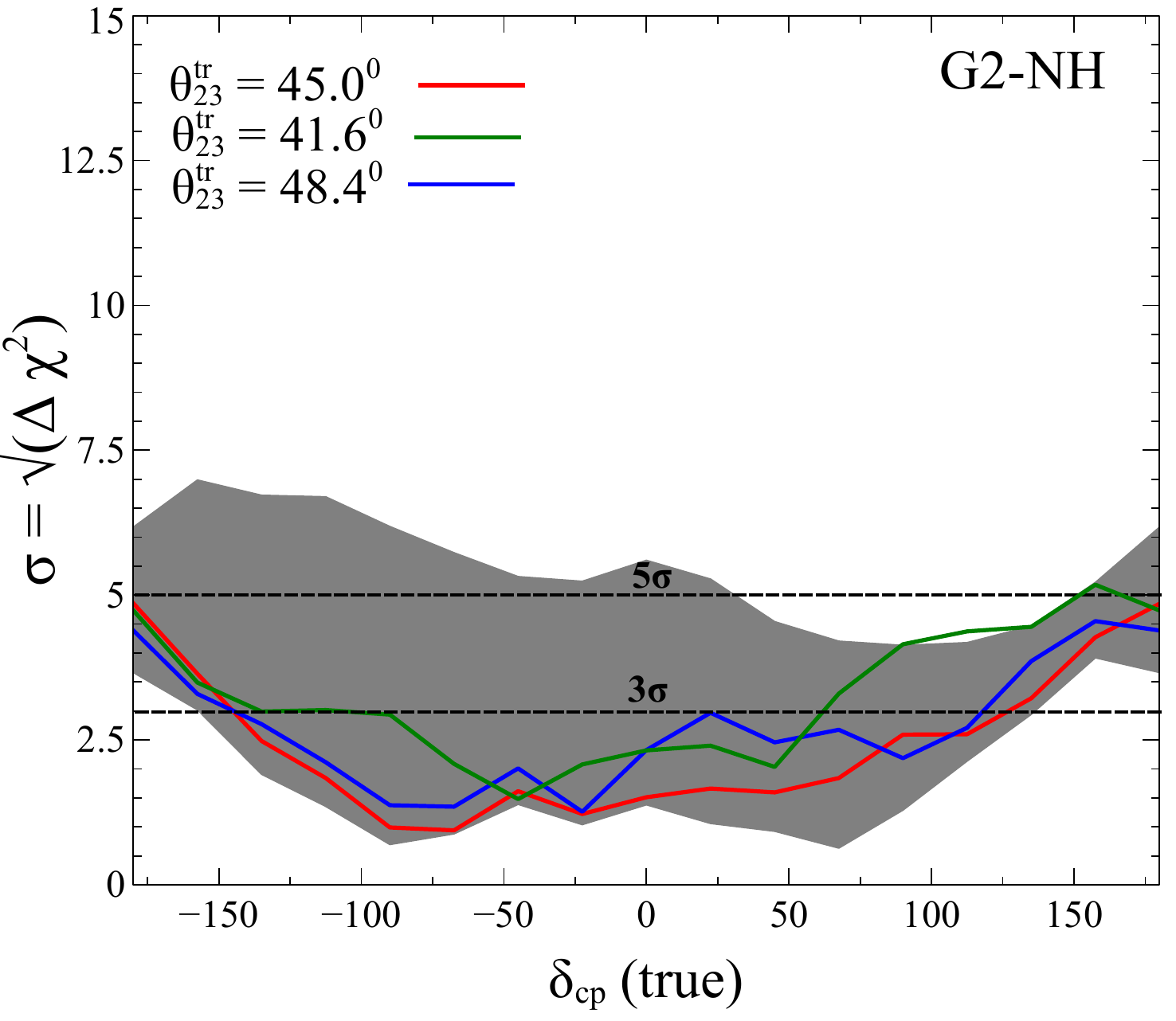}
\includegraphics[width=0.47\textwidth]{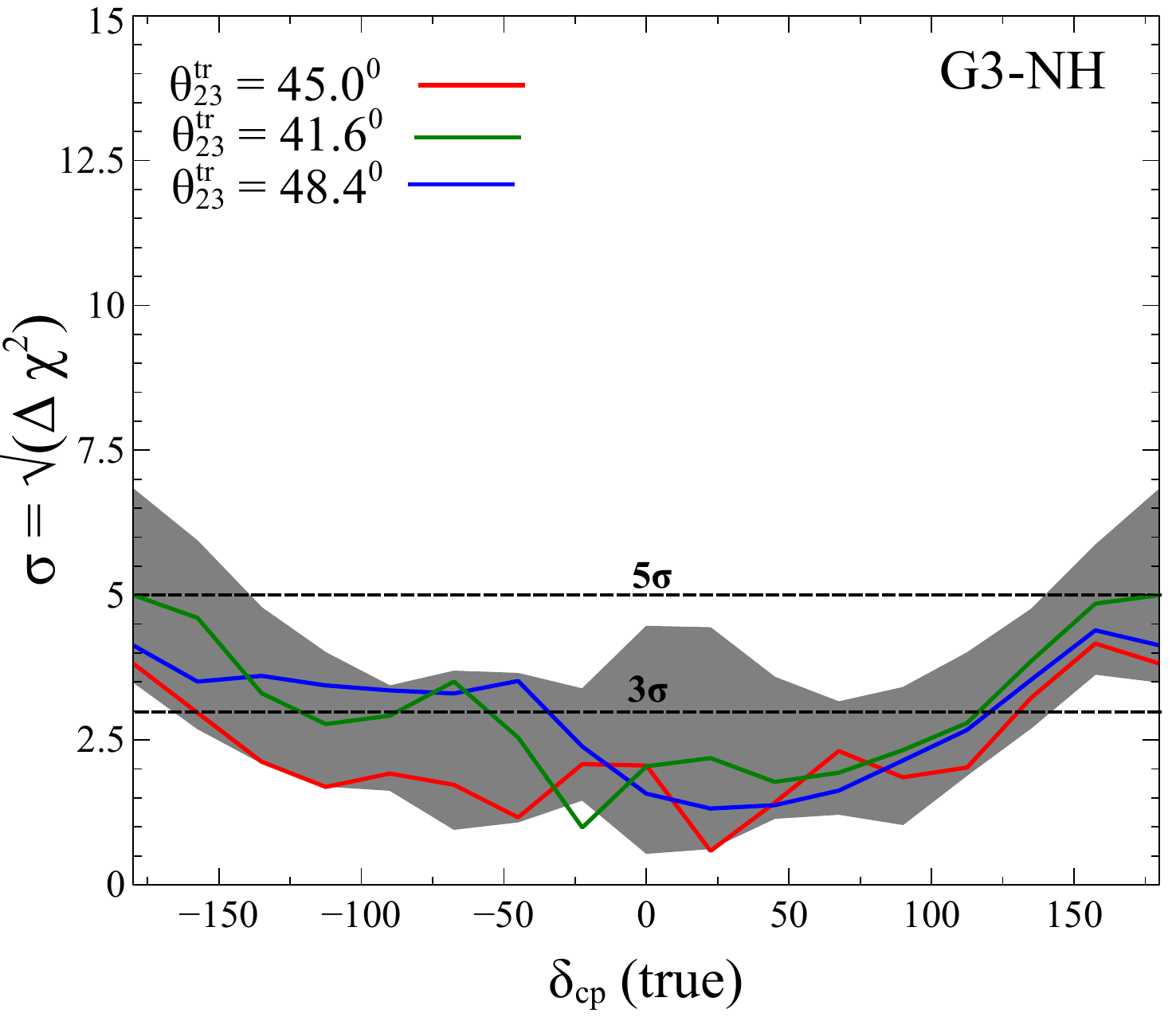}
\includegraphics[width=0.47\textwidth]{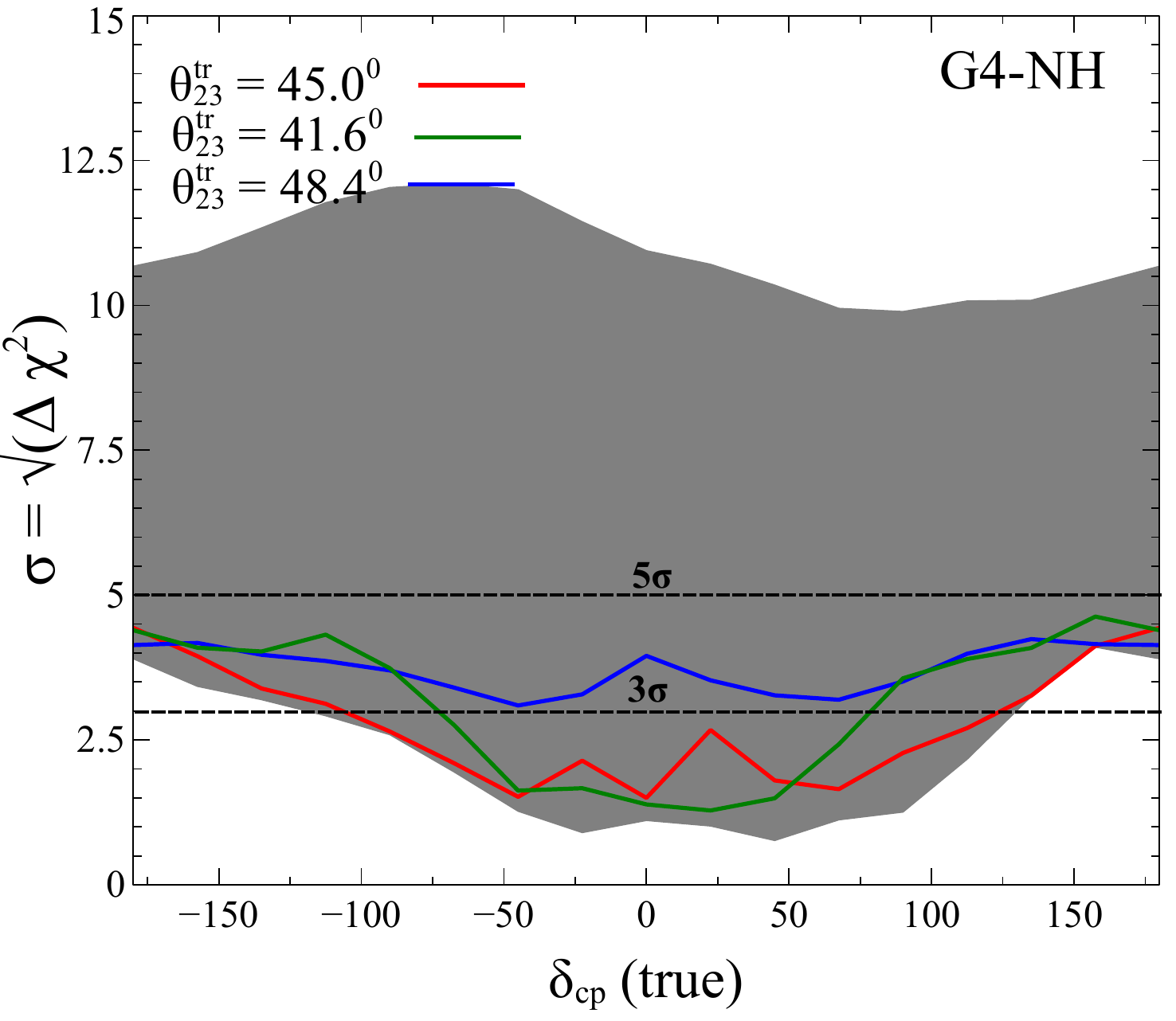}
\includegraphics[width=0.47\textwidth]{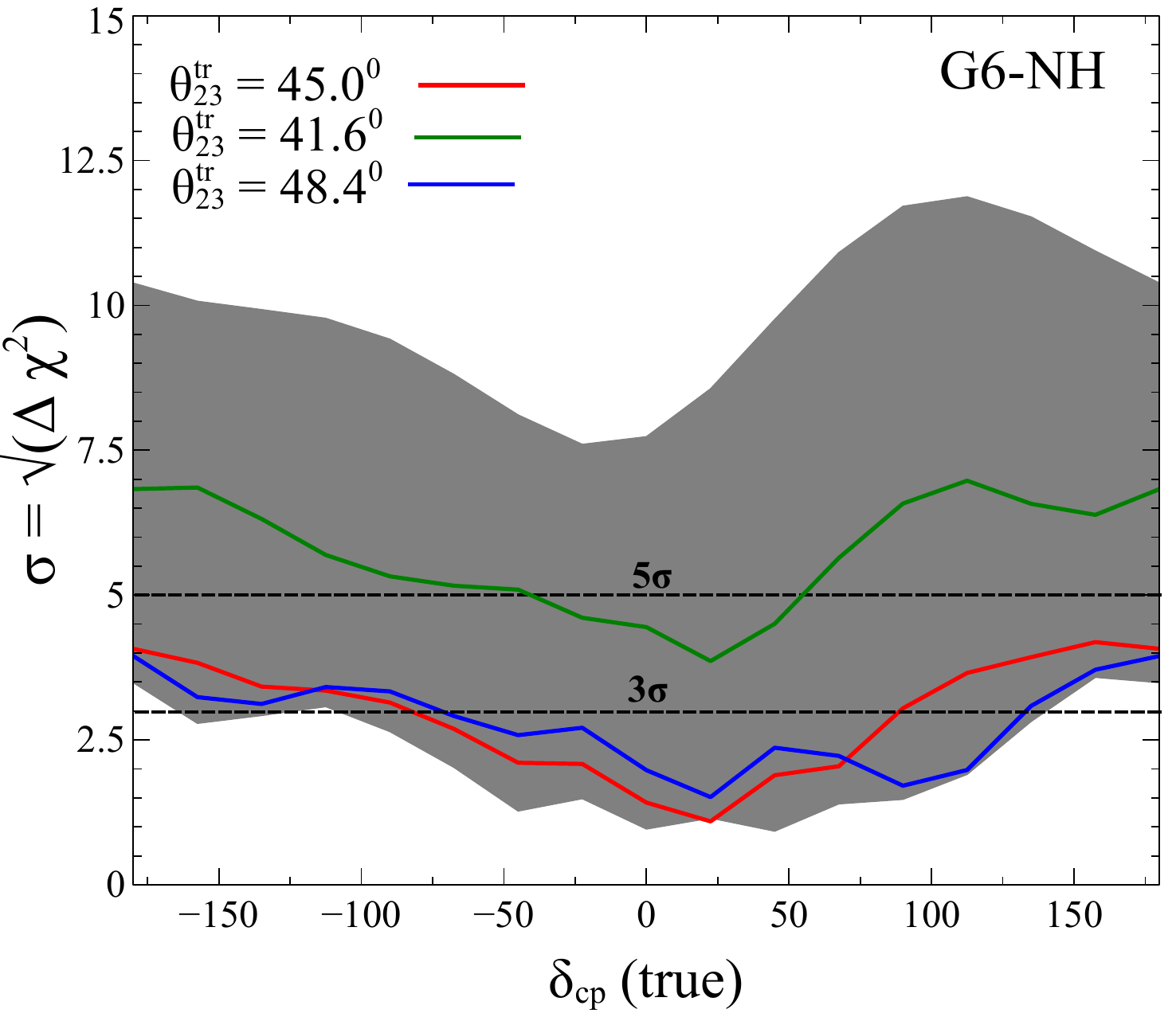}
\caption{The exclusion of different one-zero textures at DUNE assuming NH as the true hierarchy. The grey band represents the full variation of $\tz$ in its 3$\s$ allowed range. Green, blue and the red plots corresponds to three different choice of $\tz$ i.e. green plot is for the best fit value of $\tz$ in the LO while the blue plot is for $\tz$ in the HO. The red plot is for maximal $\tz$. }
\label{fig7}
\end{center}
\end{figure*}

\begin{figure*}
\begin{center}
\includegraphics[width=0.45\textwidth]{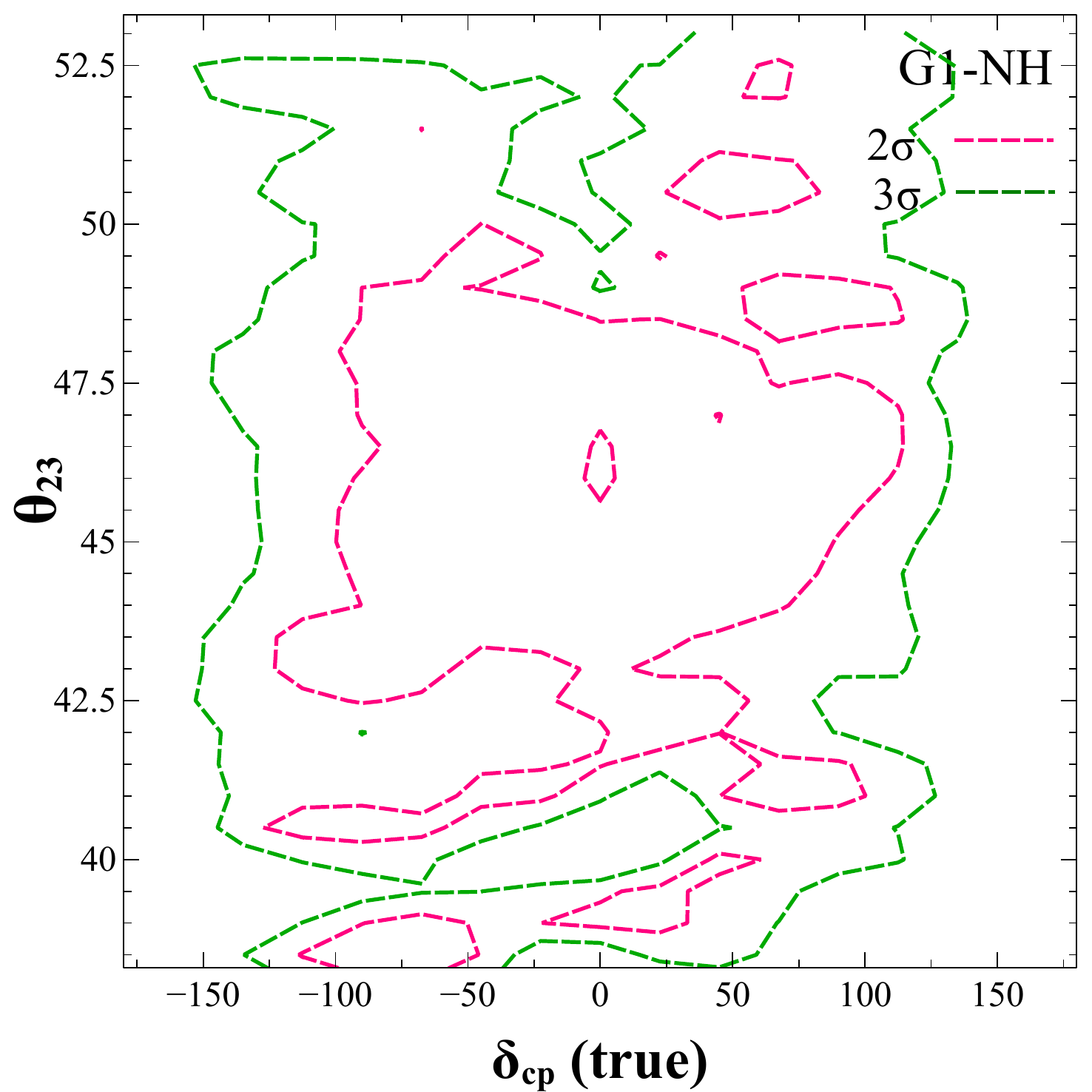}
\includegraphics[width=0.45\textwidth]{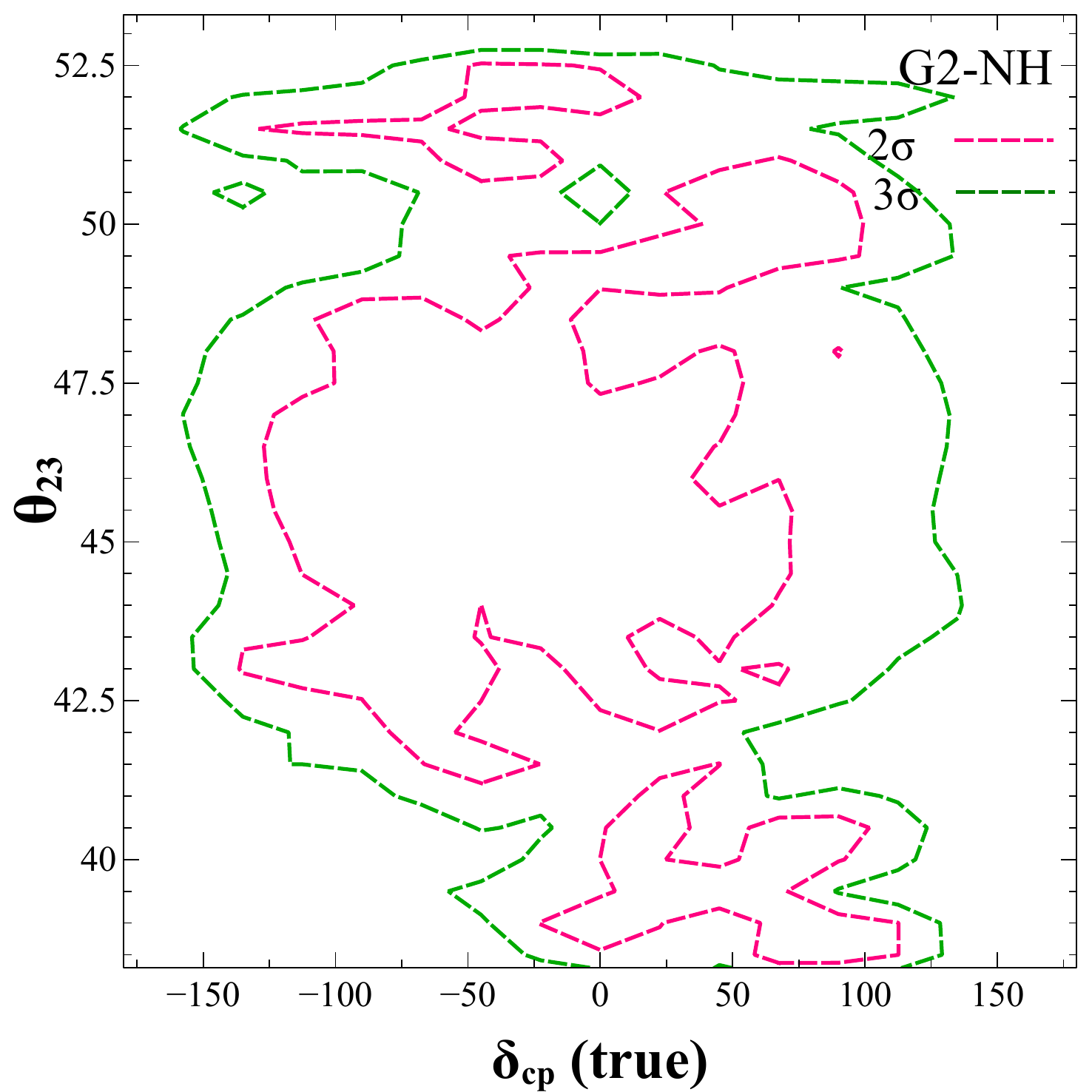}
\includegraphics[width=0.45\textwidth]{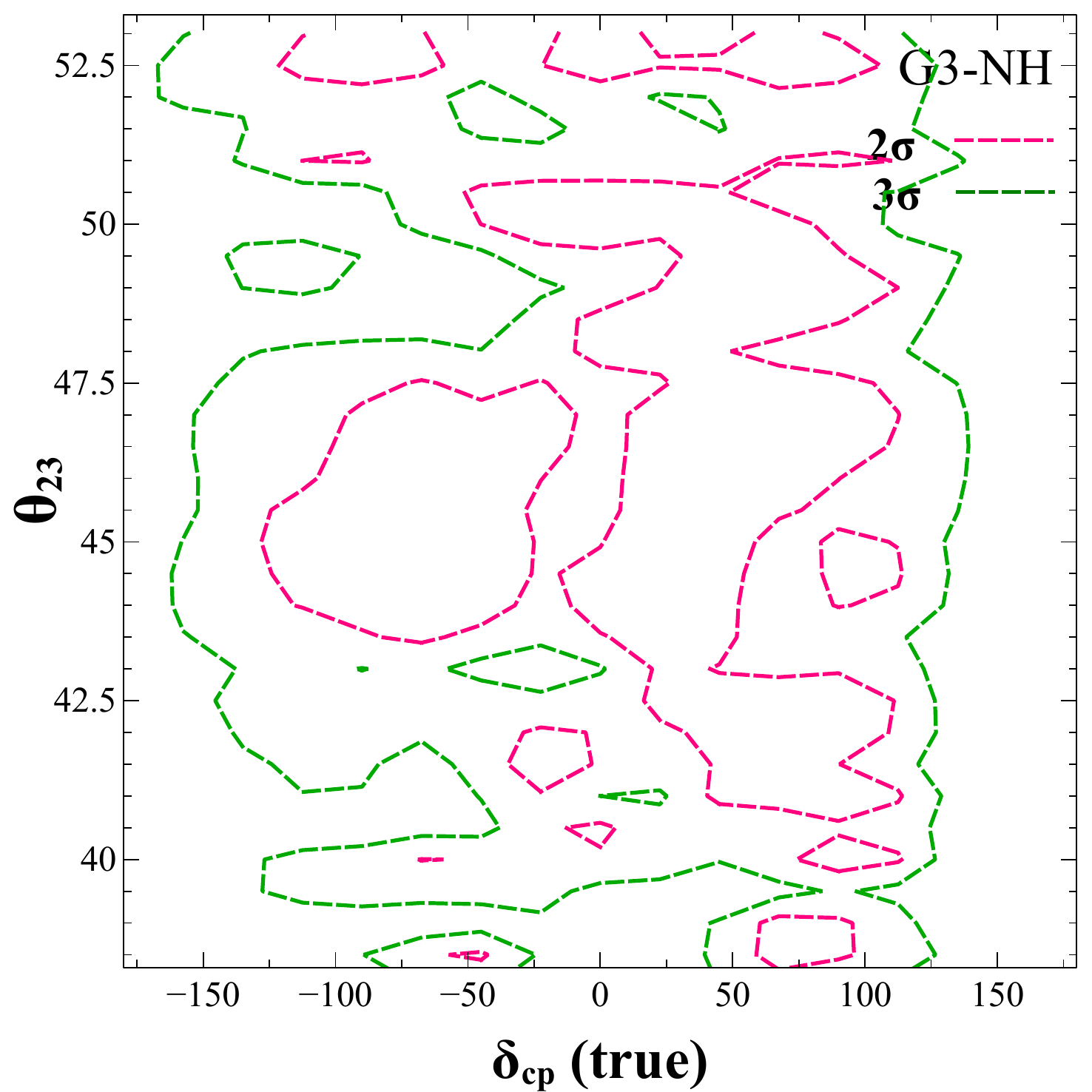}
\includegraphics[width=0.45\textwidth]{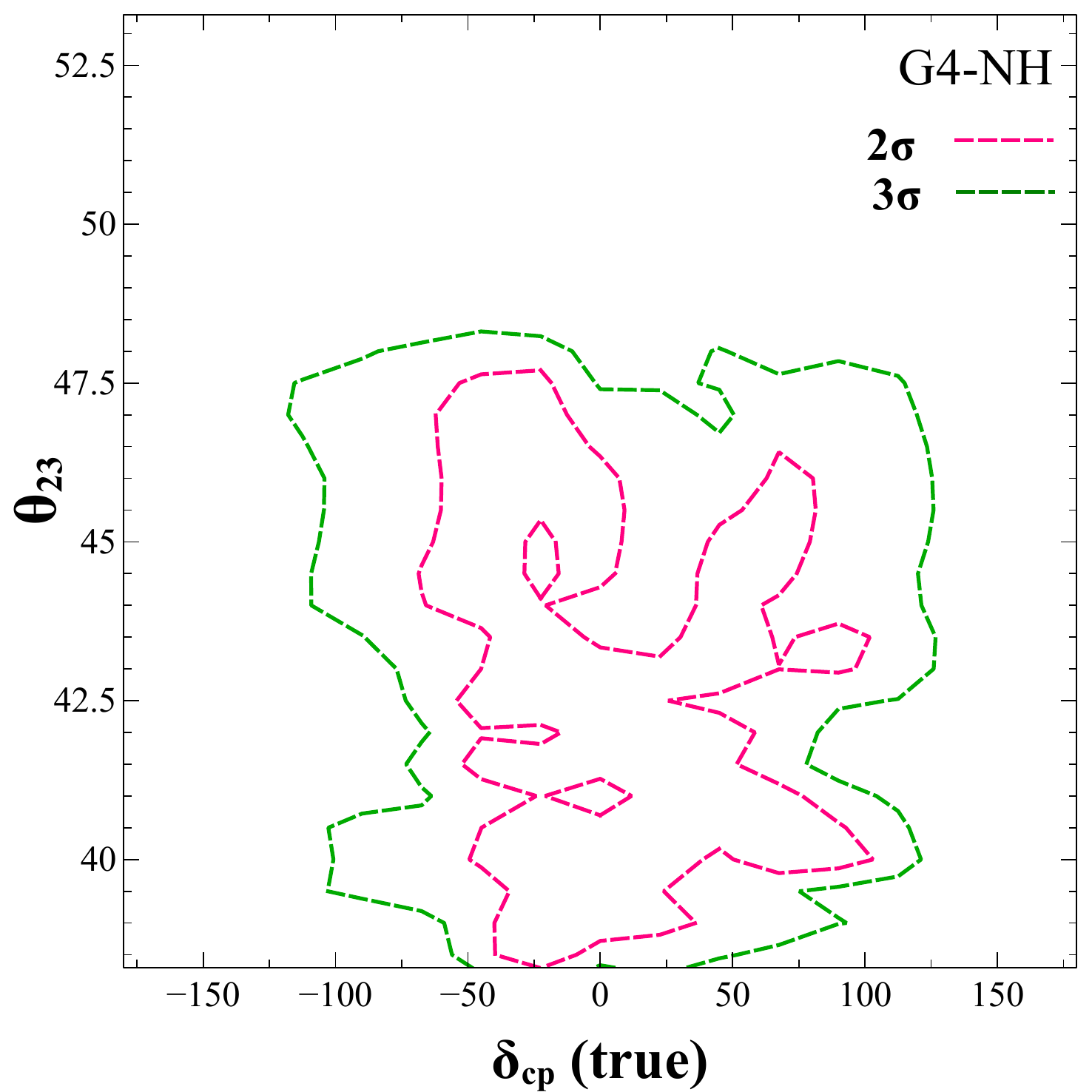}
\includegraphics[width=0.45\textwidth]{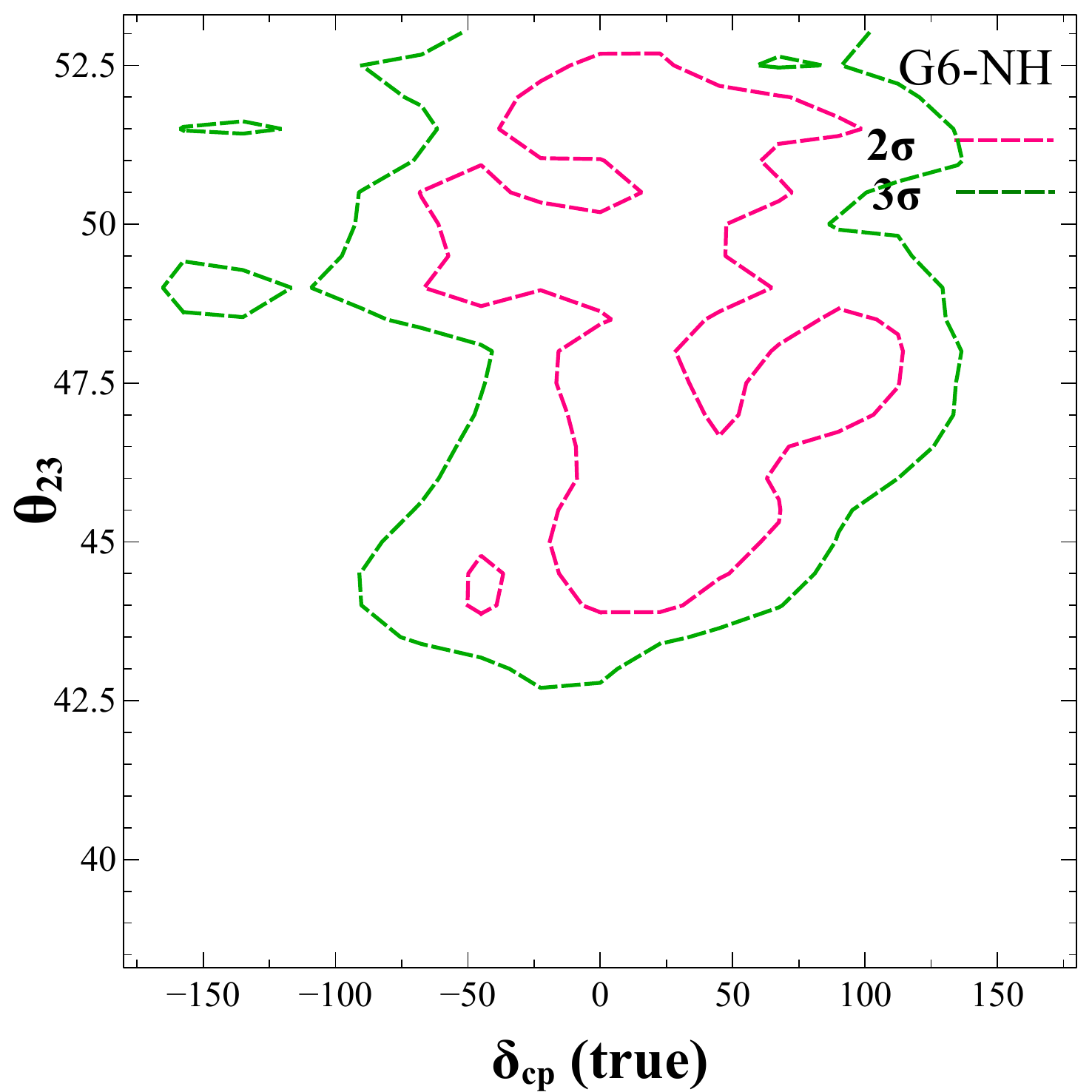}
\caption{ In the contour plots, we show the allowed regions corresponding to different one-zero textures in NH mode. The green (red) dotted line represents the 3$\s$ (2$\s$) contour at 1 d.o.f..}
\label{fig77}
\end{center}
\end{figure*}
In fig. \ref{fig7}, we show the exclusion of G1, G2, G3, G4 and G6 textures (generated assuming true NH) while in fig. \ref{fig77}, the allowed/exclusion regions are shown at DUNE in $\tz-\dcp$ parameter space. The solid green (blue) line corresponds to true $\tz = 41.6^o$ $ (48.4^o)$ i.e. $\tz$ in the LO (HO) and the red line corresponds to maximal $\tz$ i.e. $\tz = 45^o$. For each true $\dcp$, the grey band show the minimum and maximum of $\chi^2_{\rm min}$ corresponding to variation of $\tz$ in its $3\s$ allowed range. In case of G1, G2 and G3, only a small portion of the grey band lies above 5$\s$ C.L., while for G4 and G6 texture, a large fraction of $\chi^2_{\rm min}$ lies above the 5$\s$. All the region that lies above the 5$\s$ (3$\s$) black line is excluded at 5$\s$ (3$\s$). So for most of the true $\tz$, DUNE can exclude G4 and G6 at 5$\s$ C.L. for all true $\dcp$. For the three special cases, we see that except $\tz=41.6^{o}$ in G6-NH, DUNE can not exclude these textures at 5$\s$ for almost all true $\dcp$. But for $\tz=41.6^{o}$, DUNE can exclude G6-NH at 5$\s$ for some of the true $\dcp$ except a small fraction around $\dcp = 0$. The contours shown in fig. \ref{fig77} are defined by $\chi^2 >$ 4 (9) which corresponds to 2$\sigma$ ($3\sigma$) C.L.. So for a given texture and for a given set of true $\tz$ and $\dcp$, if the $\chi^2 > 4 (9)$, then we can say that the texture can be excluded at 2$\sigma$ ($3\sigma$) C.L.. It is observed from the contour plots in fig. \ref{fig77} that DUNE can exclude G4 (G6) texture in NH mode approximately for all true $\dcp$ if true $\tz > 48.5^o$ ($\tz < 43.5^o$) at 3$\s$ C.L.. It is difficult to draw such concluding remarks for G1, G2 and G3 as most of the $\tz-\dcp$ parameter space is allowed at 2$\s$/3$\s$ C.L. as seen from the contour plot in fig. \ref{fig7}.

 \begin{figure*}
\begin{center}
\includegraphics[width=0.47\textwidth]{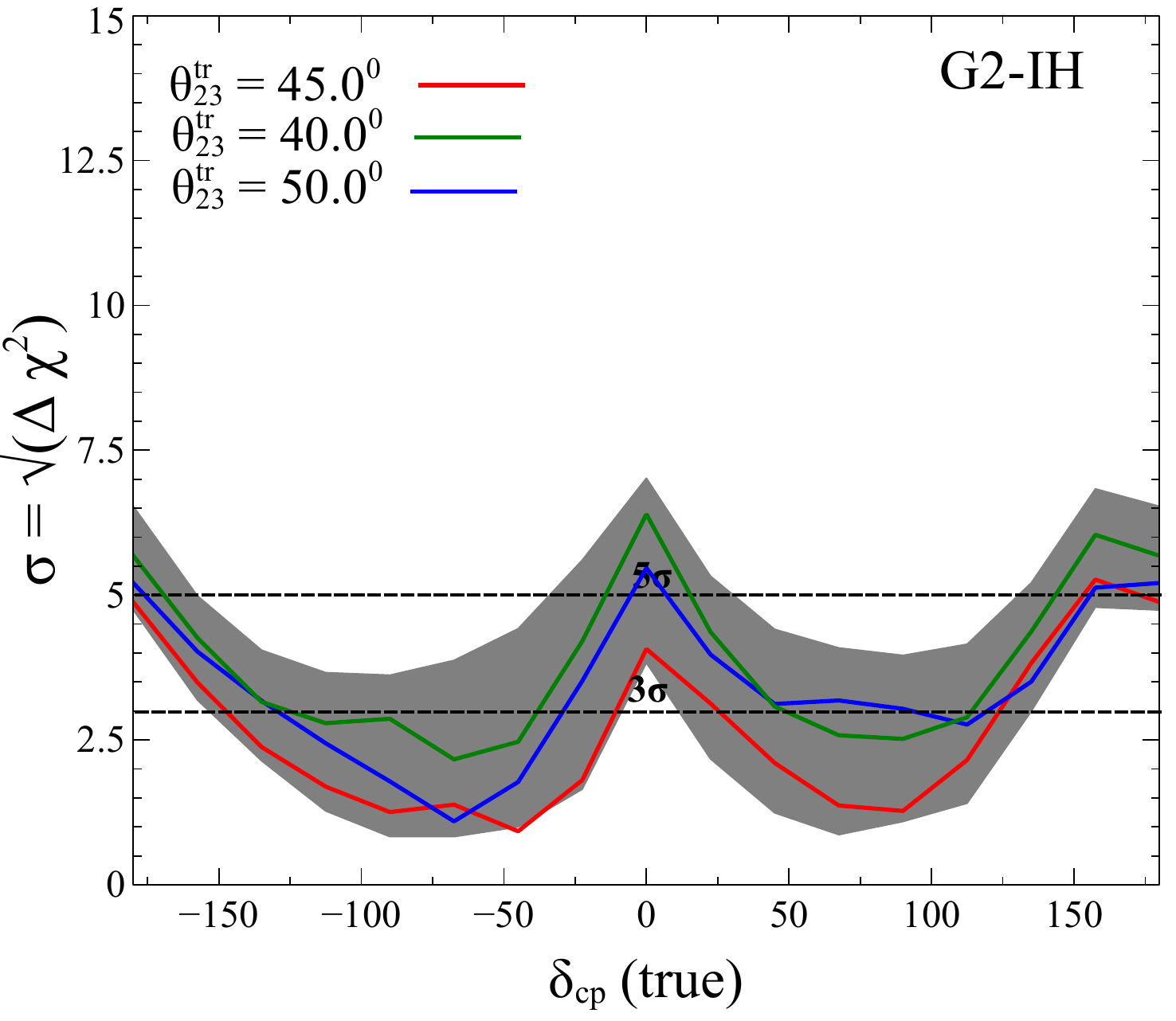}
\includegraphics[width=0.47\textwidth]{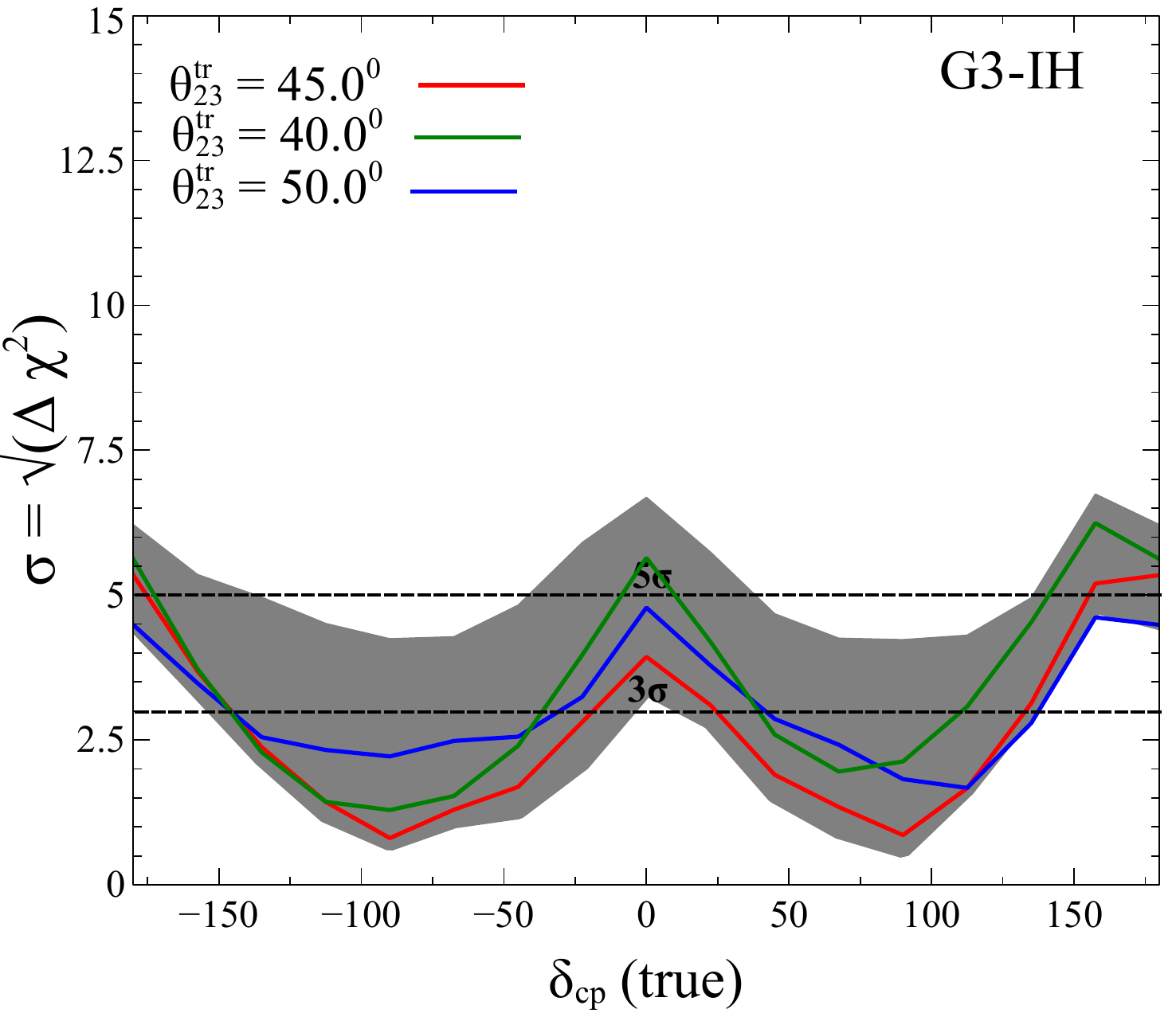}
\includegraphics[width=0.47\textwidth]{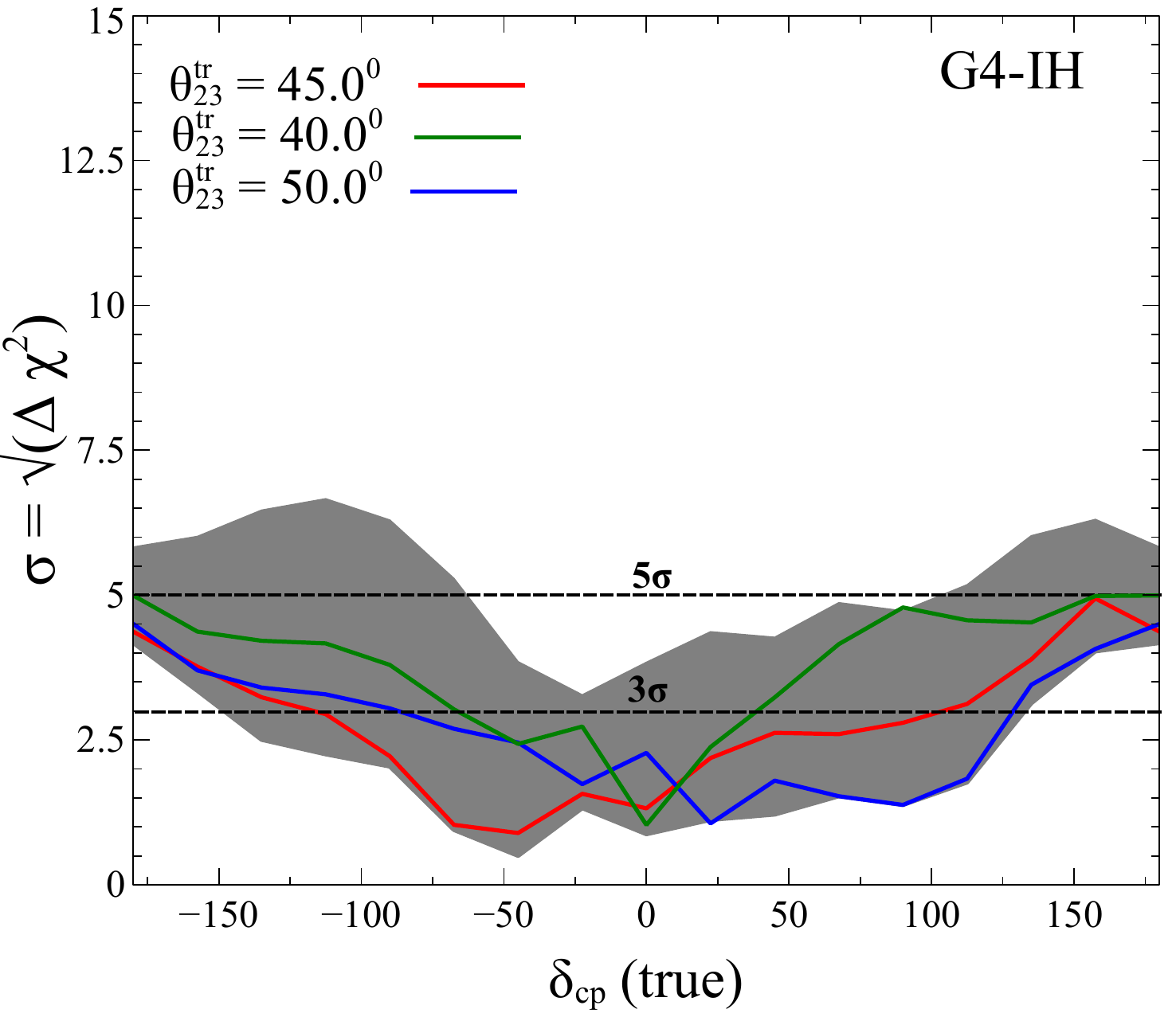}
\includegraphics[width=0.47\textwidth]{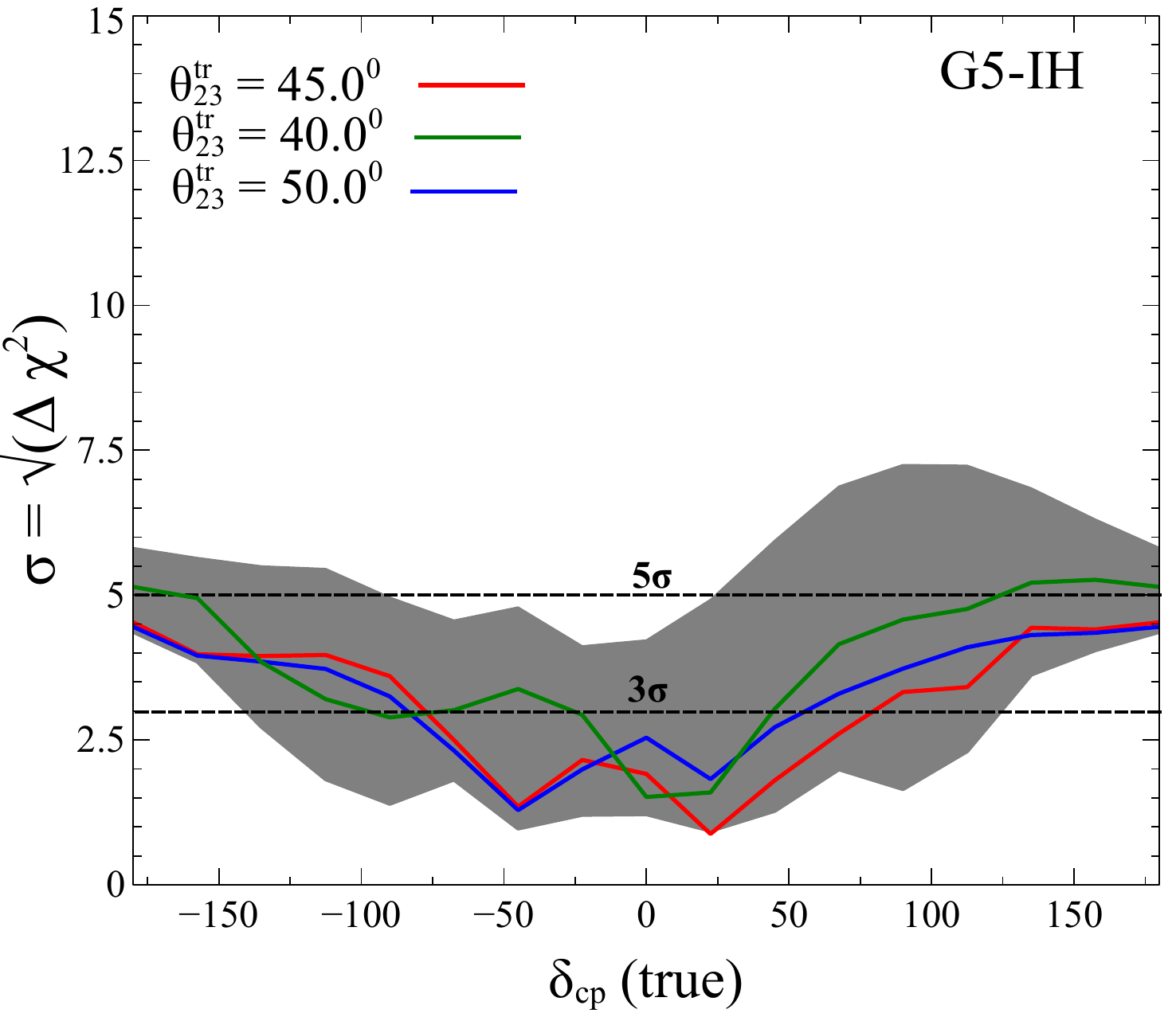}
\includegraphics[width=0.47\textwidth]{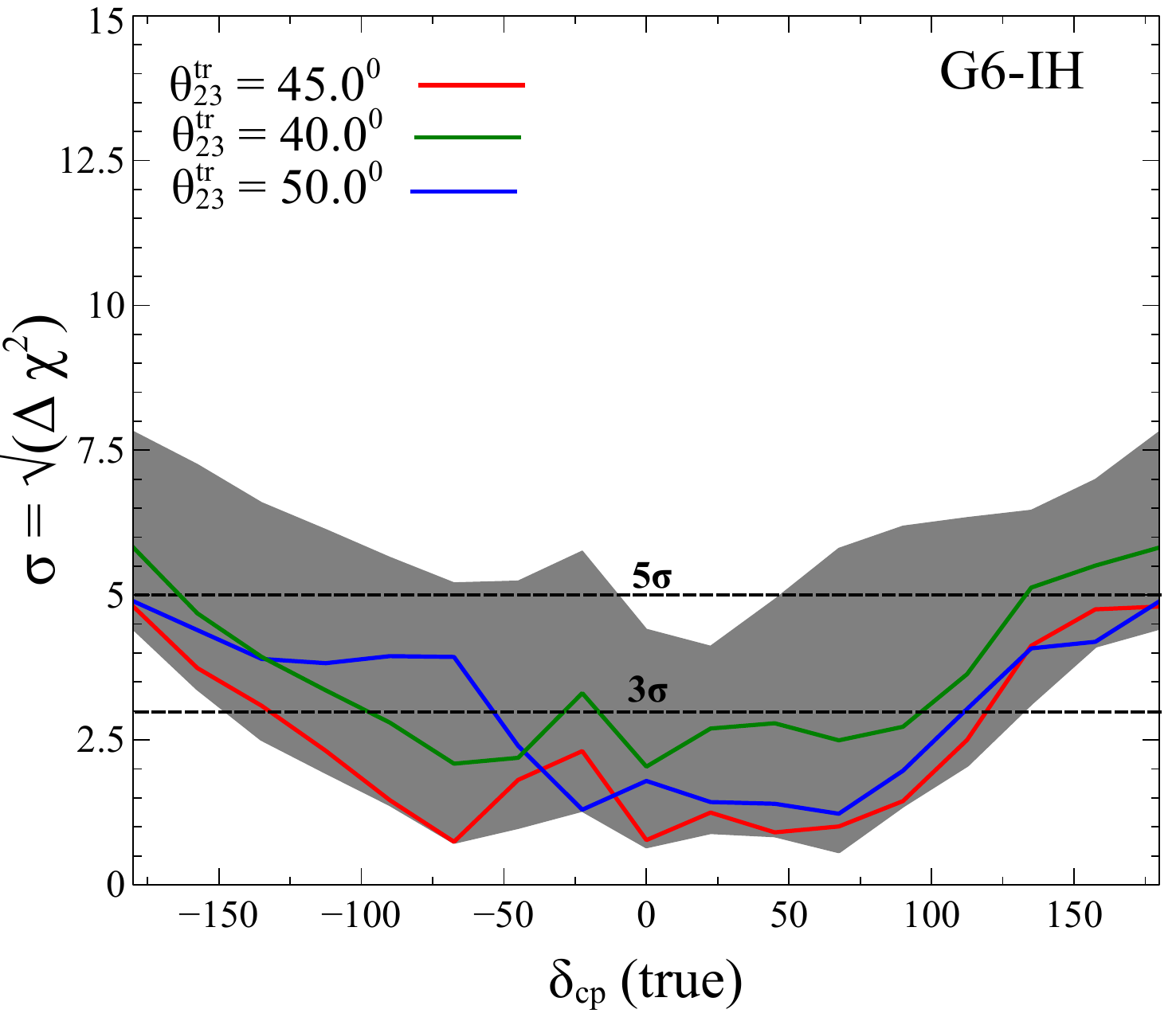}
\caption{The exclusion of different one-zero textures at DUNE assuming IH as the true hierarchy. The grey band represents the full variation of $\tz$ in its 3$\s$ allowed range. Green, blue and the red plots corresponds to three different choice of $\tz$ i.e. green plot is for the best fit value of $\tz$ in the LO while the blue plot is for $\tz$ in the HO. The red plot is for maximal $\tz$.}
\label{fig8}
\end{center}
\end{figure*}

 \begin{figure*}
\begin{center}
\includegraphics[width=0.45\textwidth]{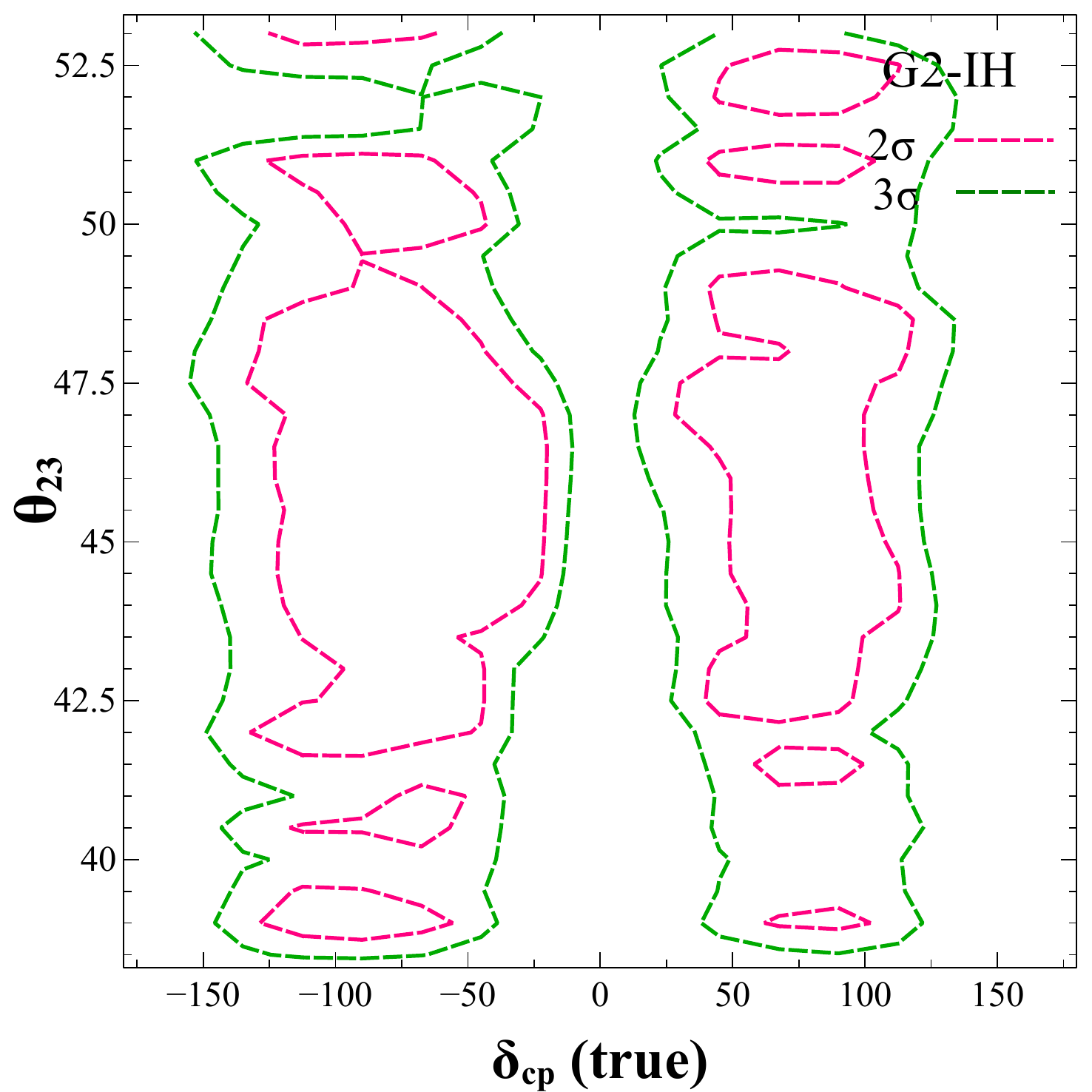}
\includegraphics[width=0.45\textwidth]{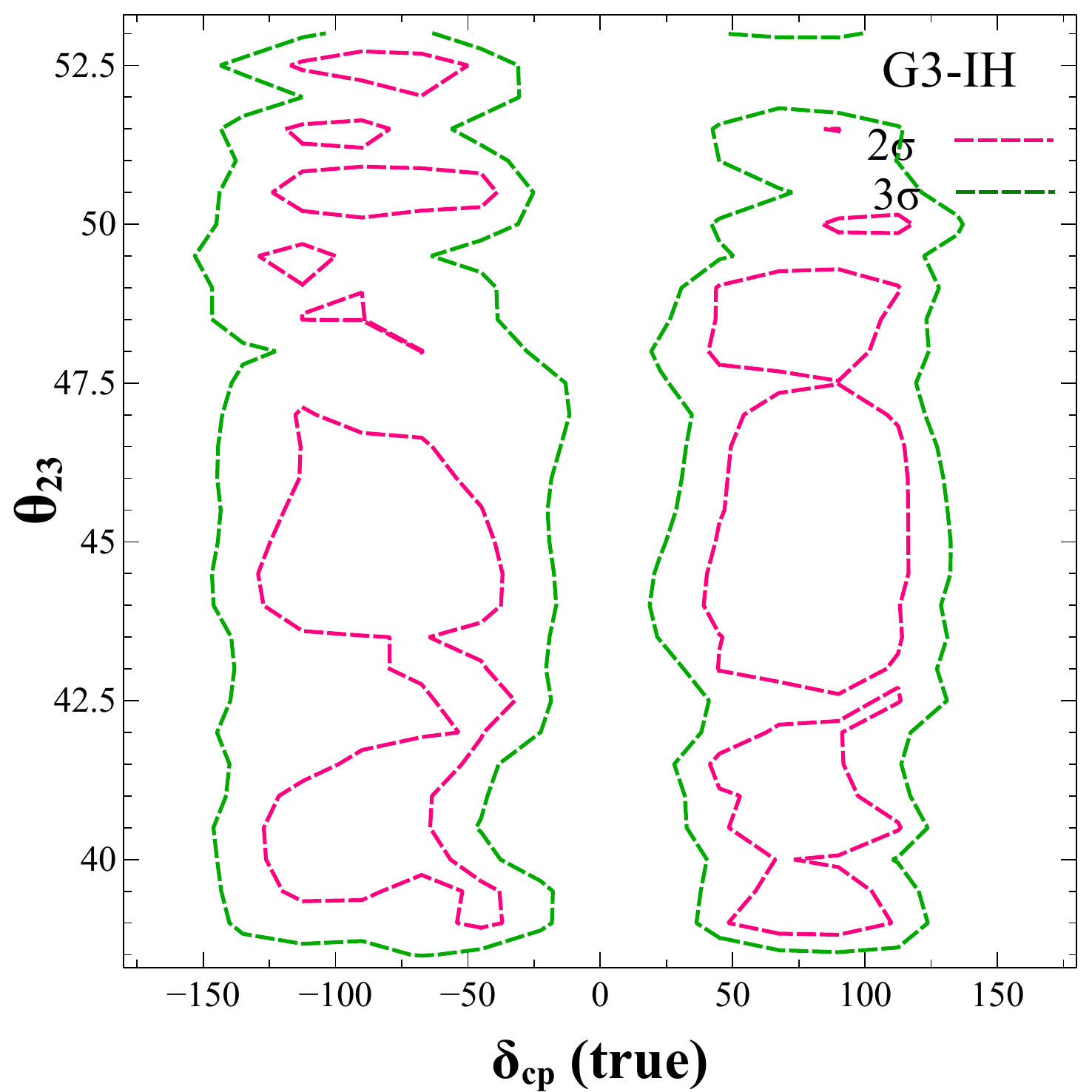}
\includegraphics[width=0.45\textwidth]{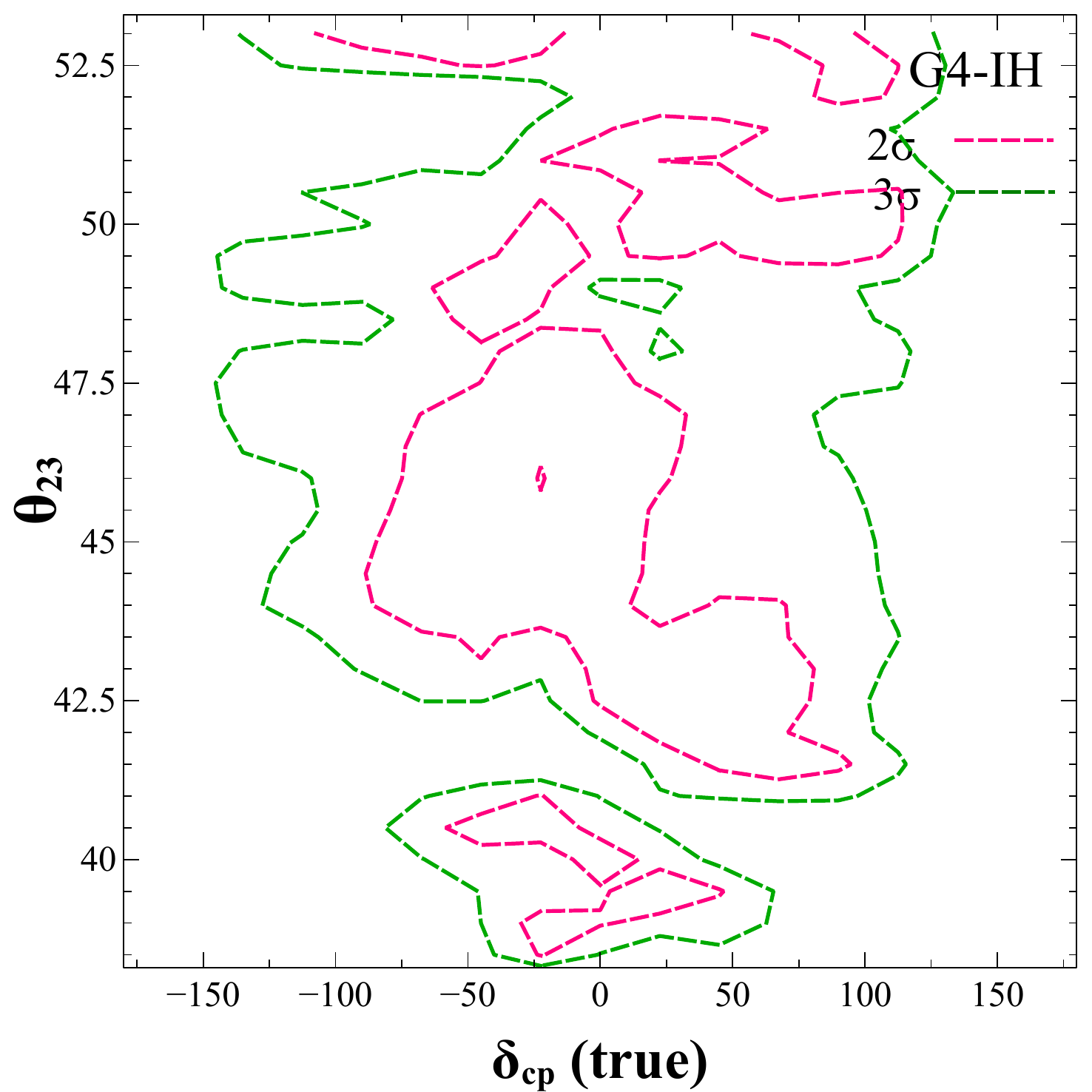}
\includegraphics[width=0.45\textwidth]{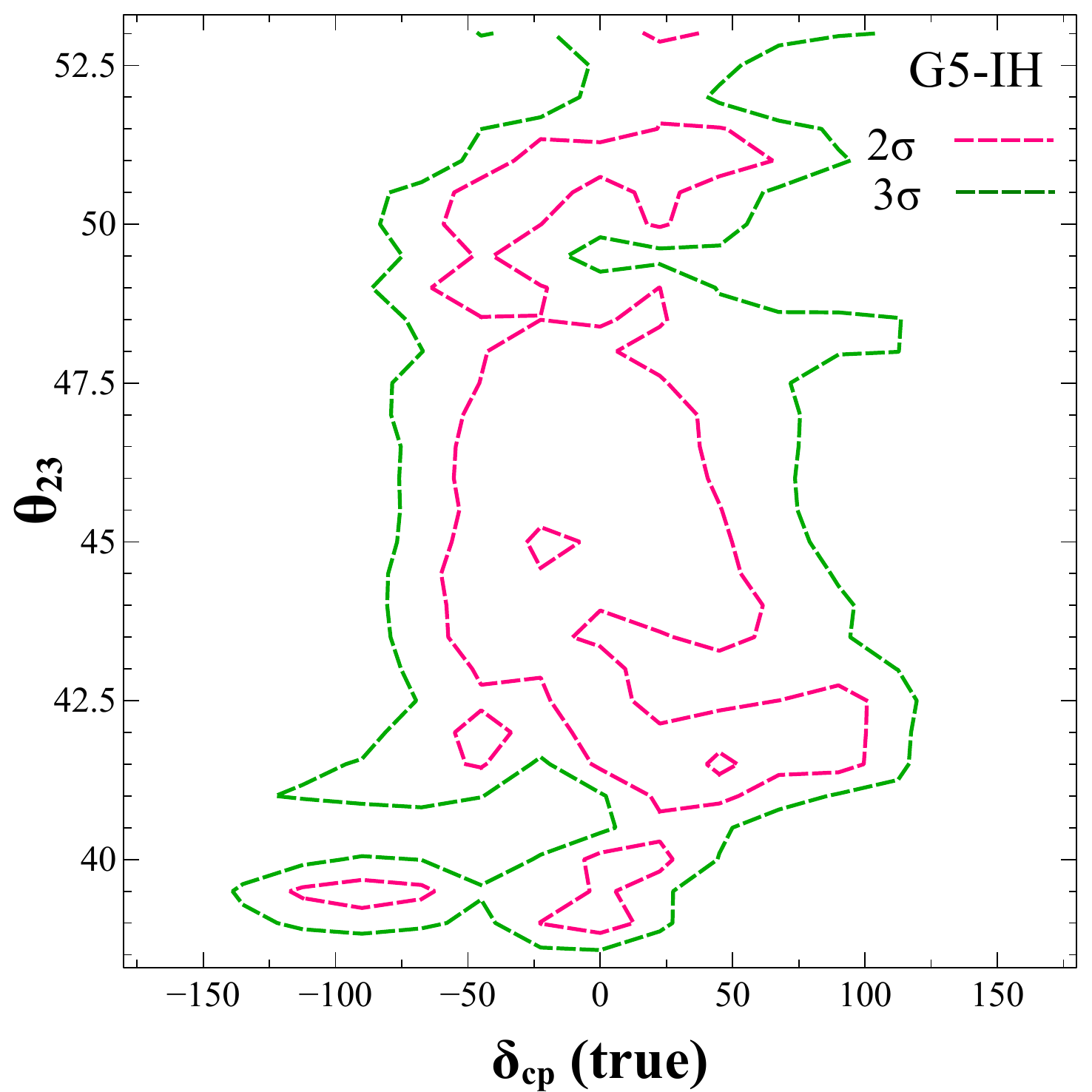}
\includegraphics[width=0.45\textwidth]{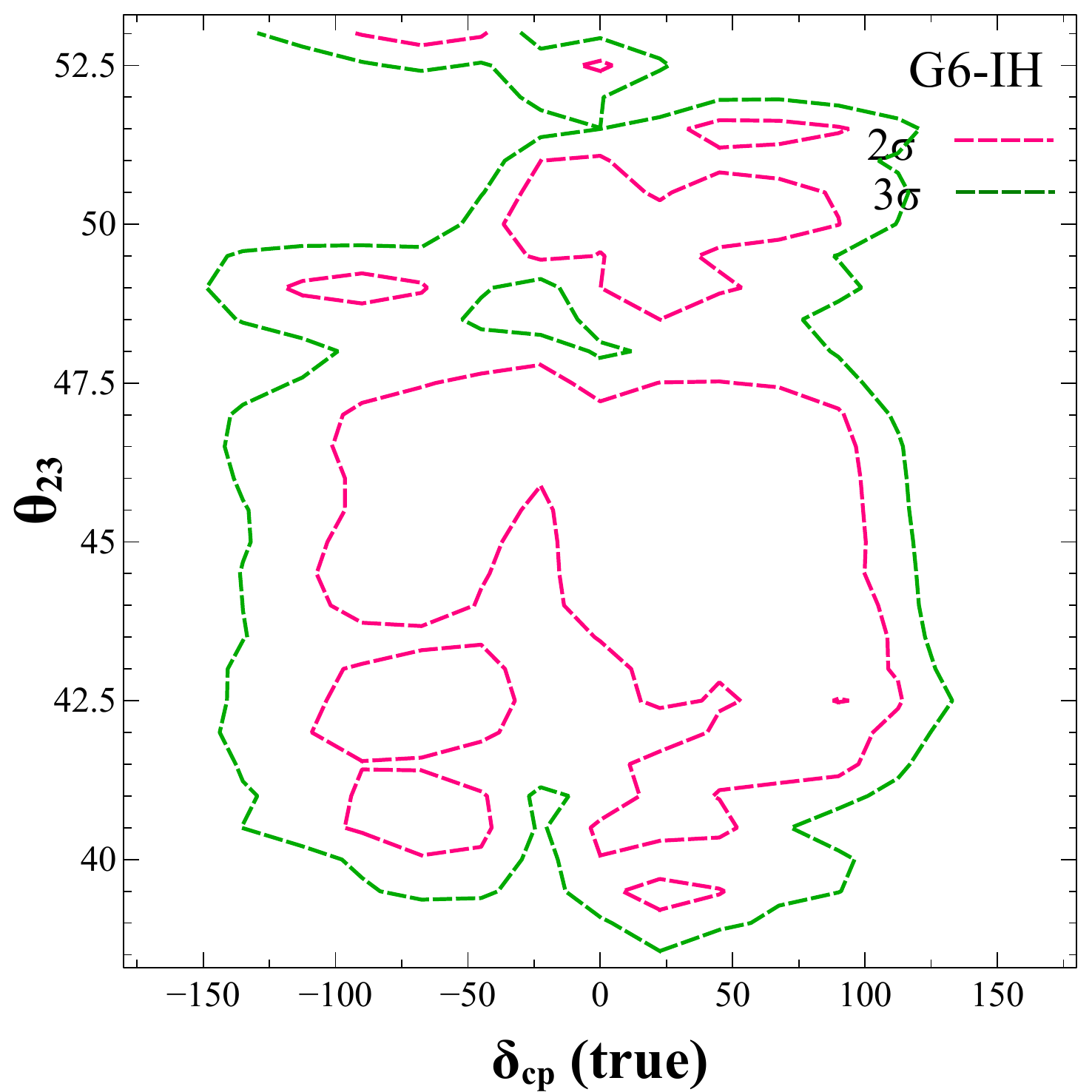}
\caption{In the contour plots, we show the allowed regions corresponding to different one-zero textures in IH mode. The green (red) dotted line represents the 3$\s$ (2$\s$) contour at 1 d.o.f.}
\label{fig88}
\end{center}
\end{figure*}

In fig. \ref{fig8} and  \ref{fig88}, we show our results for G2, G3, G4, G5 and G6 textures assuming true IH. Here also, the green (blue) line represents the best fit value of $\tz = 40.0^o$ $ (50.0^o)$ in the LO (HO) and the red line corresponds to maximal $\tz$ i.e. $\tz = 45^o$. The predictivity of these textures are less as most of the $\tz-\dcp$ parameter space is allowed at 3$\s$. But DUNE can exclude G2 and G3 at 3$\s$ for any $\tz$ at the CP conserving values of $\dcp$ while G4, G5 and G6 are excluded for $\dcp = \pm \pi$ at 3$\s$ for all $\tz$. 
 \begin{figure*}
\begin{center}
\includegraphics[width=0.47\textwidth]{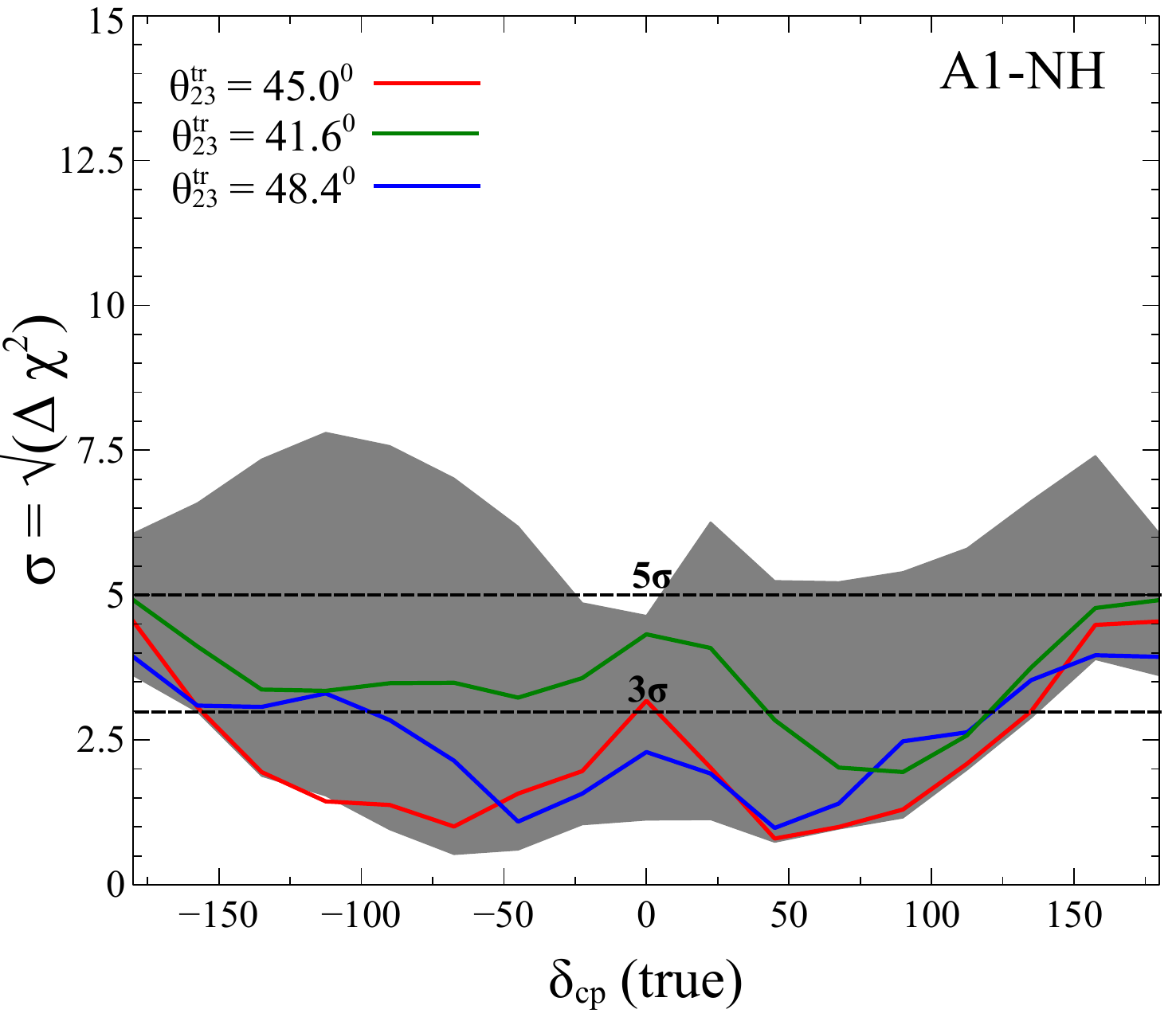}
\includegraphics[width=0.47\textwidth]{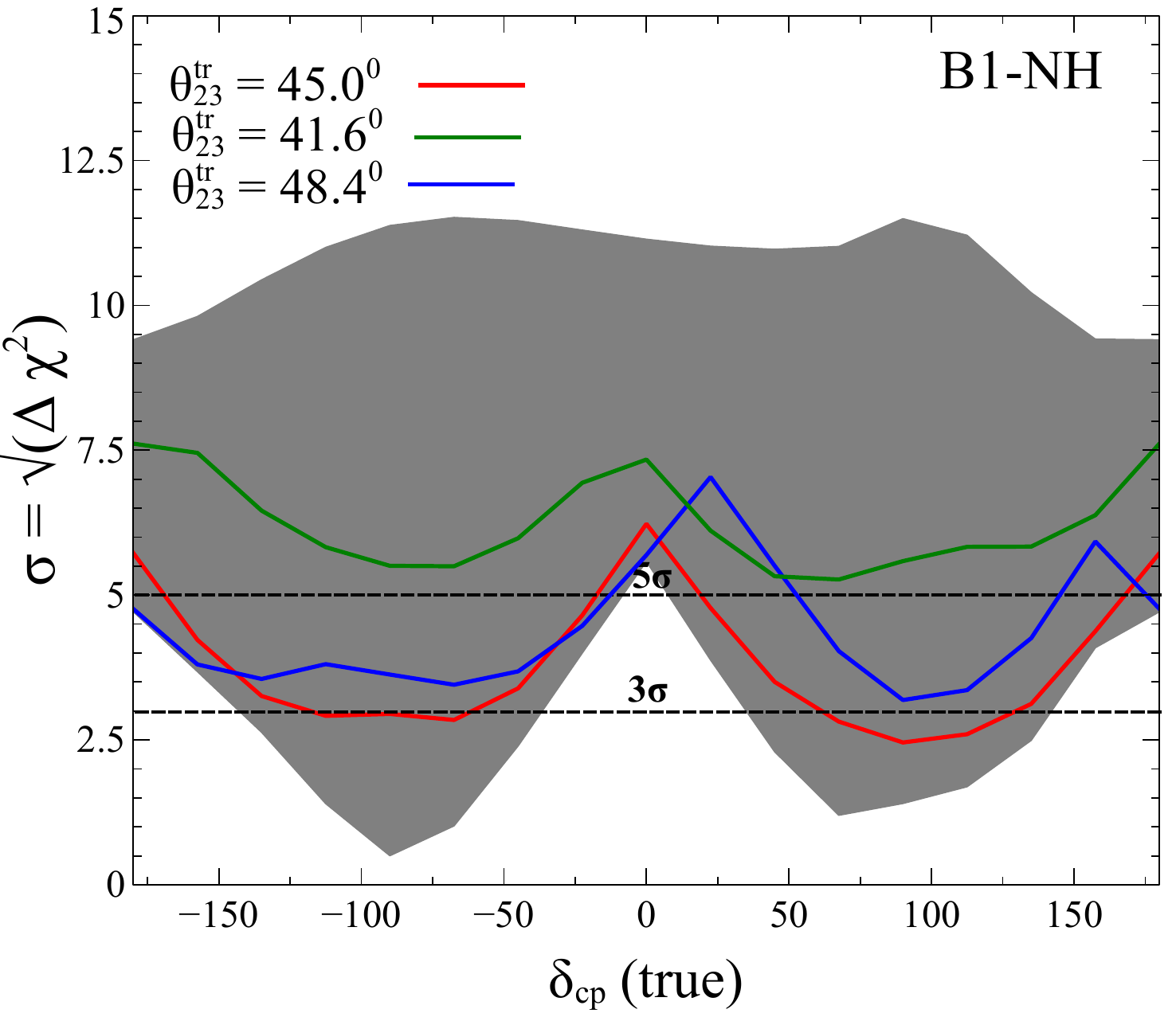}
\includegraphics[width=0.47\textwidth]{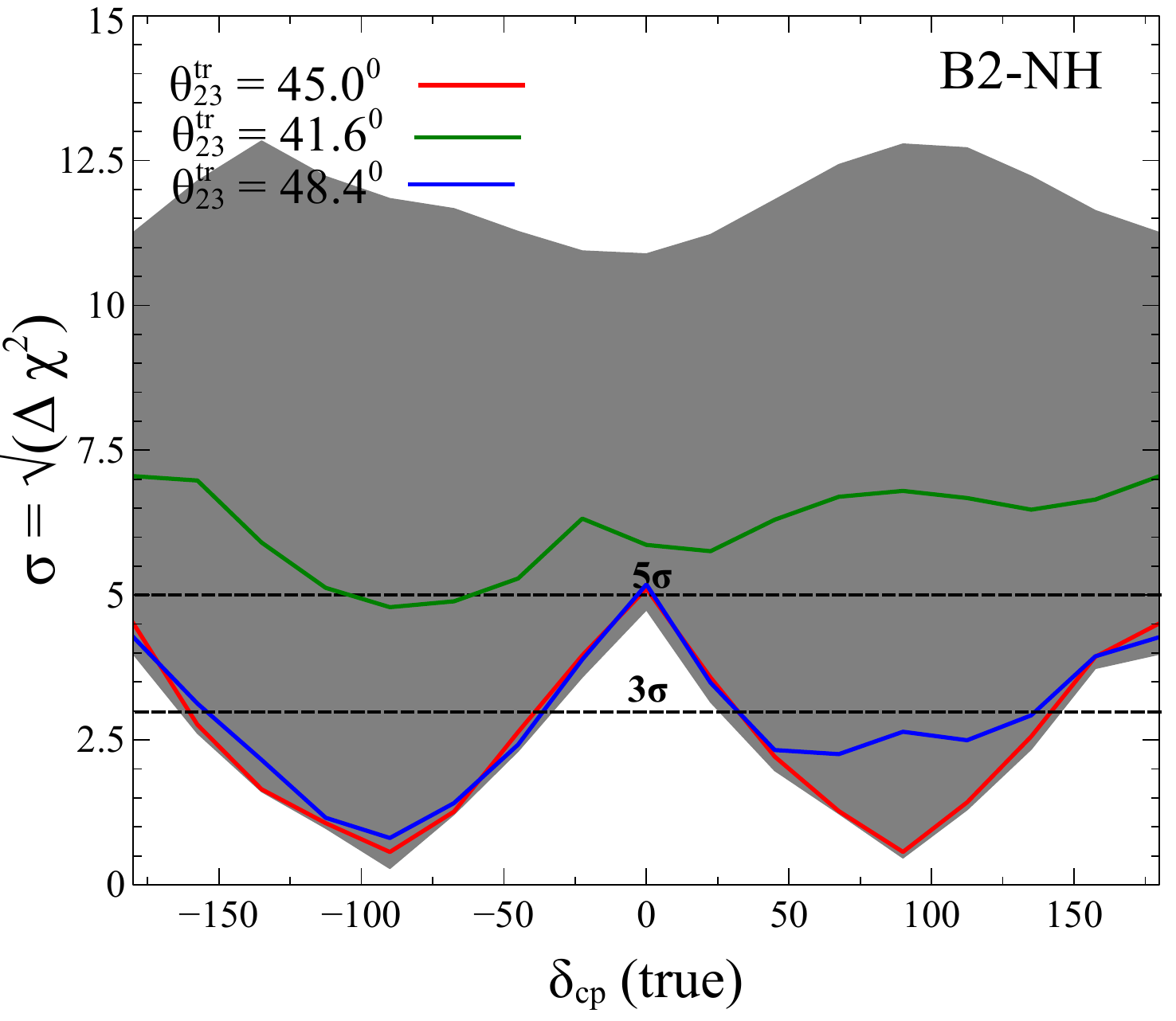}
\includegraphics[width=0.47\textwidth]{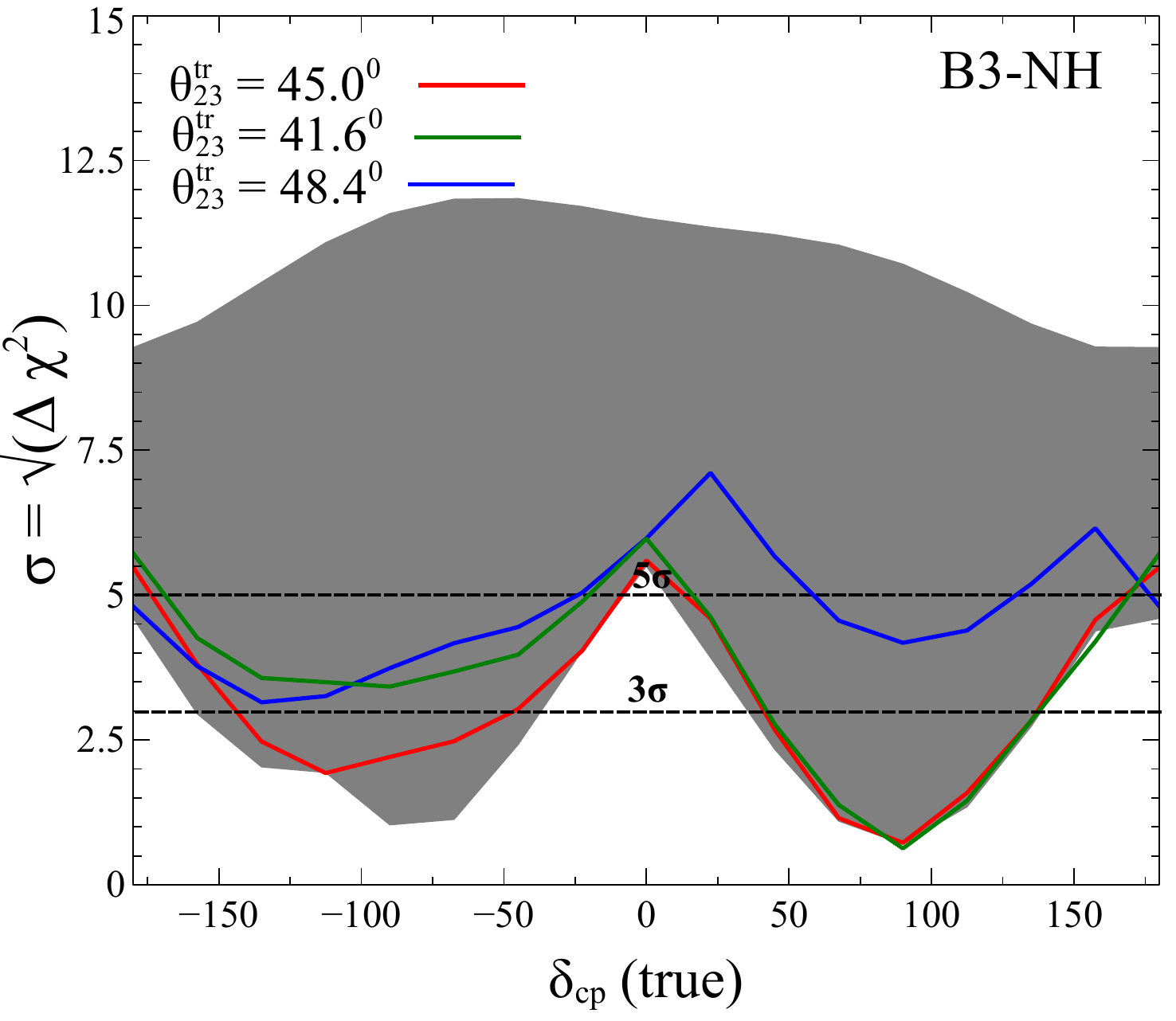}
\includegraphics[width=0.47\textwidth]{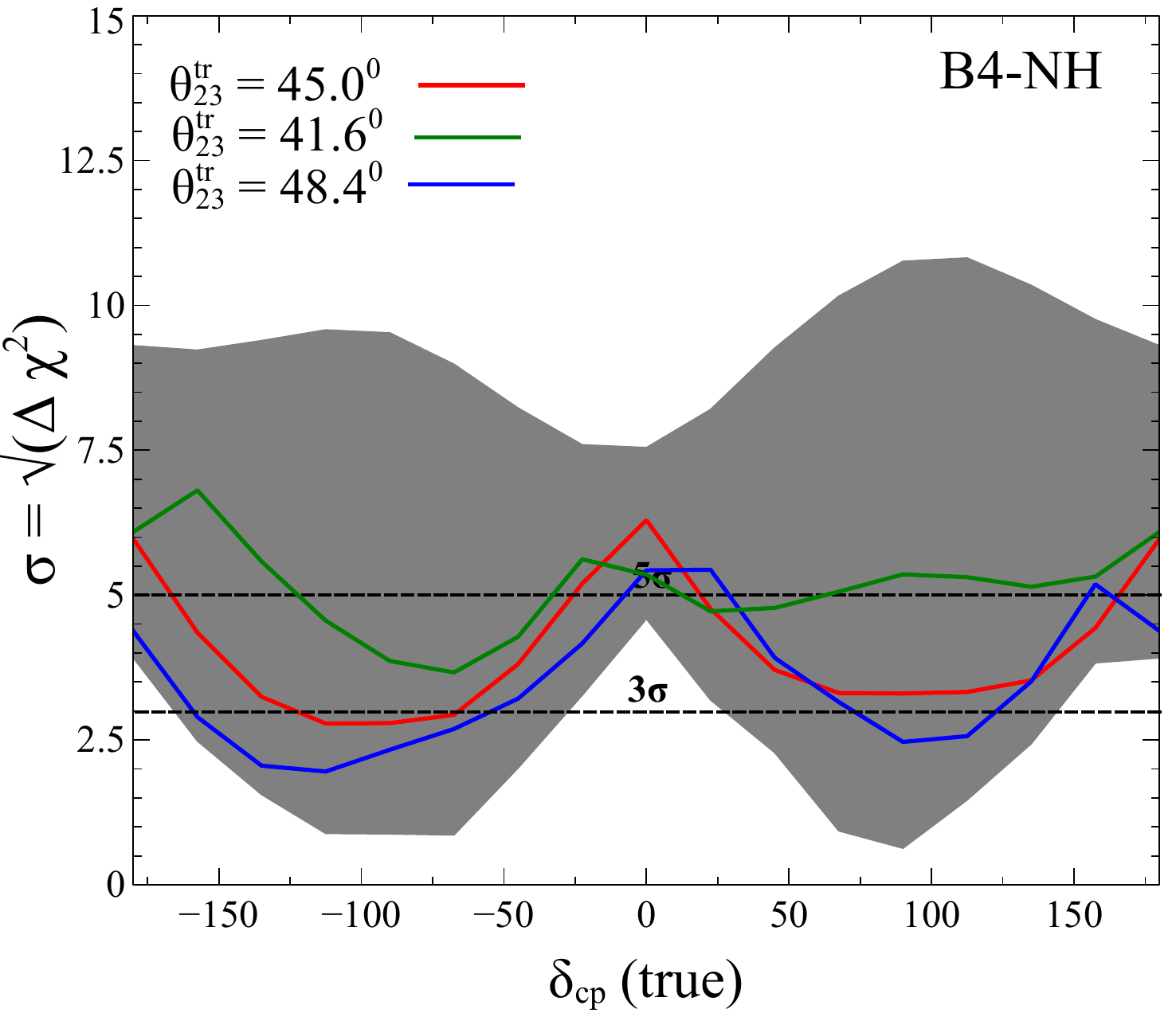}
\caption{The exclusion of different two-zero textures at DUNE assuming NH as the true hierarchy. In the upper panel, the grey band represents the full variation of $\tz$ in its 3$\s$ allowed range. Green, blue and the red plots corresponds to three different choice of $\tz$ i.e. green plot is for the best fit value of $\tz$ in the LO while the blue plot is for $\tz$ in the HO. The red plot is for maximal $\tz$. }
\label{figk}
\end{center}
\end{figure*}

 \begin{figure*}
\begin{center}
\includegraphics[width=0.45\textwidth]{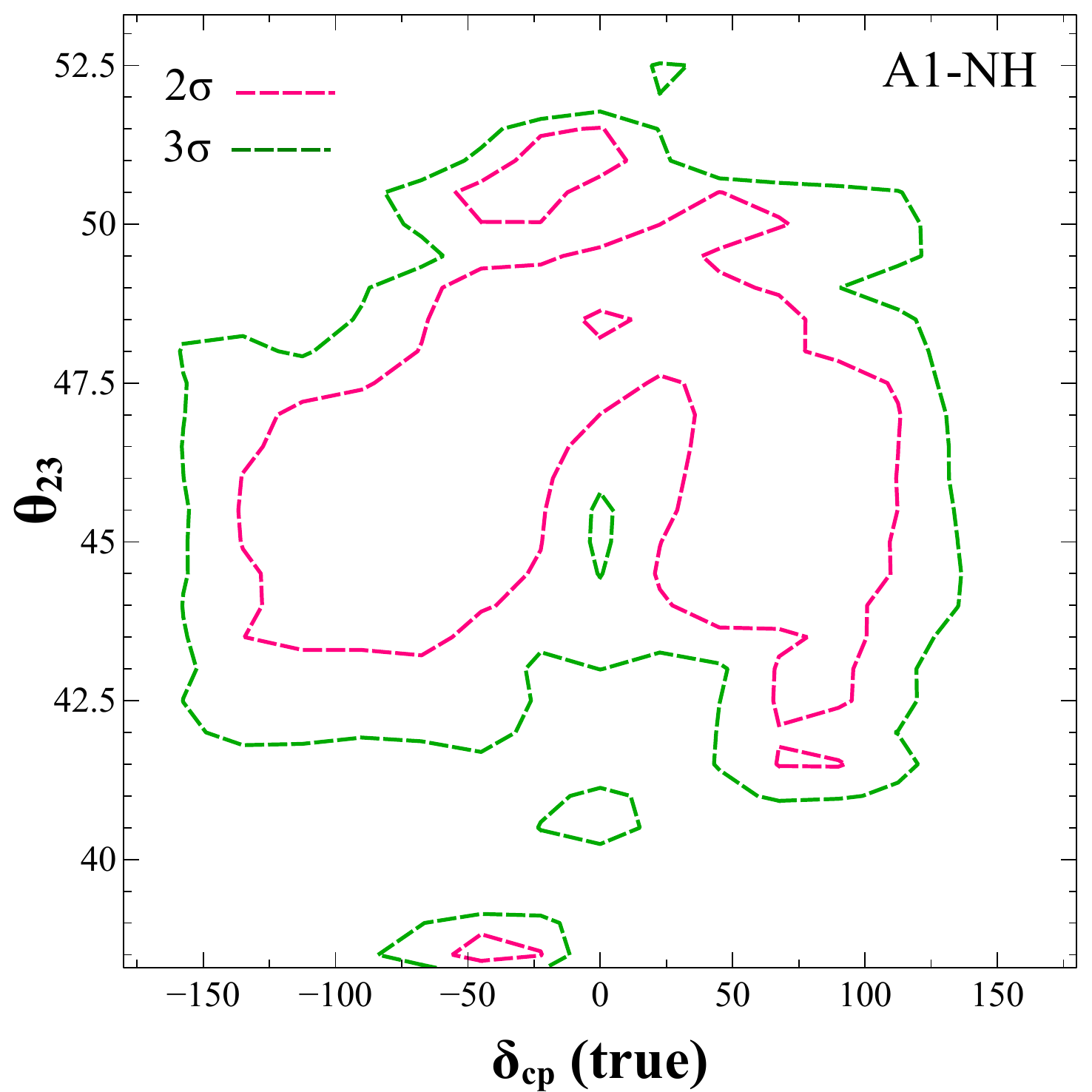}
\includegraphics[width=0.45\textwidth]{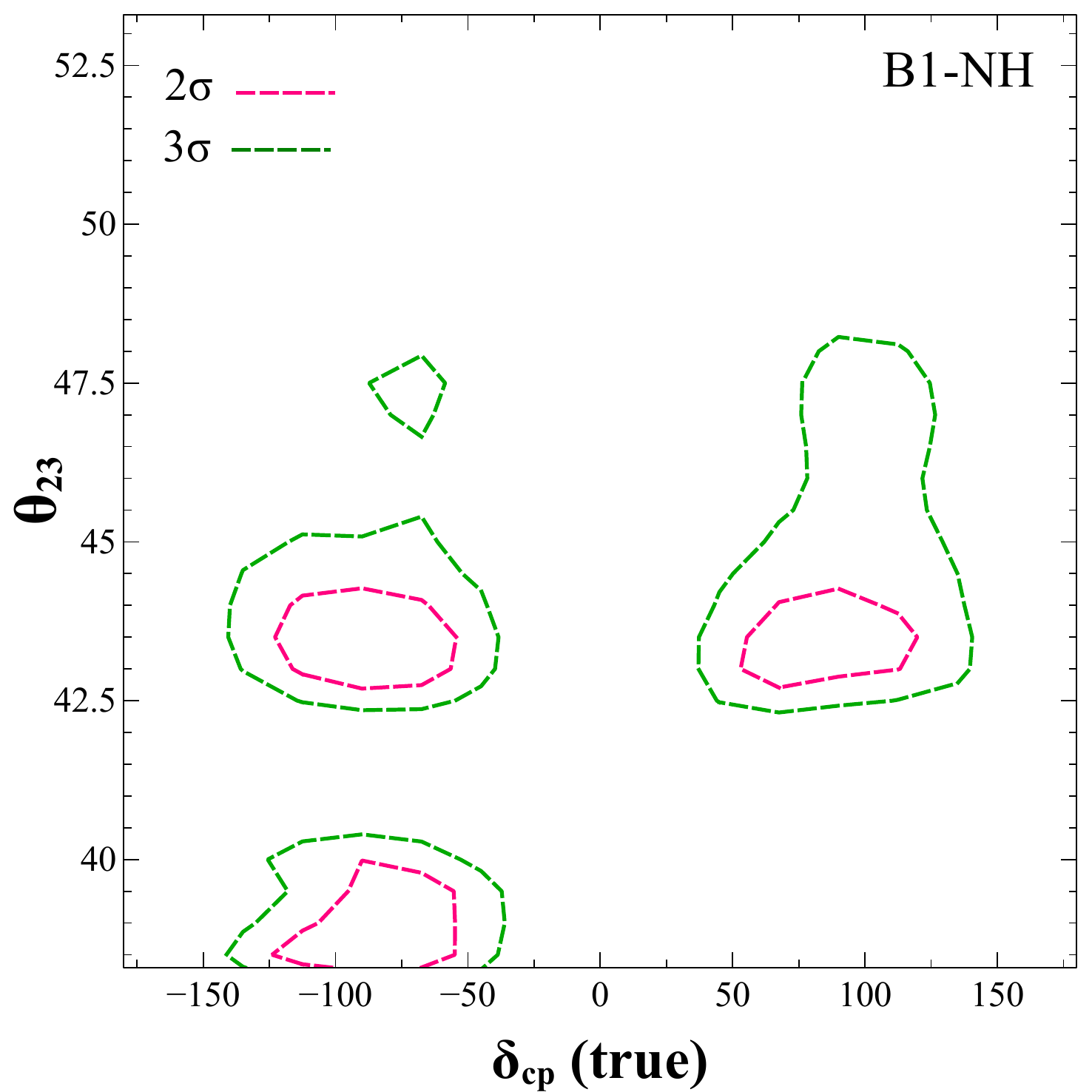}
\includegraphics[width=0.45\textwidth]{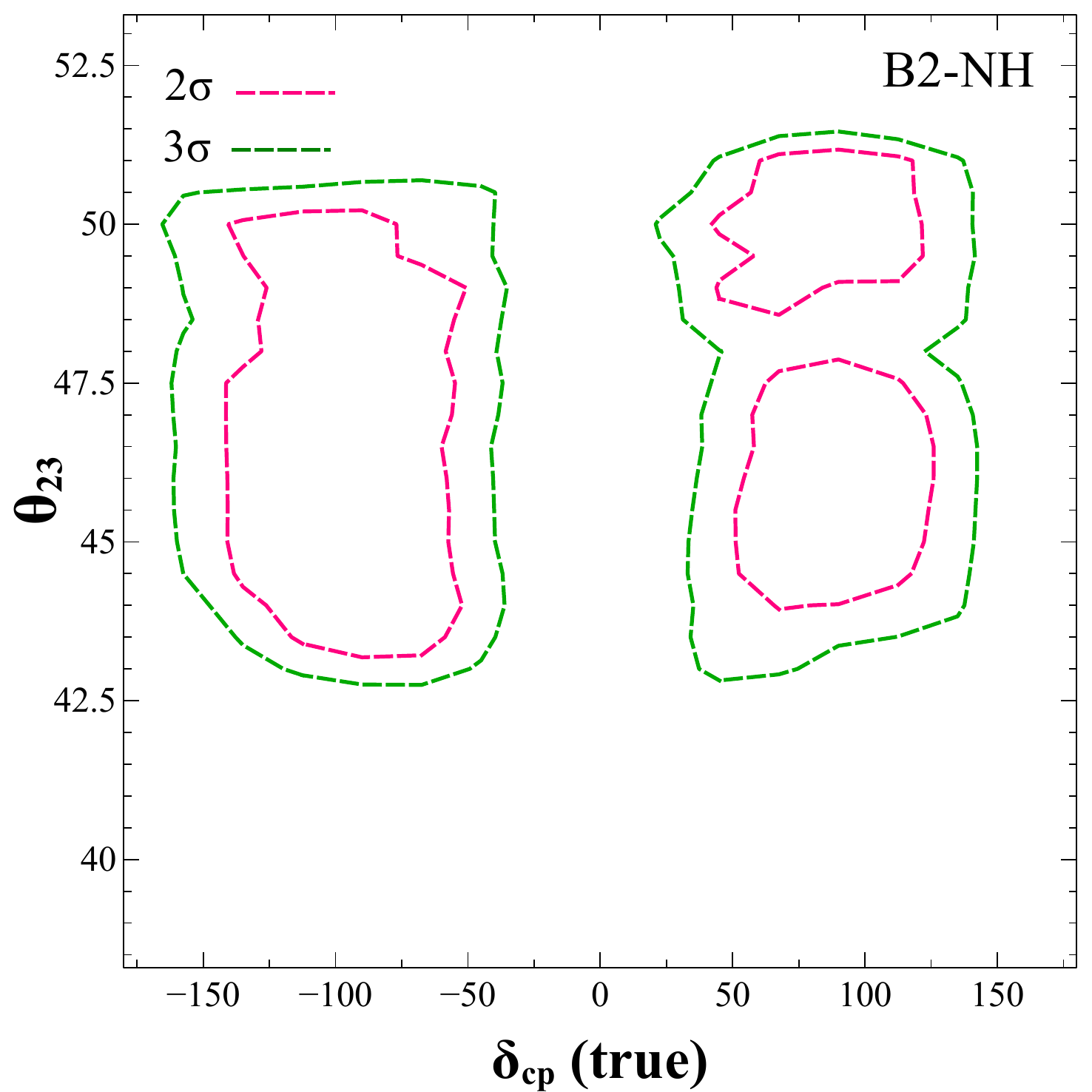}
\includegraphics[width=0.45\textwidth]{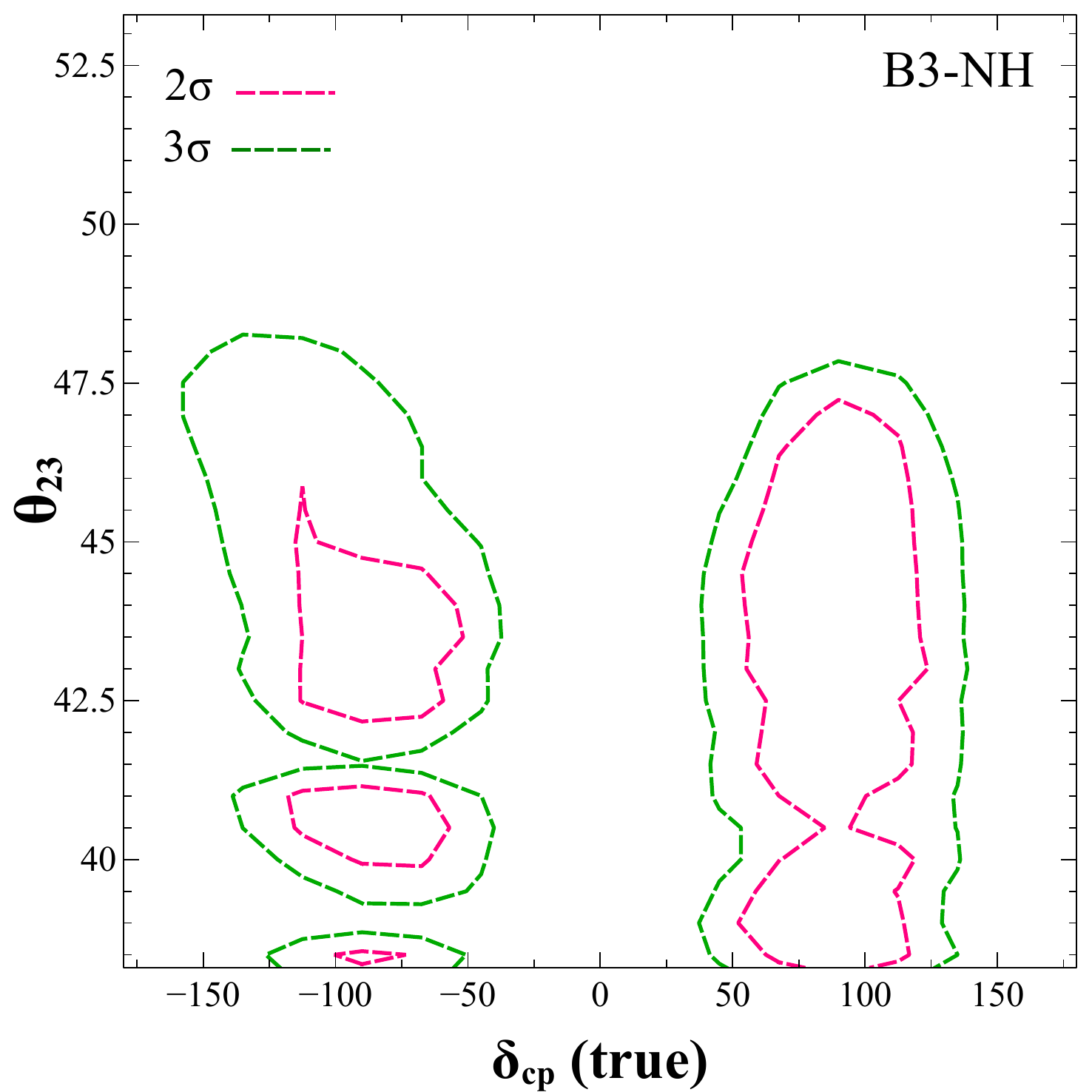}
\includegraphics[width=0.45\textwidth]{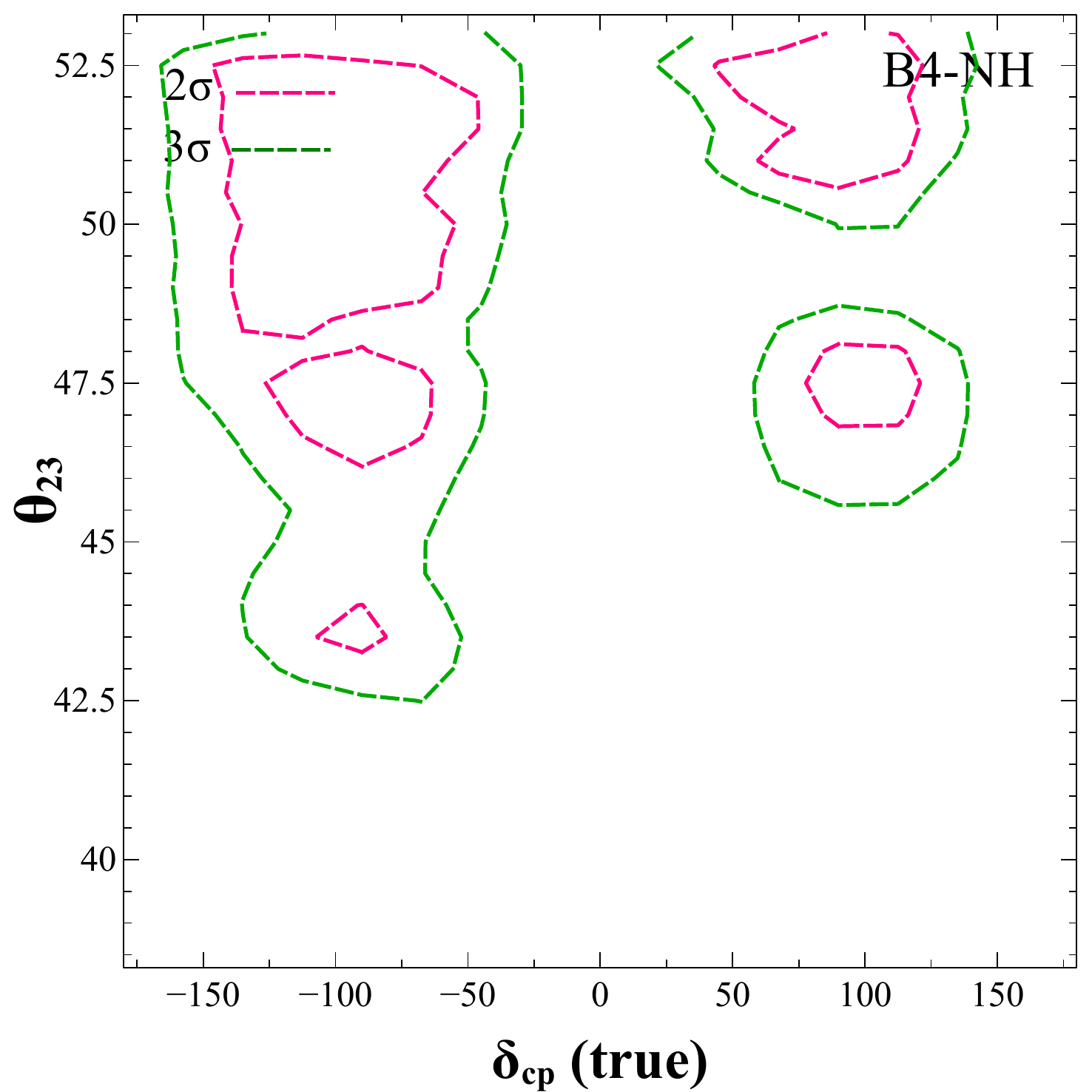}
\caption{In the contour plots, we show the allowed regions corresponding to different two-zero textures in NH mode. The green (red) dotted line represents the 3$\s$ (2$\s$) contour at 1 d.o.f.}
\label{figkk}
\end{center}
\end{figure*}

 \begin{figure*}
\begin{center}
\includegraphics[width=0.47\textwidth]{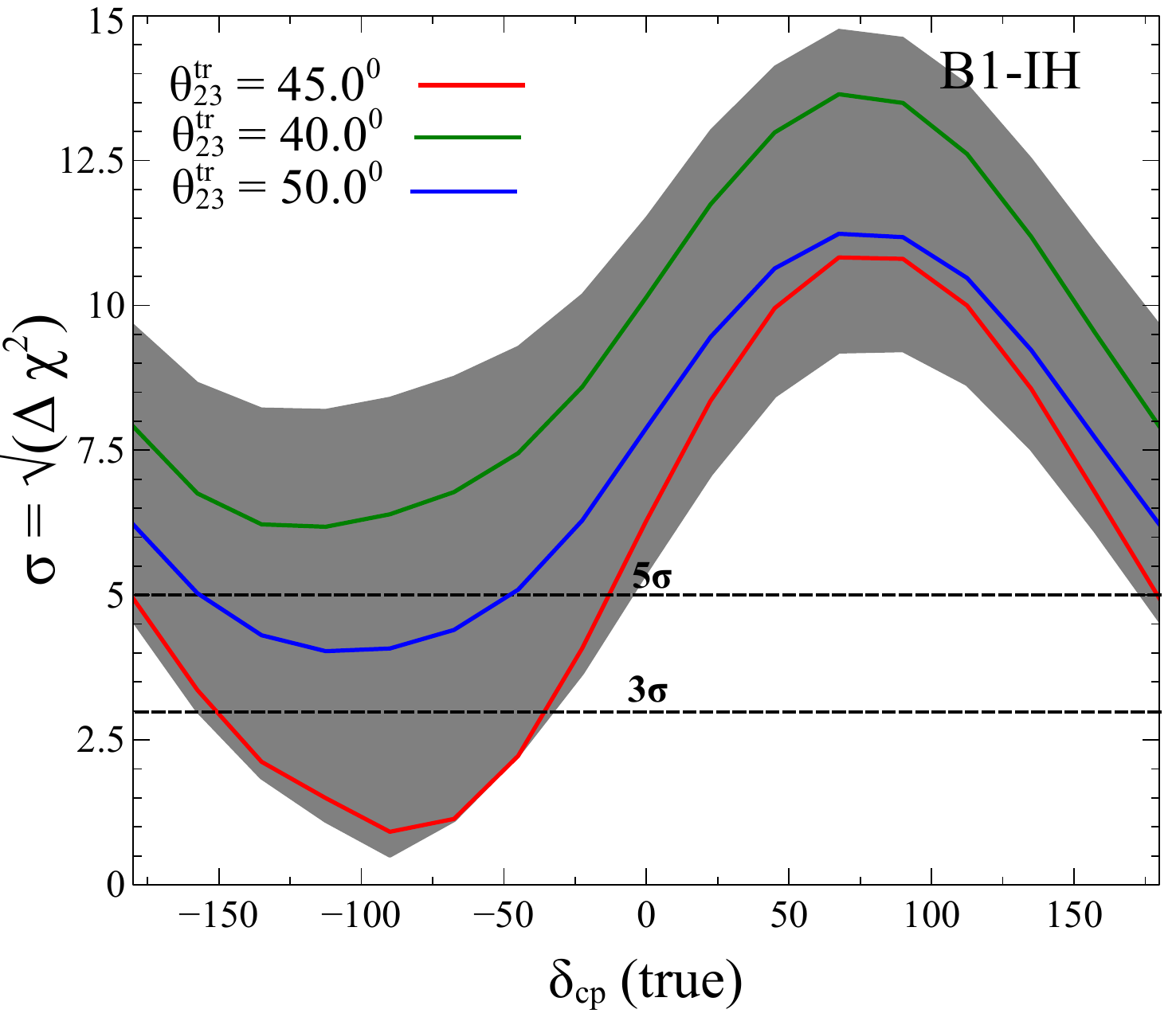}
\includegraphics[width=0.47\textwidth]{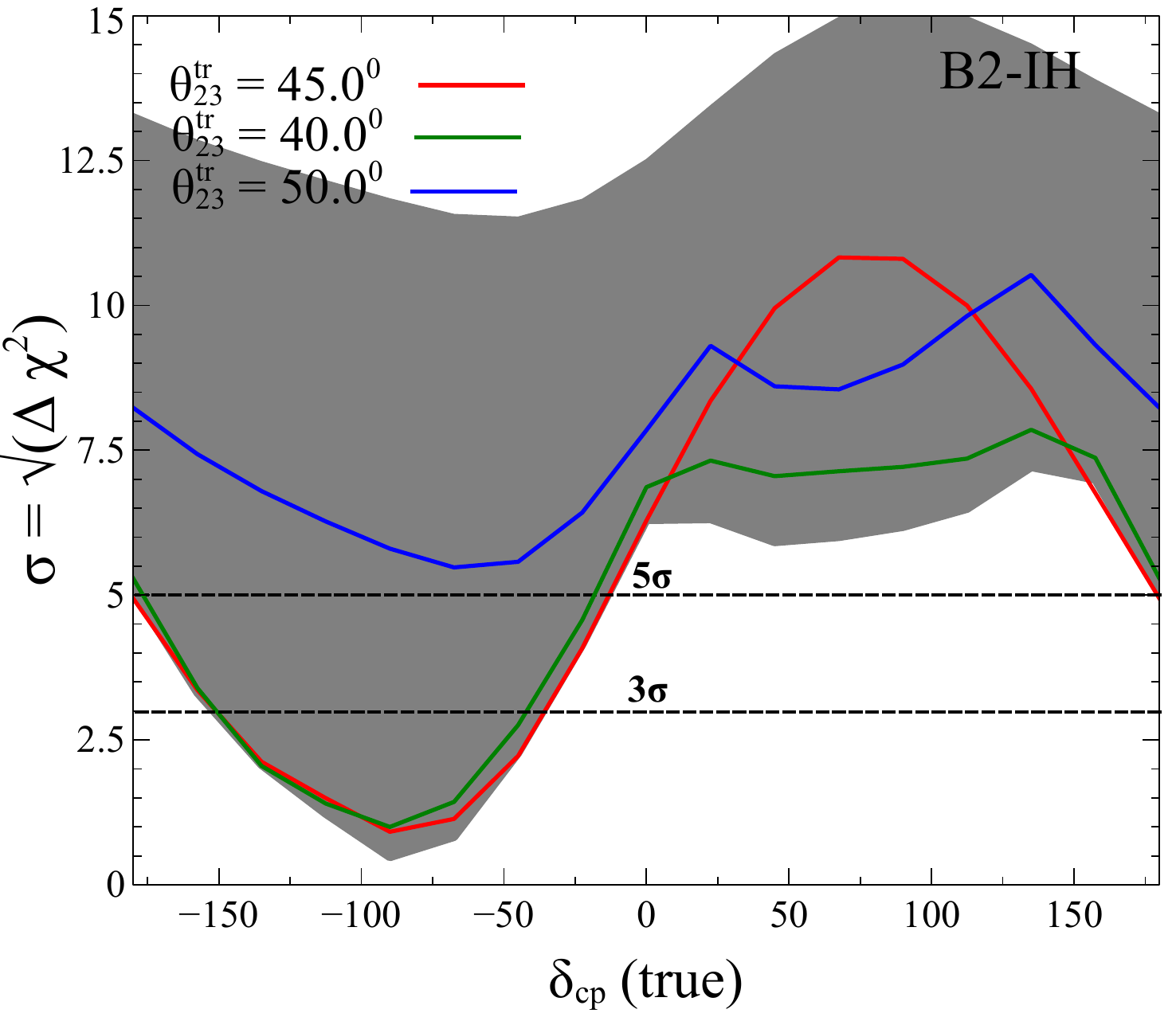}
\includegraphics[width=0.47\textwidth]{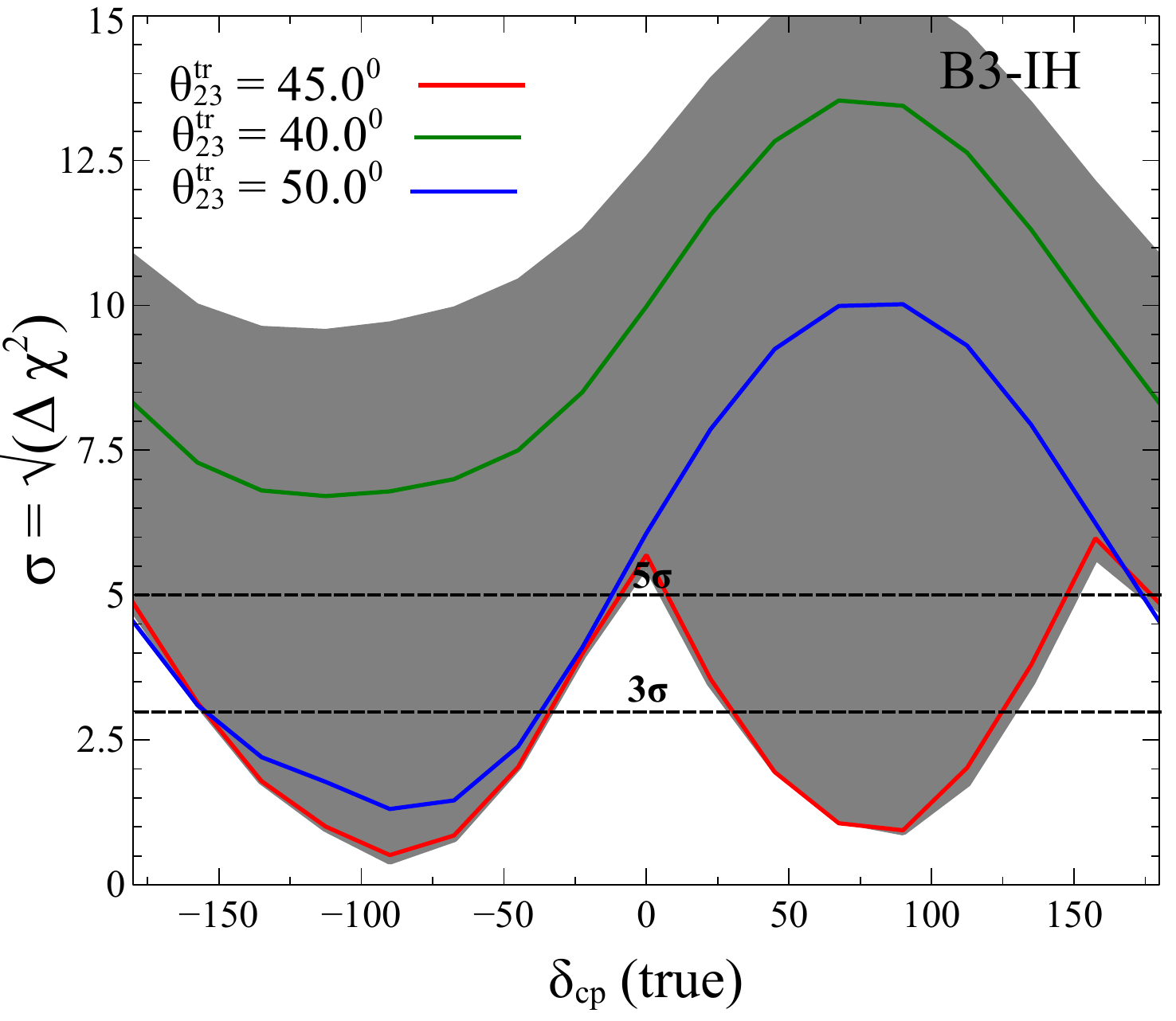}
\caption{The exclusion of different two-zero textures at DUNE assuming IH as the true hierarchy. In the upper panel, the grey band represents the full variation of $\tz$ in its 3$\s$ allowed range. Green, blue and the red plots corresponds to three different choice of $\tz$ i.e. green plot is for the best fit value of $\tz$ in the LO while the blue plot is for $\tz$ in the HO. The red plot is for maximal $\tz$.}
\label{fig6}
\end{center}
\end{figure*}

 \begin{figure*}
\begin{center}
\includegraphics[width=0.47\textwidth]{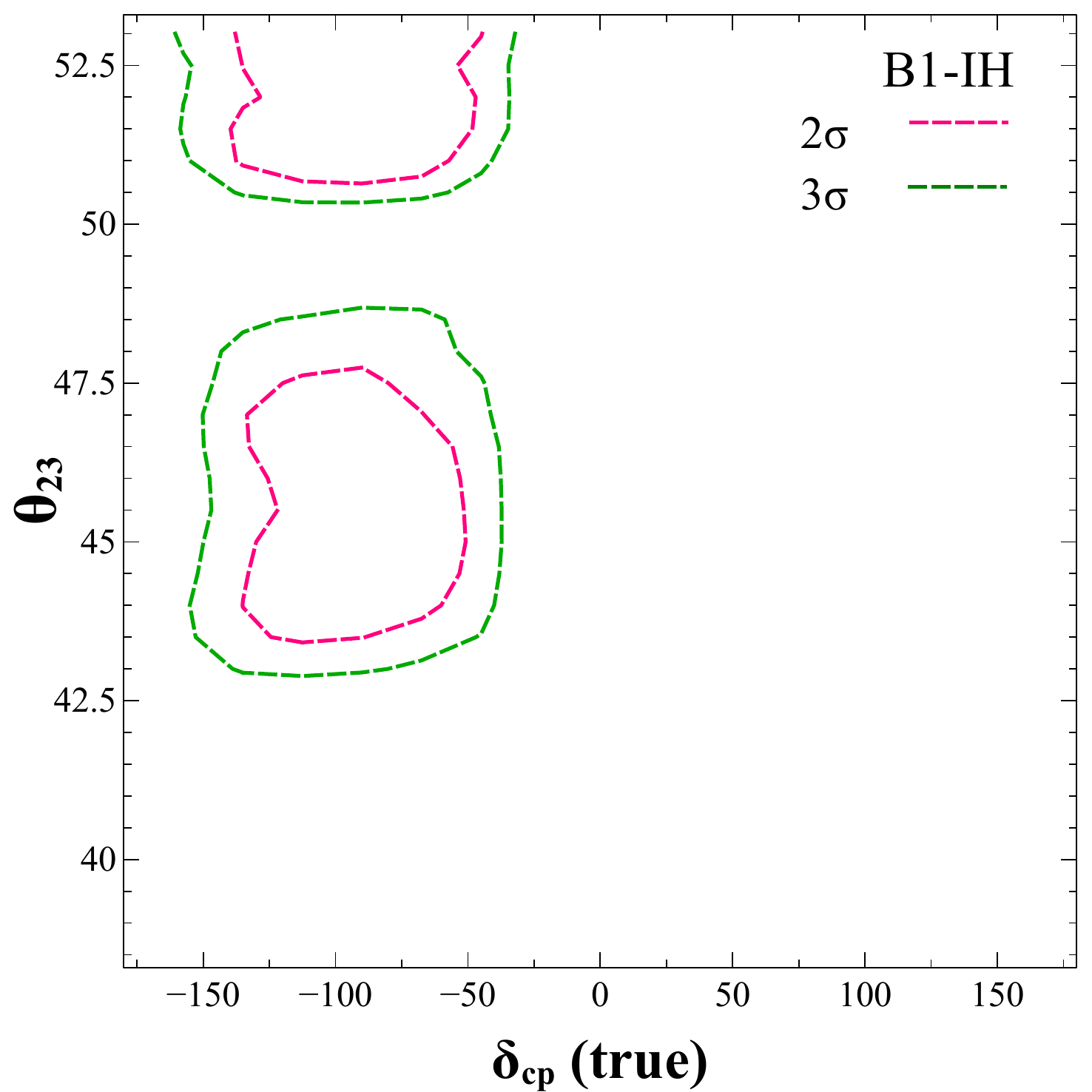}
\includegraphics[width=0.47\textwidth]{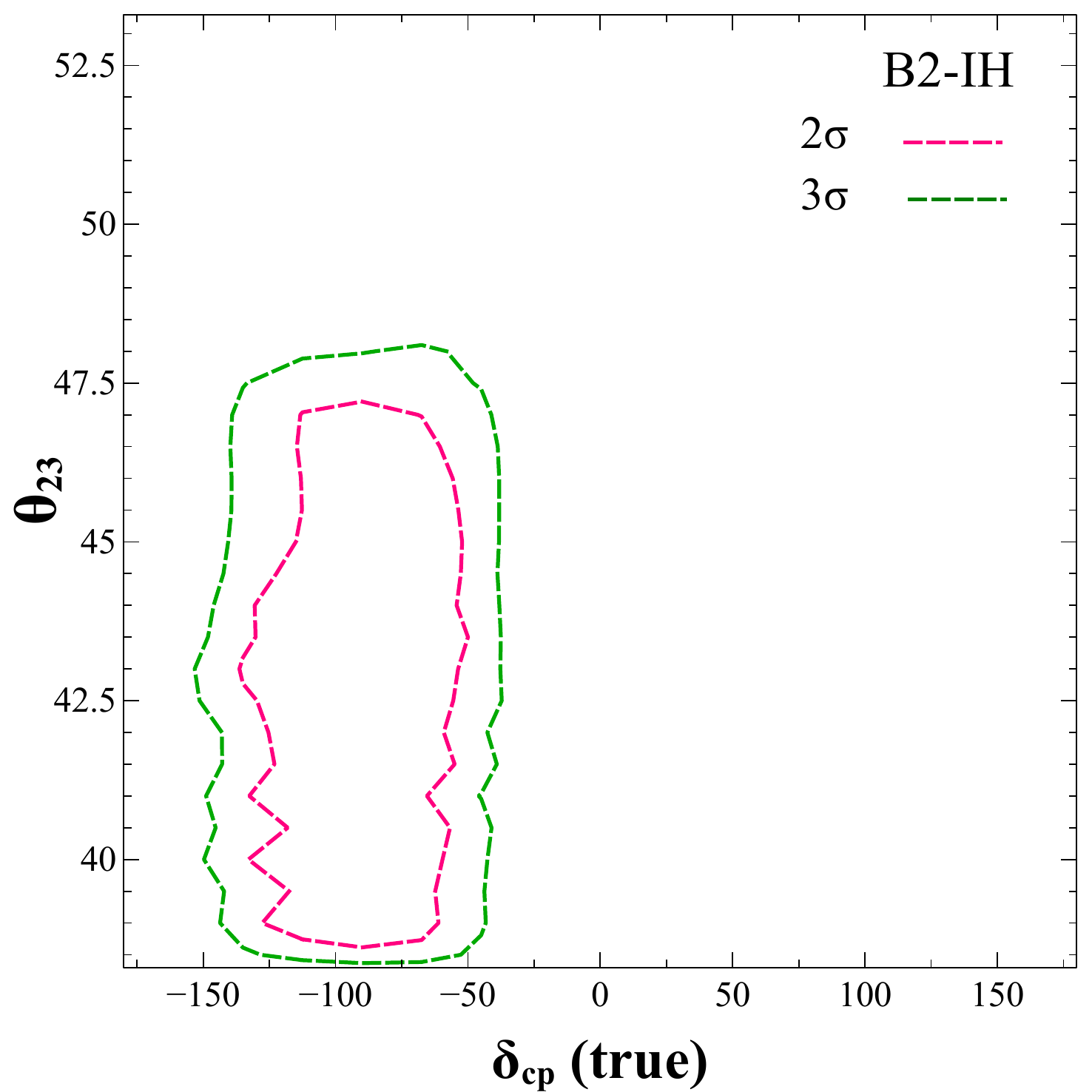}
\includegraphics[width=0.47\textwidth]{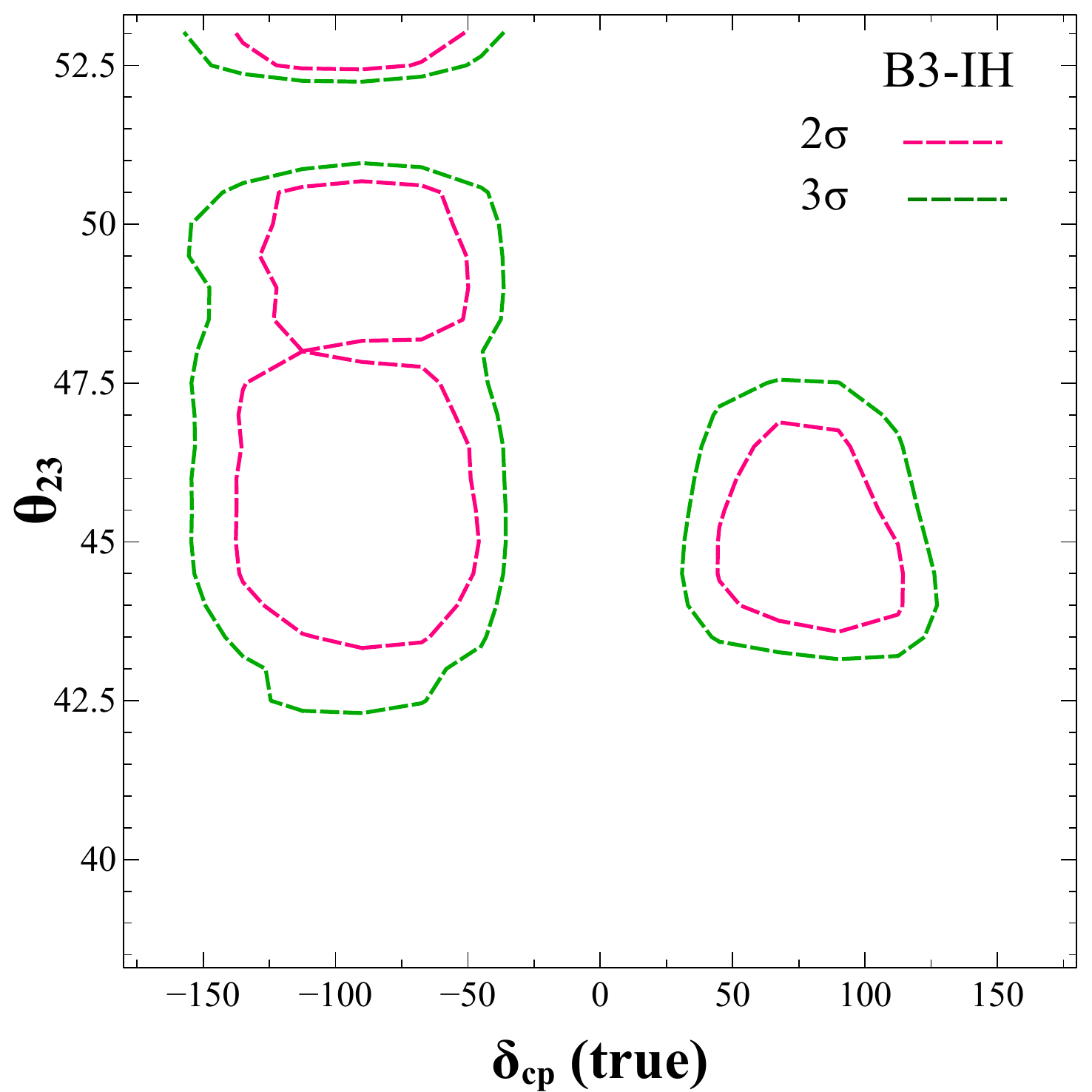}
\caption{In the contour plots, we show the allowed regions corresponding to different two-zero textures in IH mode. The green (red) dotted line represents the 3$\s$ (2$\s$) contour at 1 d.o.f.}
\label{fig66}
\end{center}
\end{figure*}

In fig. \ref{figk}, we present our results for the two-zero textures: A1, B1, B2, B3 and B4 assuming true NH. The behaviour of the grey band is very much similar in case of B1, B2 , B3 and B4 and the most of the $\tz-\dcp$ parameter space is excluded at DUNE at 3$\s$. For $\tz=41.6^o$, DUNE can exclude B1, B2 and B4 at 3$\s$ for all true $\dcp$. Similarly, B1 and B3 is possible to exclude at 3$\sigma$ for all $\dcp$ if $\tz=48.4$. As seen from the contour plots in fig. \ref{figkk}, B1 is more predictive compared to all other two-zero texture in NH mode, as most of the  $\tz-\dcp$ parameter space is excluded at 3$\s$. Almost, for any $\tz$ in HO, DUNE can exclude B1 for all true values of $\dcp$ at 2$\s$ as seen from fig. \ref{figkk}. For any $\tz>51.5^o$ and $\tz<42.5^o$, DUNE can exclude B2 for all true $\dcp$ at 3$\s$ C.L.. Similarly, DUNE can exclude both B3 and B4 for all true $\dcp$ at 3$\s$, if $\tz >48.5^o$ and $\tz<42.5^o$ respectively. A1 is less predictive as it allows most of the $\tz-\dcp$ parameter space at 3$\s$. Except A1, all other two-zero textures in NH mode are excluded at DUNE for the CP conserving values almost at 5$\s$ C.L.(see fig. \ref{figk}).

In fig. \ref{fig6} and \ref{fig66}, we have shown our results for B1, B2 and B3 textures assuming IH as the true hierarchy. It is observed that B1 and B2 textures can be excluded at DUNE for all true $\dcp$ in the upper half plan (UHP, $\dcp$ from $0^o$ to $180^0$) for any values of $\tz$ at 5$\s$ C.L. If  $\tz = 40.0^o$, then DUNE can exclude both B1 and B3 at 5$\s$ irrespective of any true $\dcp$. For this $\tz$, DUNE can exclude B2 for all $\dcp$ in the UHP at 5$\s$. Similarly, for maximal $\tz$, both B1 and B2 can be excluded at 5$\s$ in the UHP. If true $\tz = 50.0^o$, then DUNE have the potential to exclude both B1 and B2 at 3$\s$ for all true $\dcp$ and it can rule out B3 at 5$\s$ for all true $\dcp$ in the UHP. We observe from fig. \ref{fig66} that most of the $\tz-\dcp$ parameter space is ruled out at 3$\s$ in case of B1 and B2. All $\tz$ are excluded at 3$\s$ for all true $\dcp$ in the UHP. In the lower half plane of $\dcp$ (LHP, $\dcp$ from $-180^o$ to $0^o$), DUNE can exclude both B1 and B2 at 3$\s$ if $\tz<43.0^o$ and $\tz>48^o$ respectively. DUNE can exclude  B3 at 3$\s$ for all true $\dcp$ if Nature chooses $\tz$ such that $\tz<42.5^o$. DUNE allows B3 almost for all $\dcp$ except some fractions around the CP conserving values. All these textures in IH mode are excluded at DUNE for the CP conserving values at 5$\s$ irrespective of any true $\tz$.

 \begin{figure*}
\begin{center}
\includegraphics[width=0.45\textwidth]{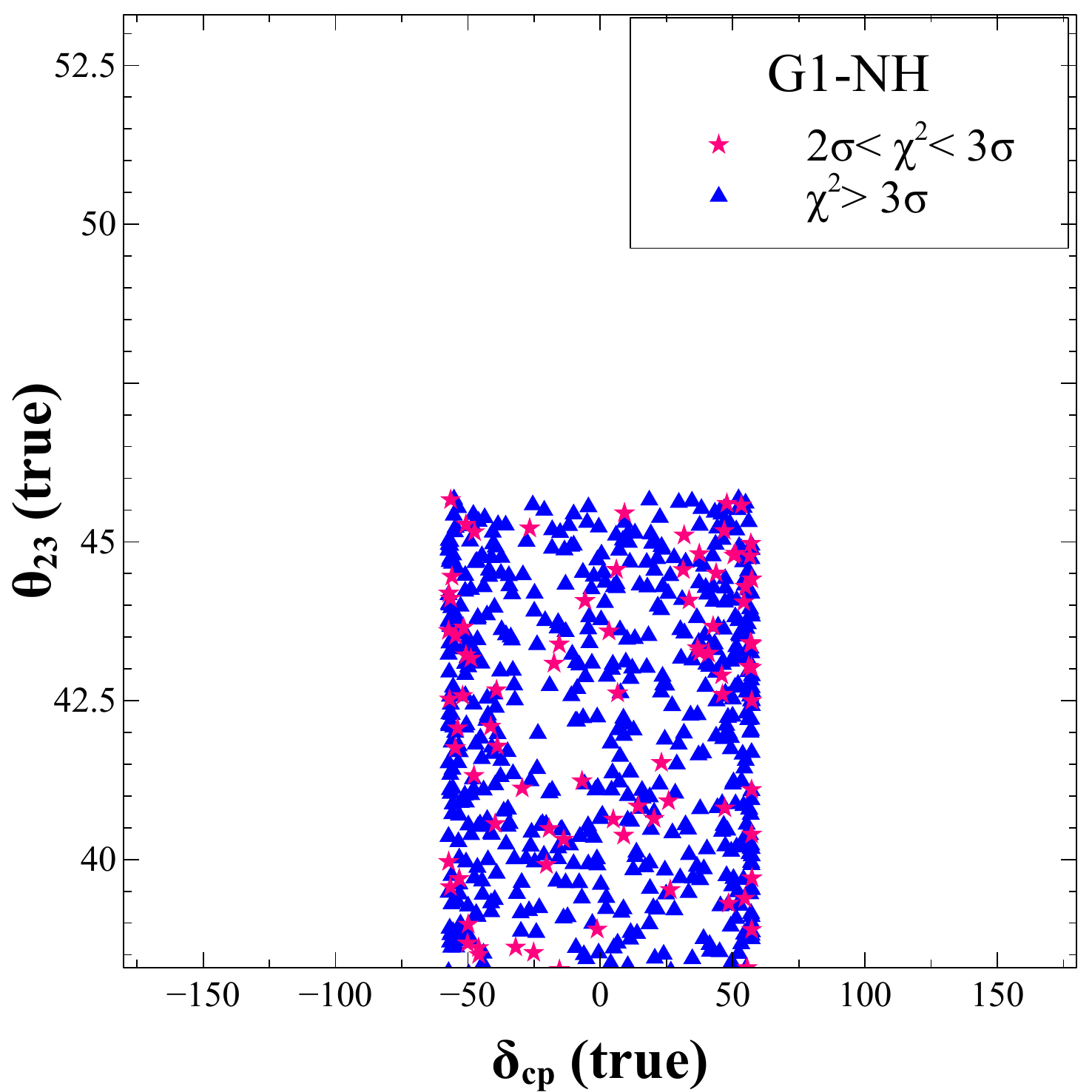}
\includegraphics[width=0.45\textwidth]{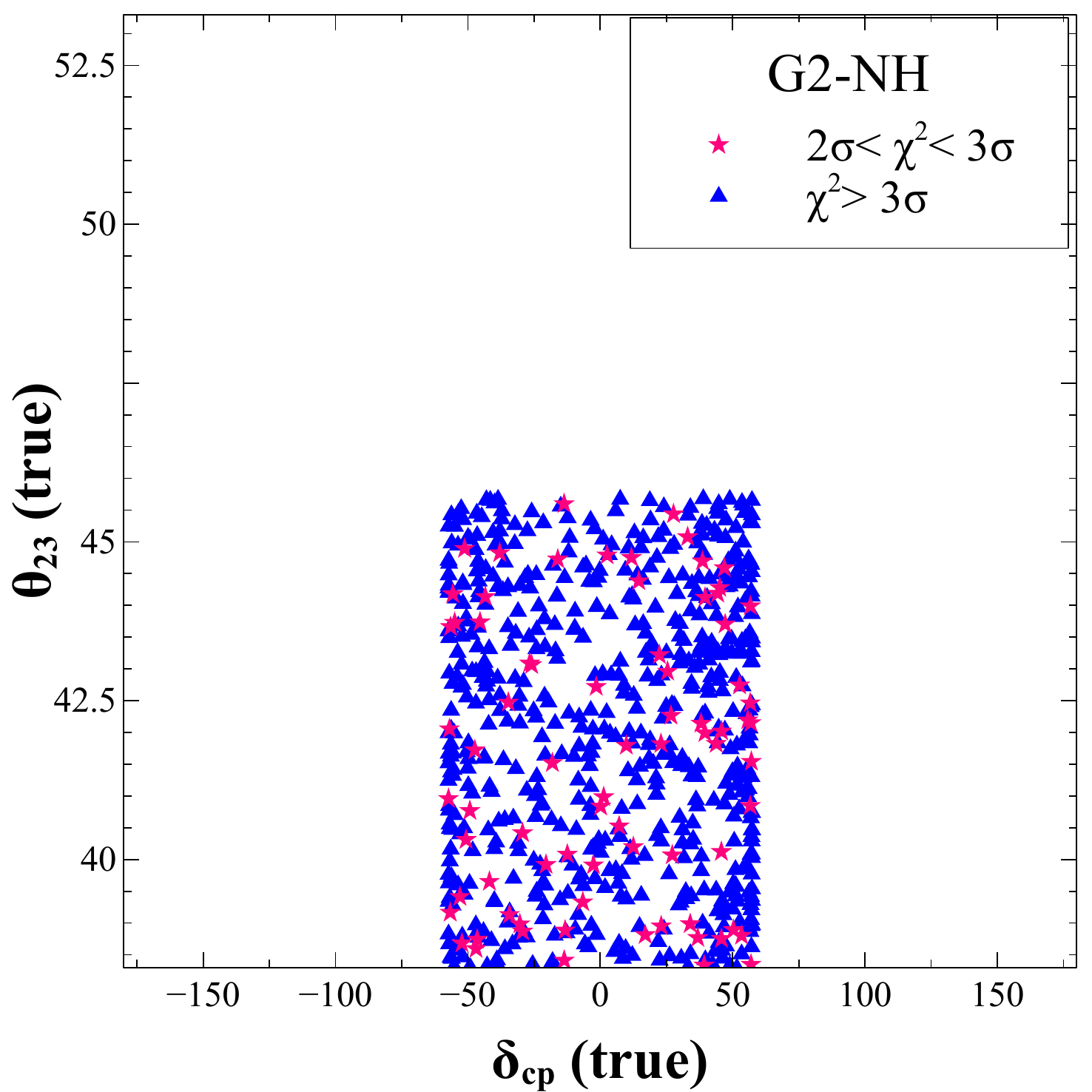}
\includegraphics[width=0.45\textwidth]{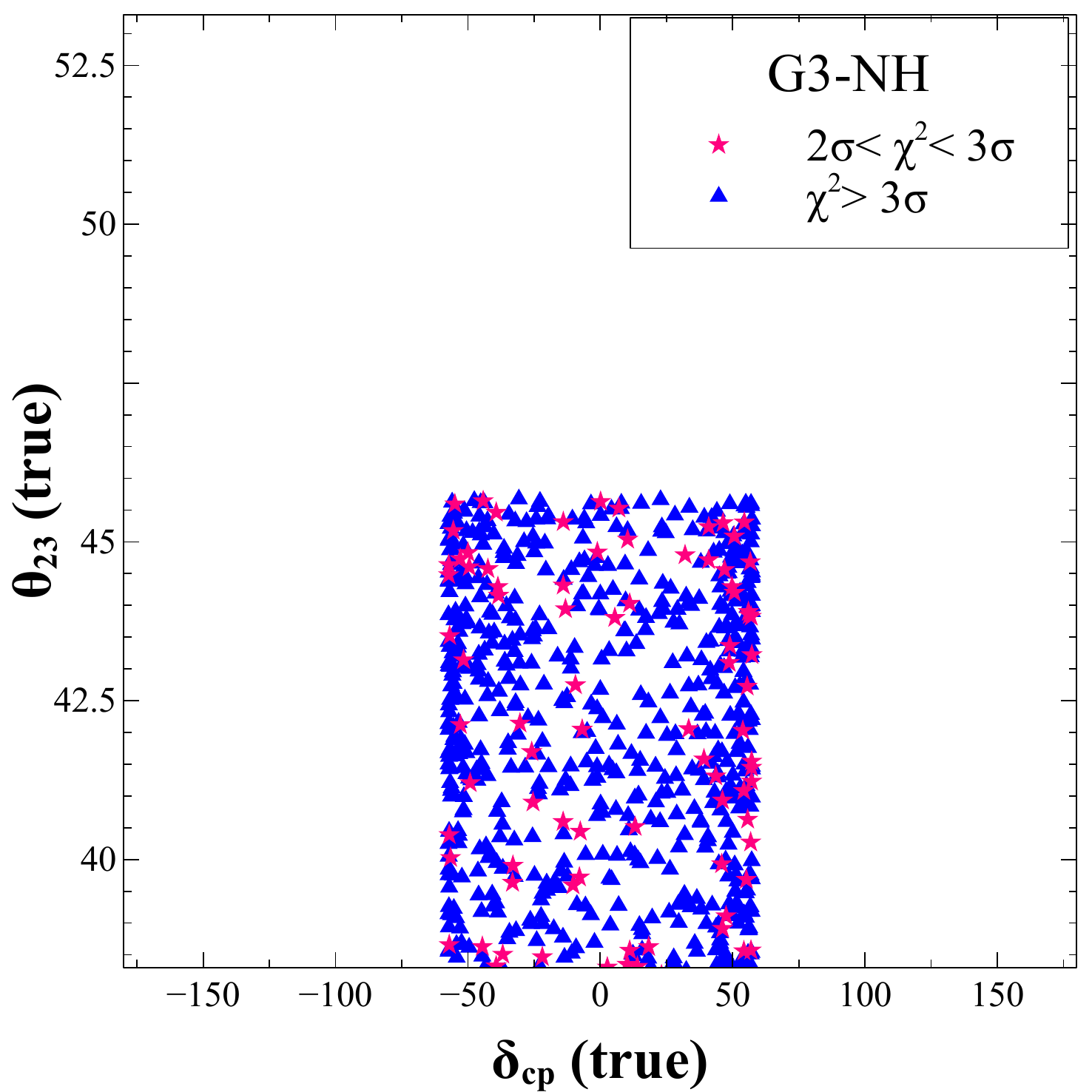}
\includegraphics[width=0.45\textwidth]{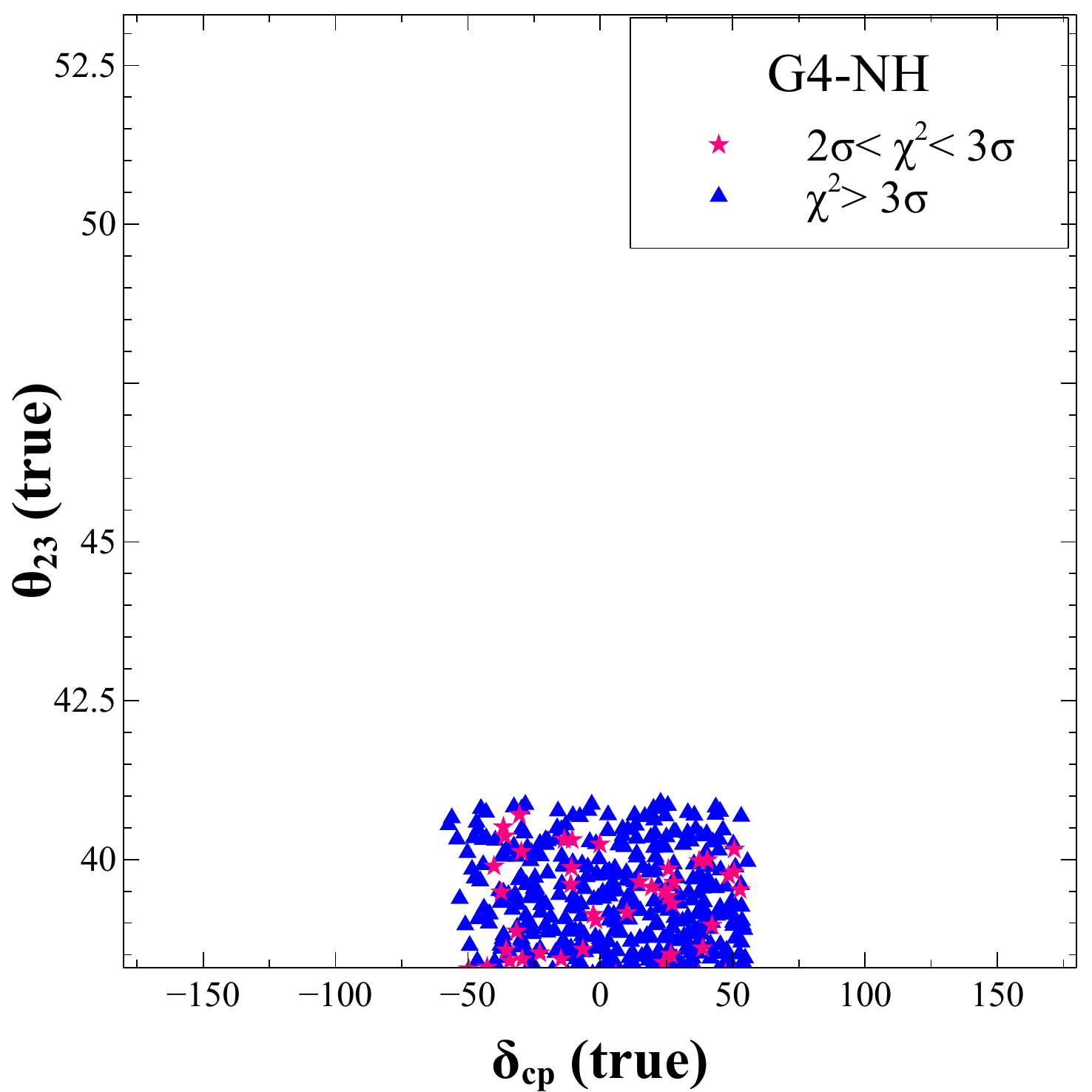}\\
\includegraphics[width=0.45\textwidth]{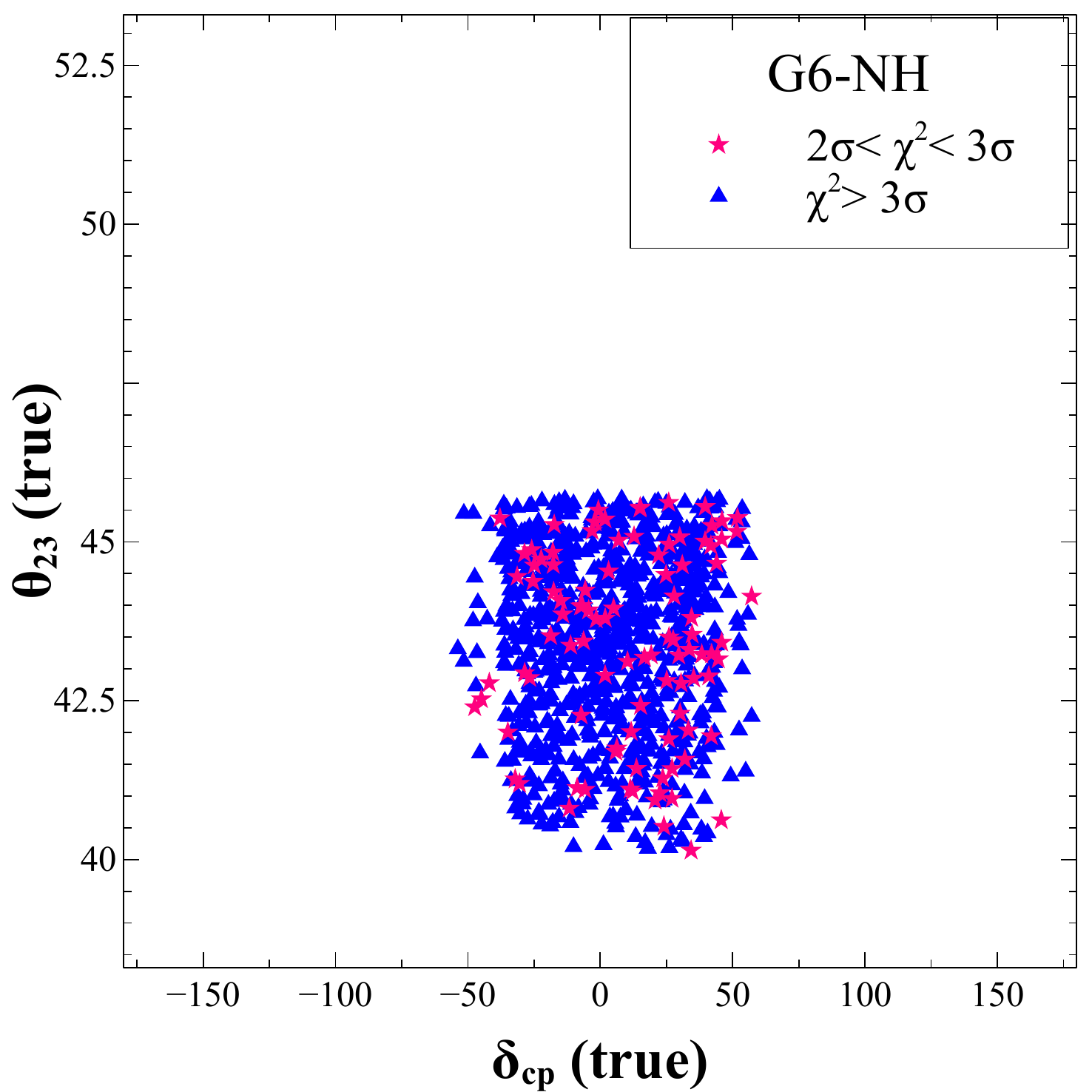}
\caption{Region of the $\theta_{23}-\delta_{cp}$ parameter space for which DUNE can establish one-zero texture against the present oscillation scenario for assumed true NH}
\label{fig91}
\end{center}
\end{figure*}

 \begin{figure*}
\begin{center}
\includegraphics[width=0.45\textwidth]{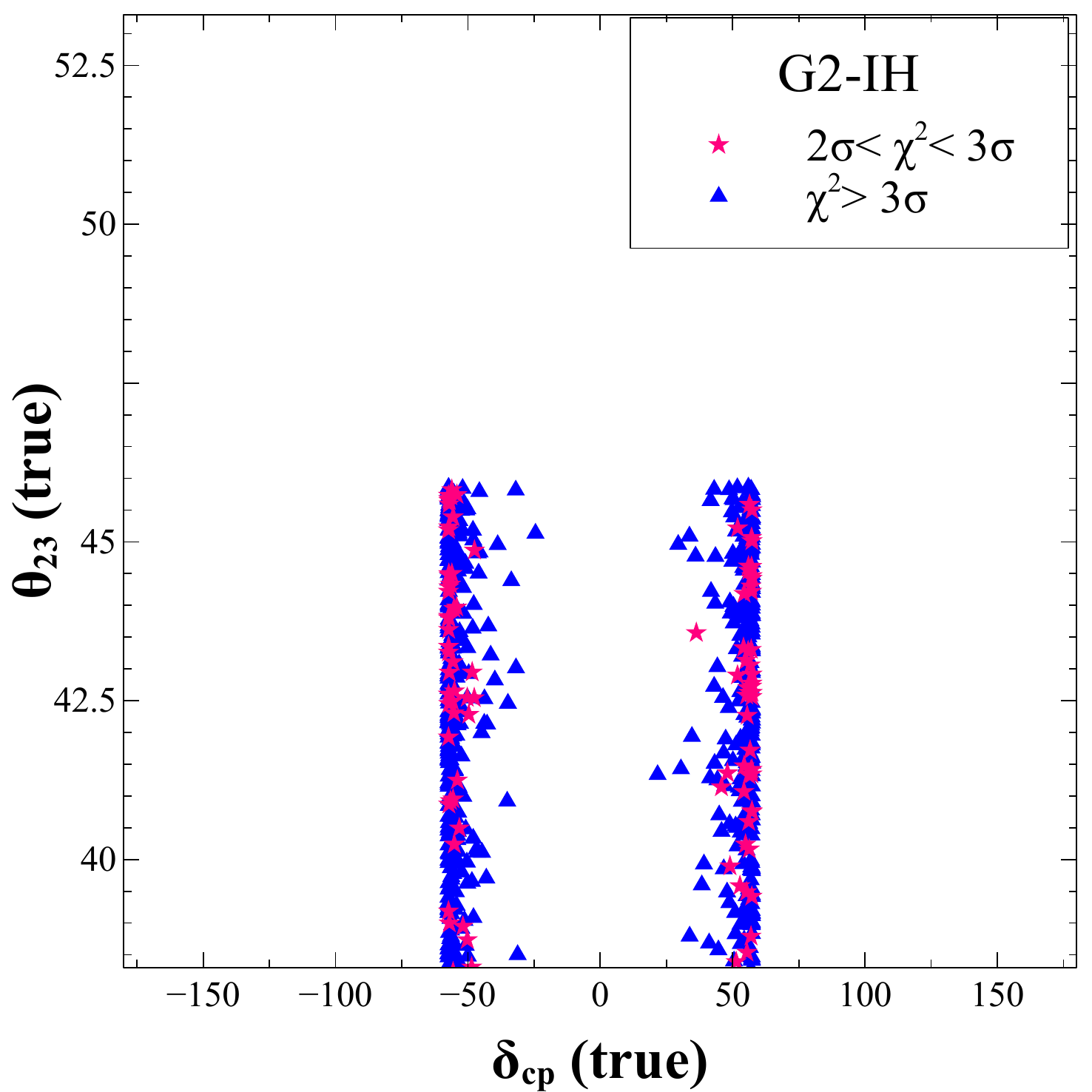}
\includegraphics[width=0.45\textwidth]{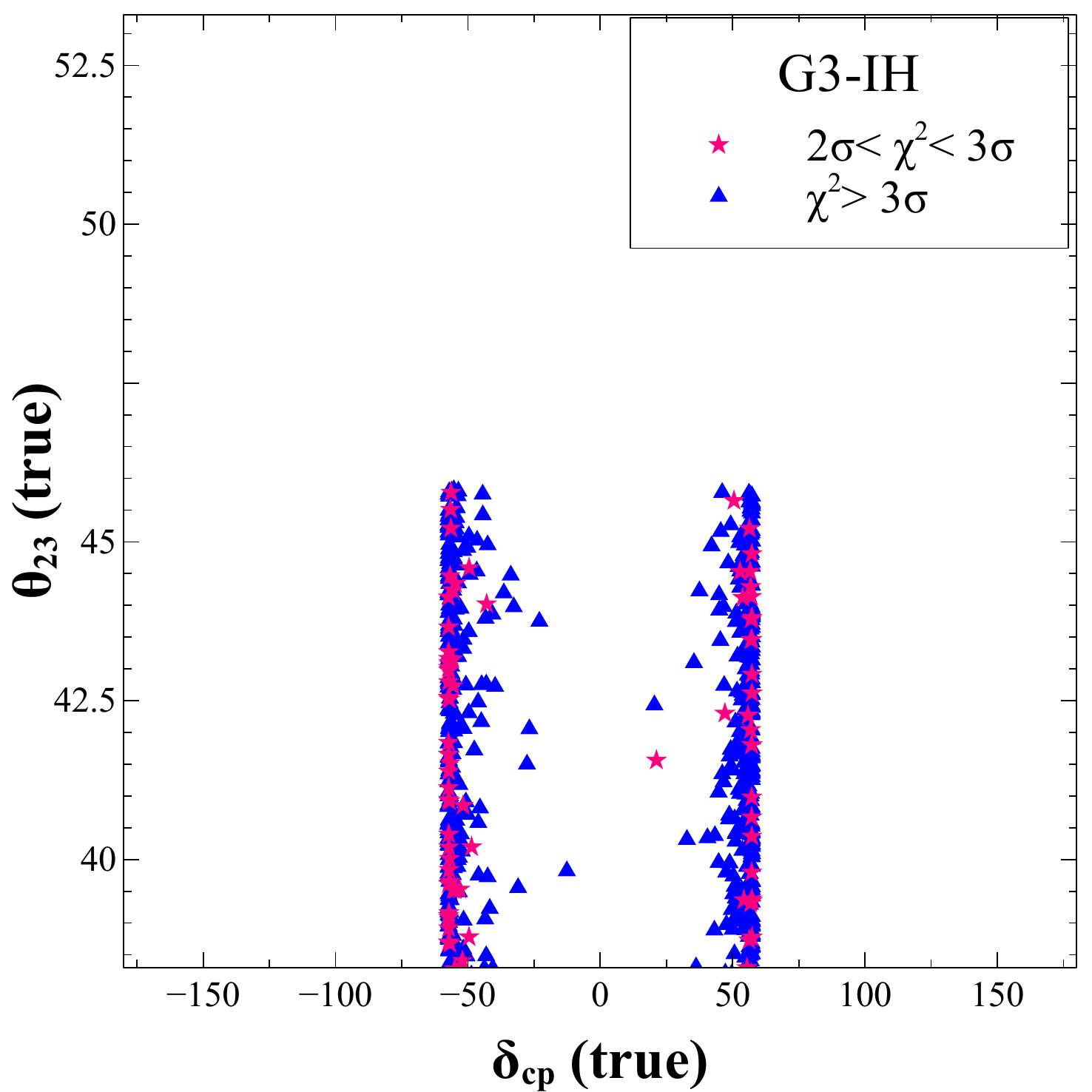}\\
\includegraphics[width=0.45\textwidth]{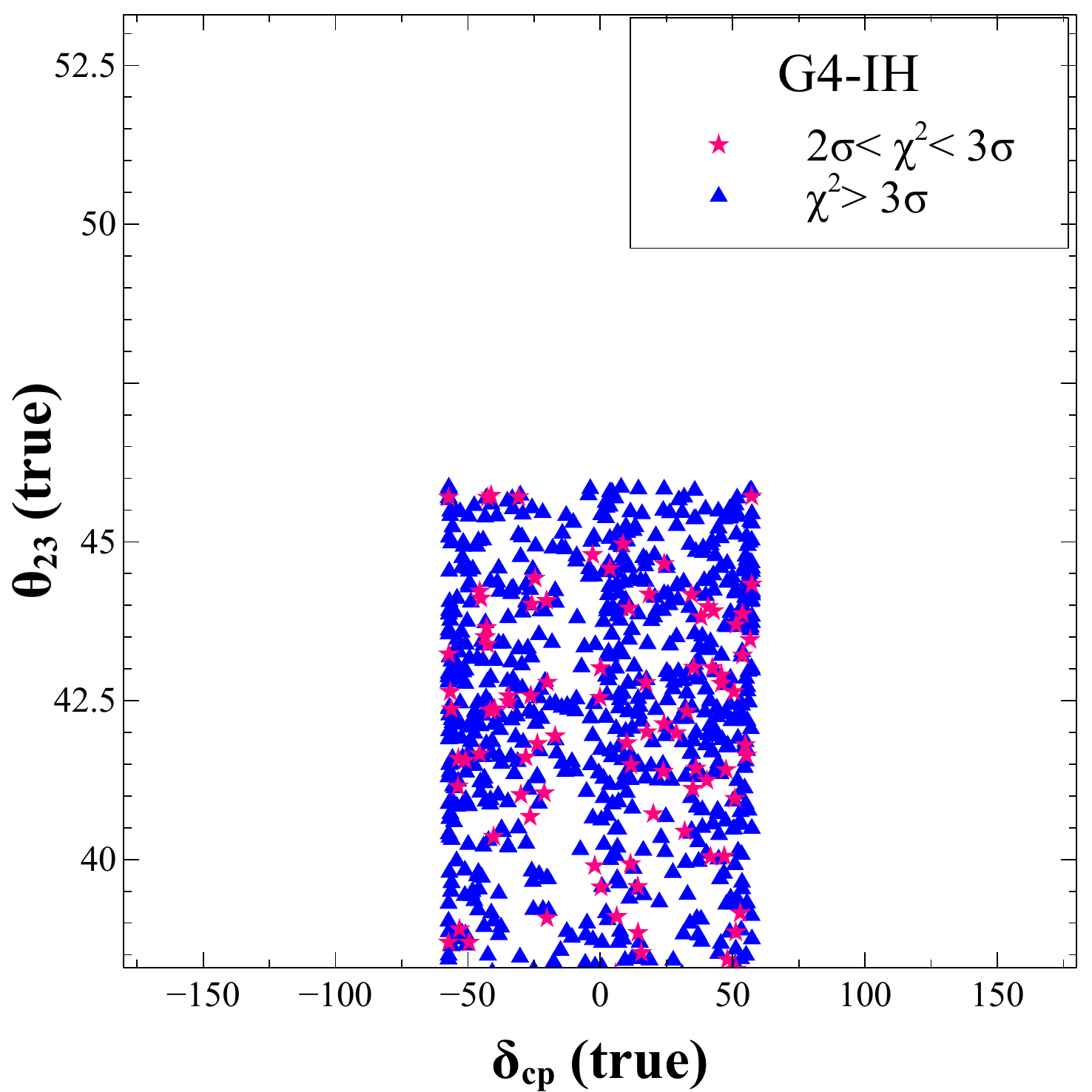}
\includegraphics[width=0.45\textwidth]{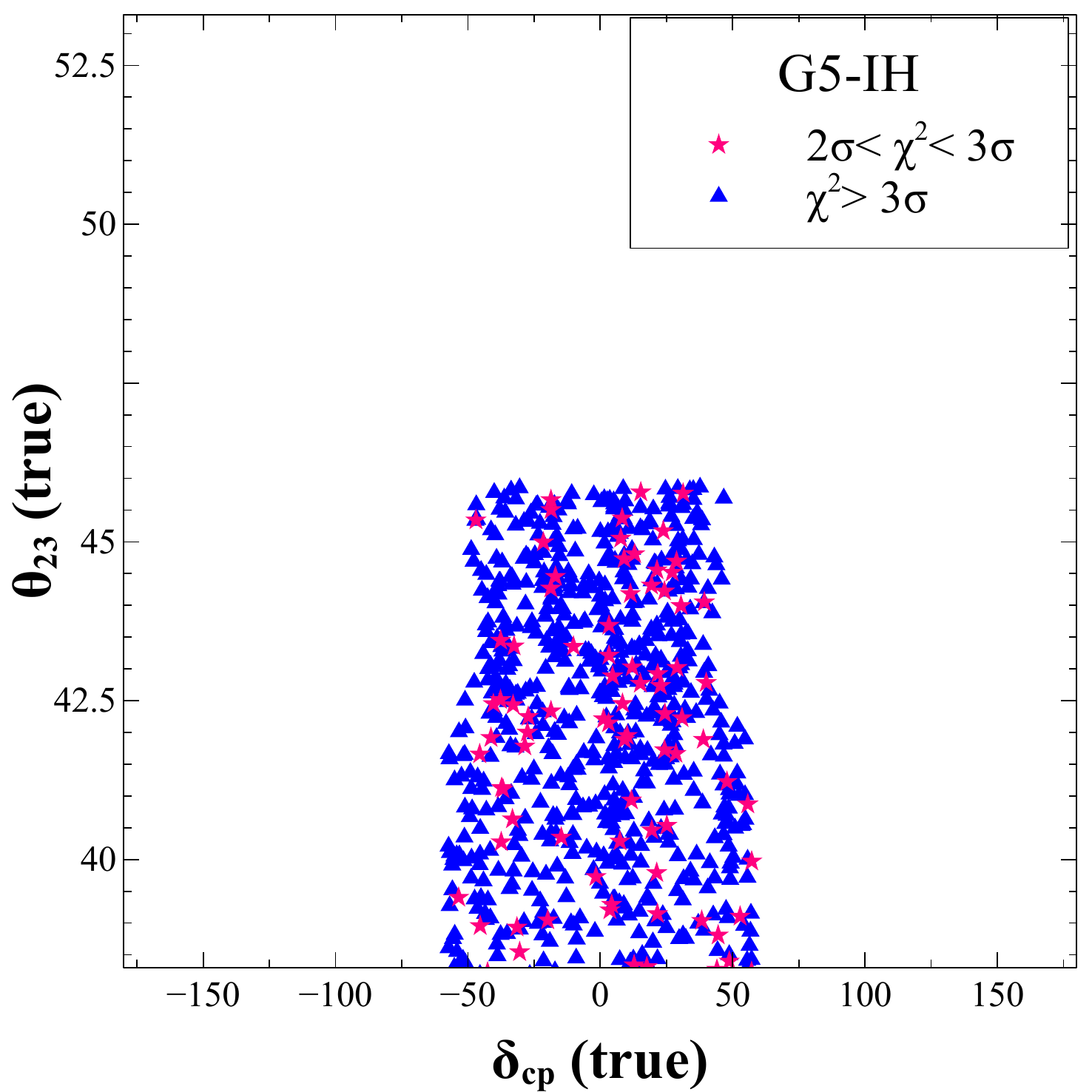}\\
\includegraphics[width=0.45\textwidth]{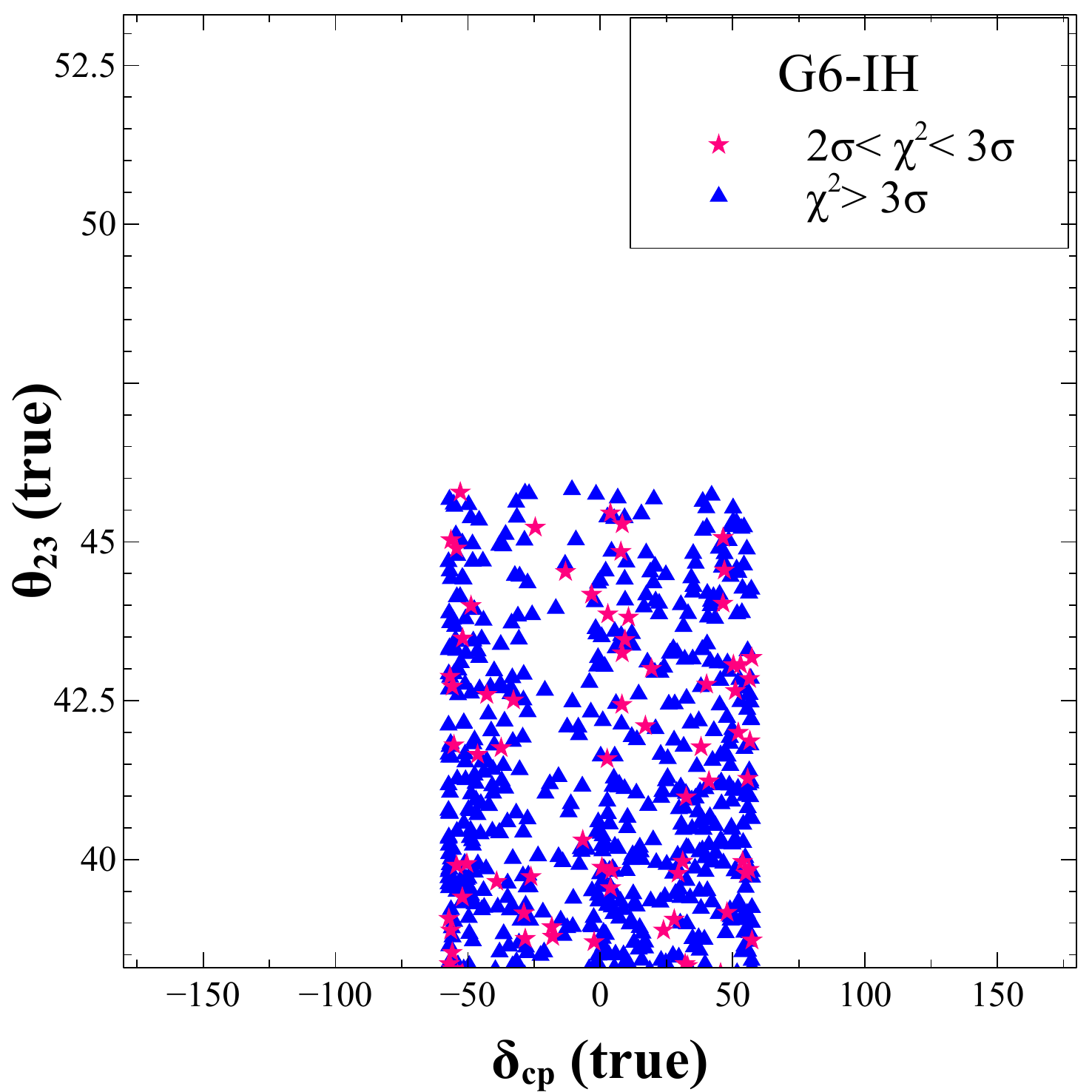}
\caption{Region of the $\theta_{23}-\delta_{cp}$ parameter space for which DUNE can establish one-zero texture against the present oscillation scenario for assumed true IH}
\label{fig92}
\end{center}
\end{figure*}

\begin{figure*}
\begin{center}
\includegraphics[width=0.45\textwidth]{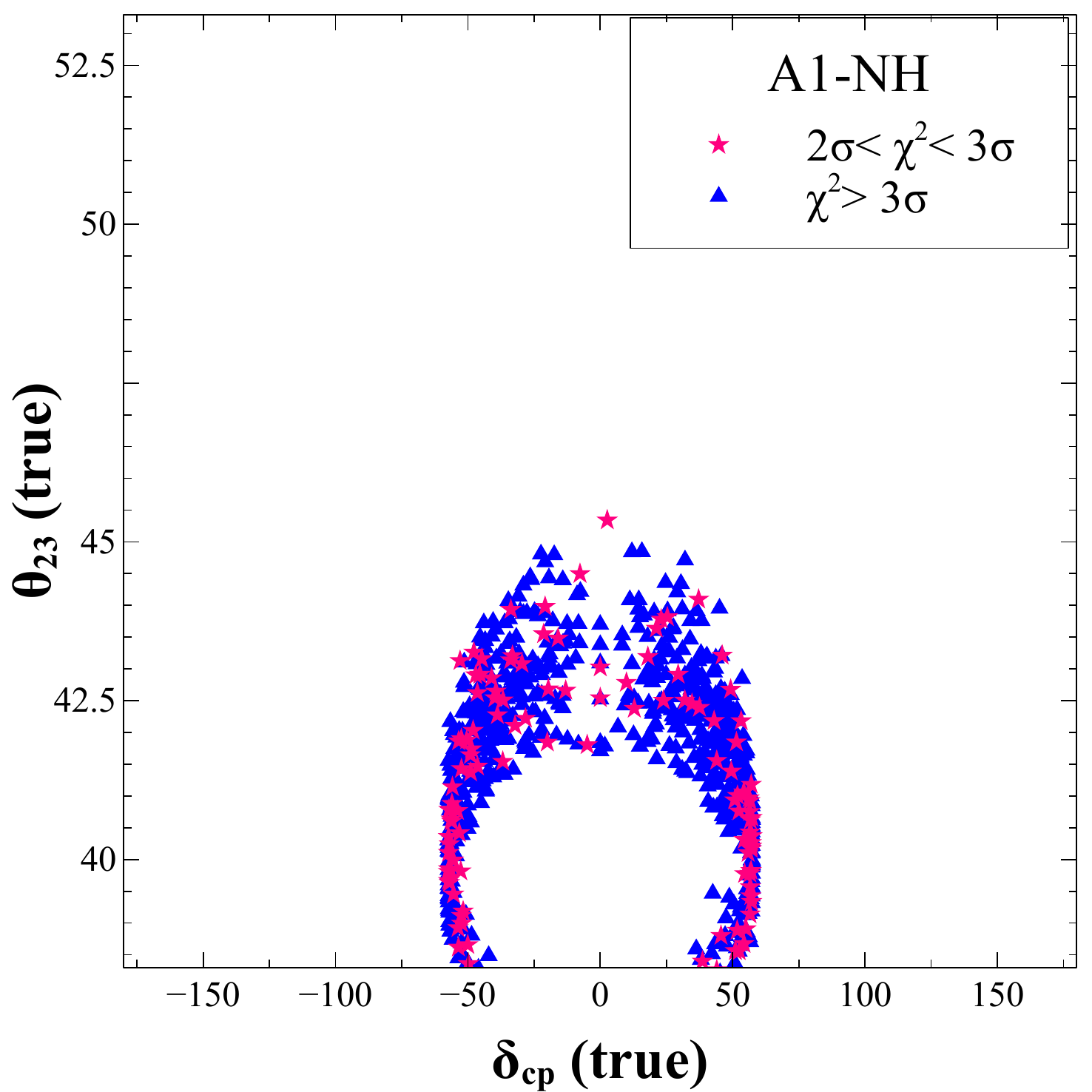}\\
\includegraphics[width=0.45\textwidth]{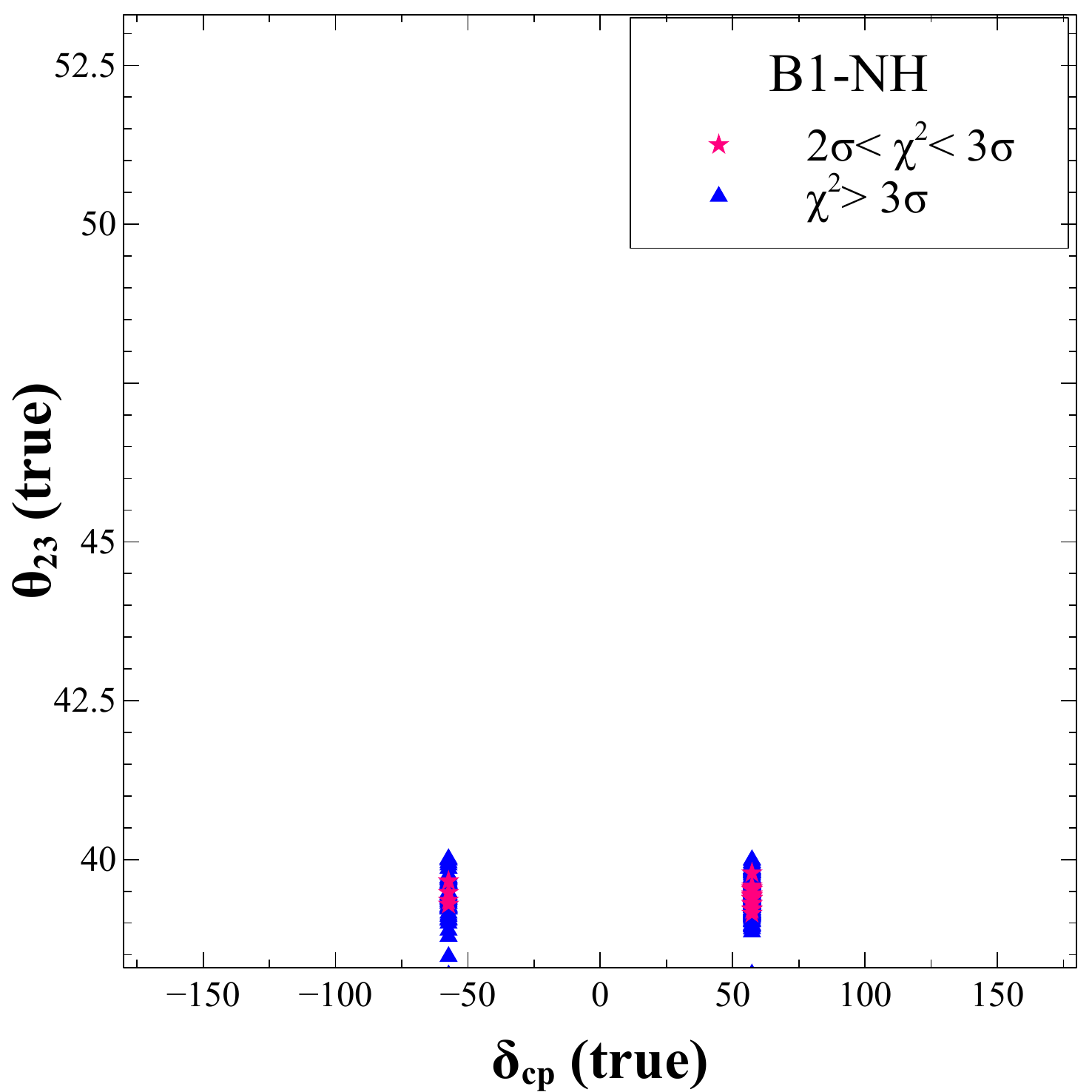}
\includegraphics[width=0.45\textwidth]{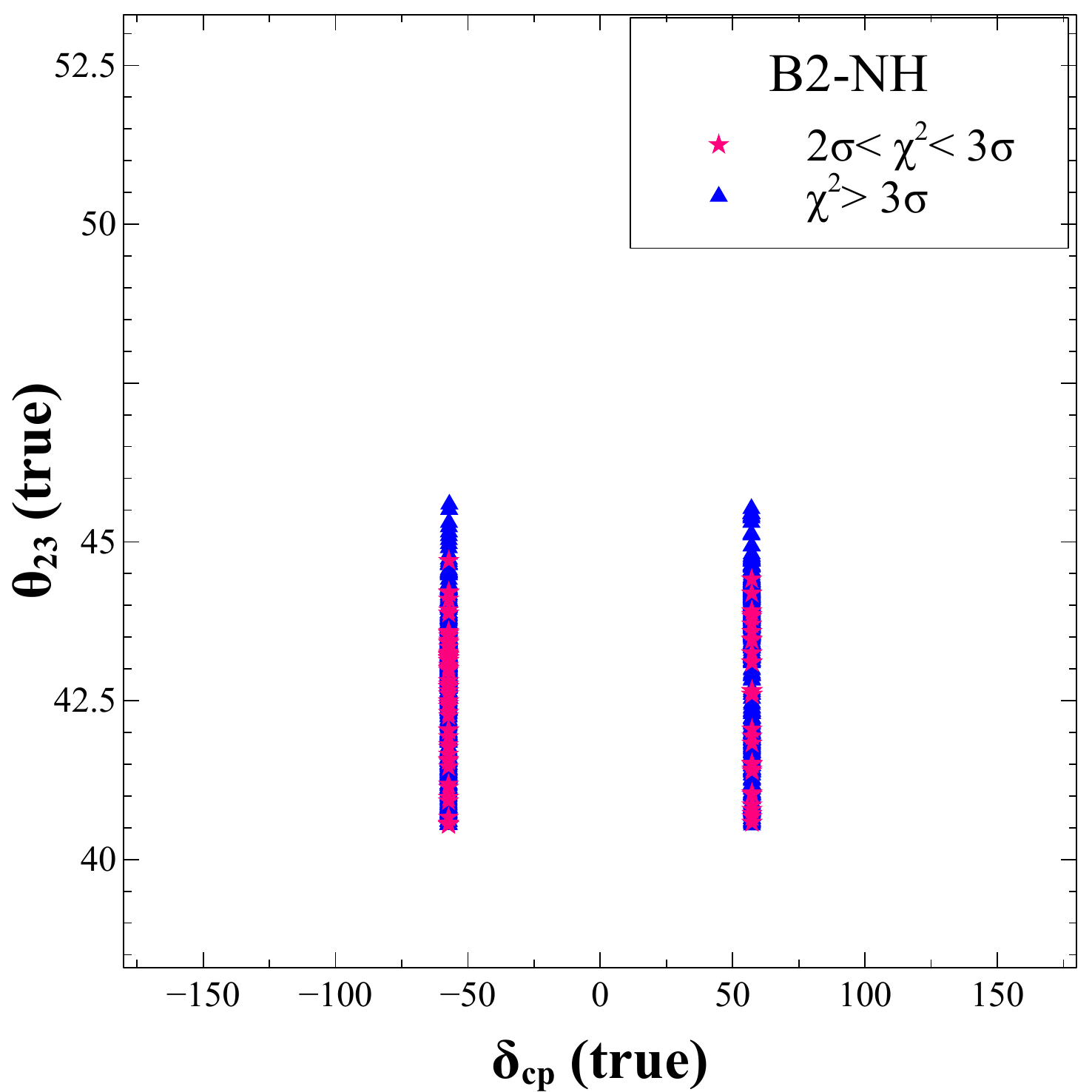}\\
\includegraphics[width=0.45\textwidth]{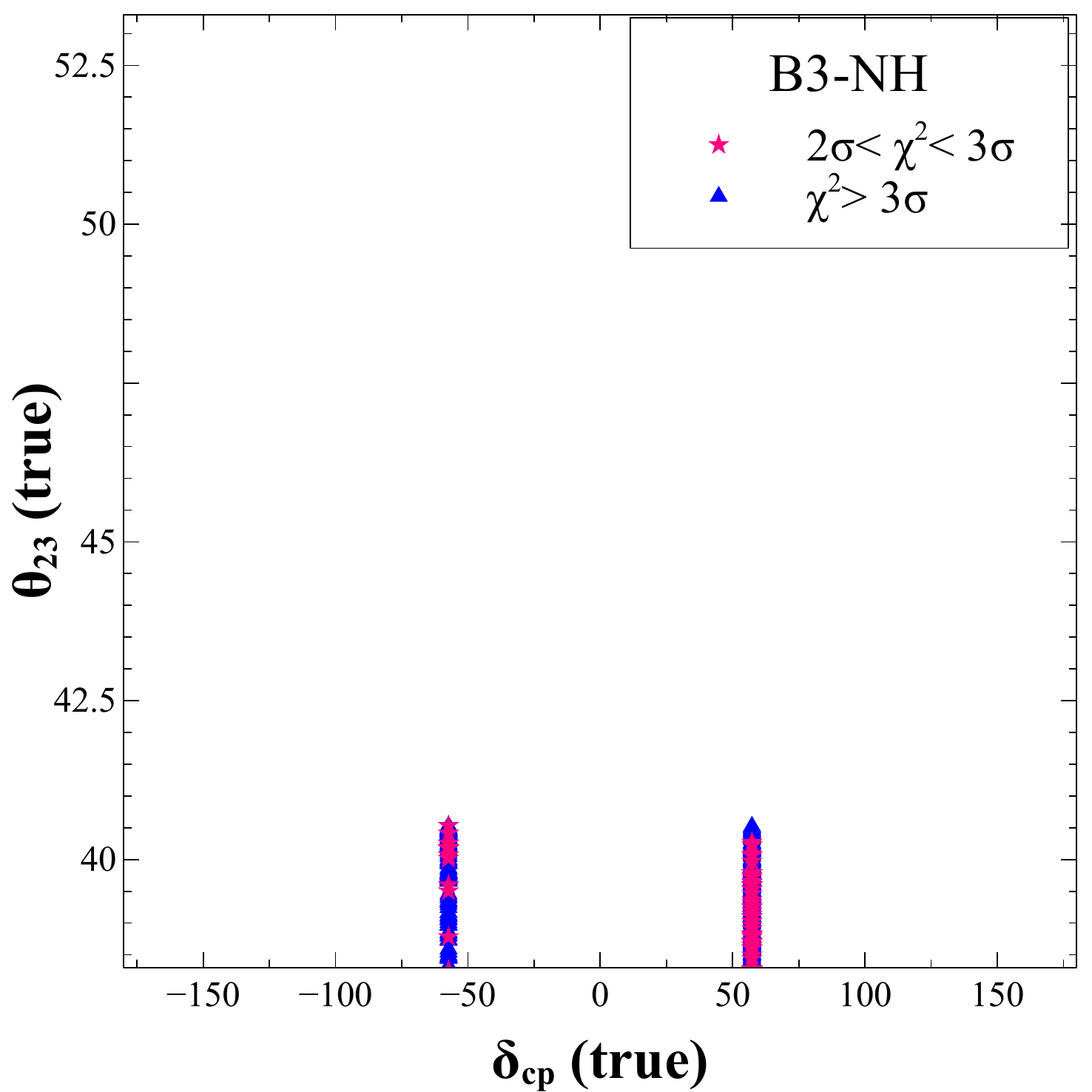}
\includegraphics[width=0.45\textwidth]{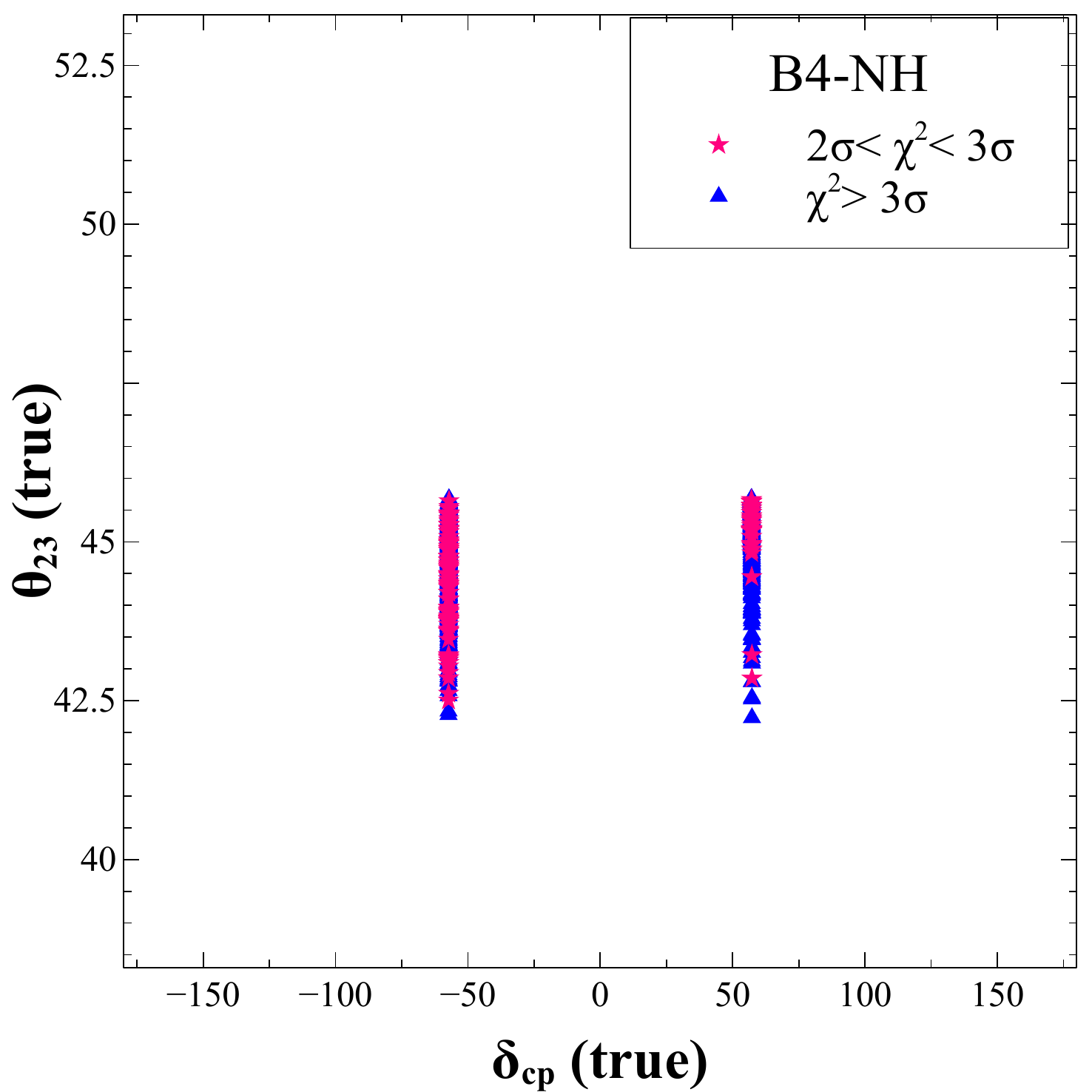}
\caption{Region of the $\theta_{23}-\delta_{cp}$ parameter space for which DUNE can establish two-zero texture against the present oscillation scenario for assumed true NH}
\label{fig93}
\end{center}
\end{figure*}

\begin{figure*}
\begin{center}
\includegraphics[width=0.45\textwidth]{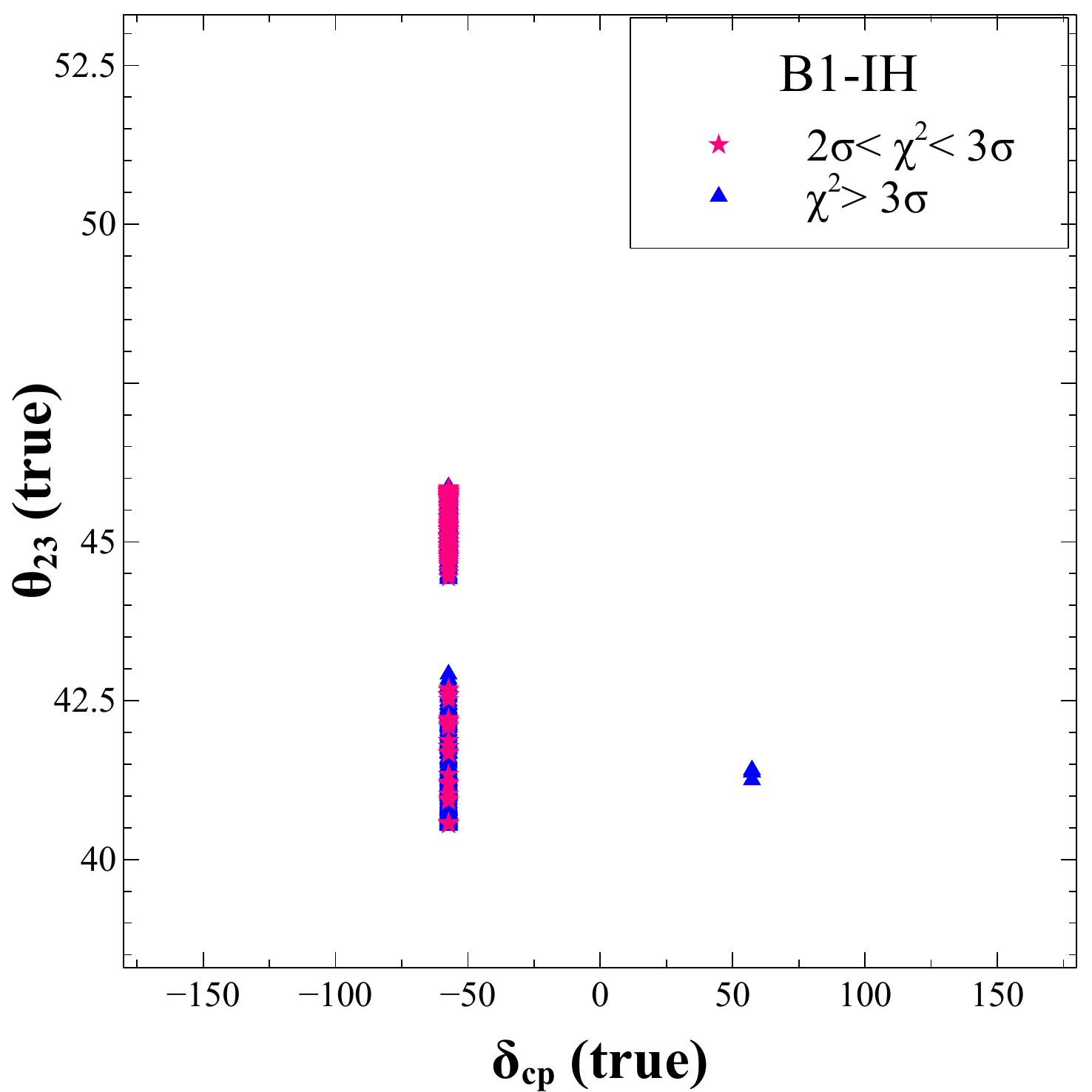}
\includegraphics[width=0.45\textwidth]{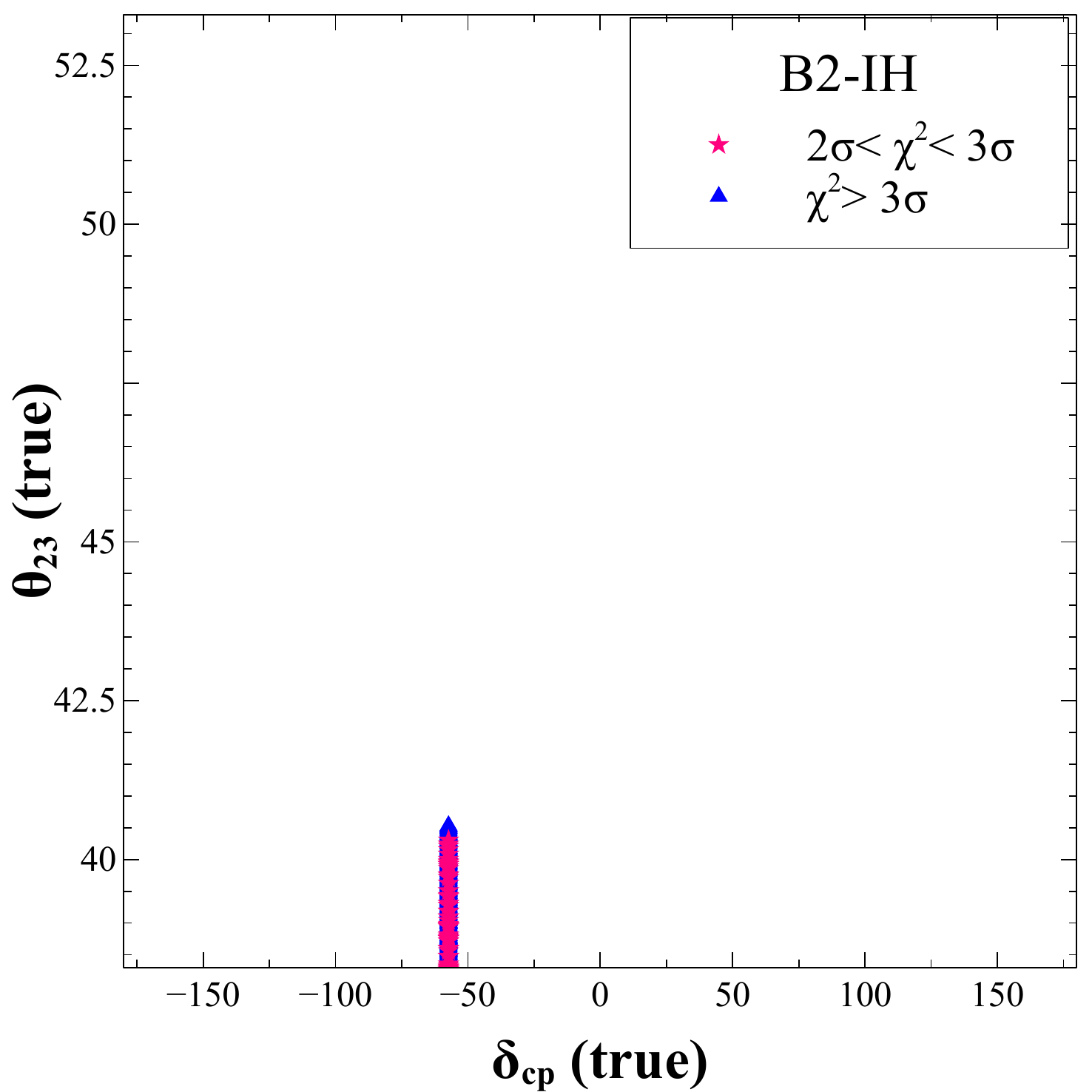}\\
\includegraphics[width=0.45\textwidth]{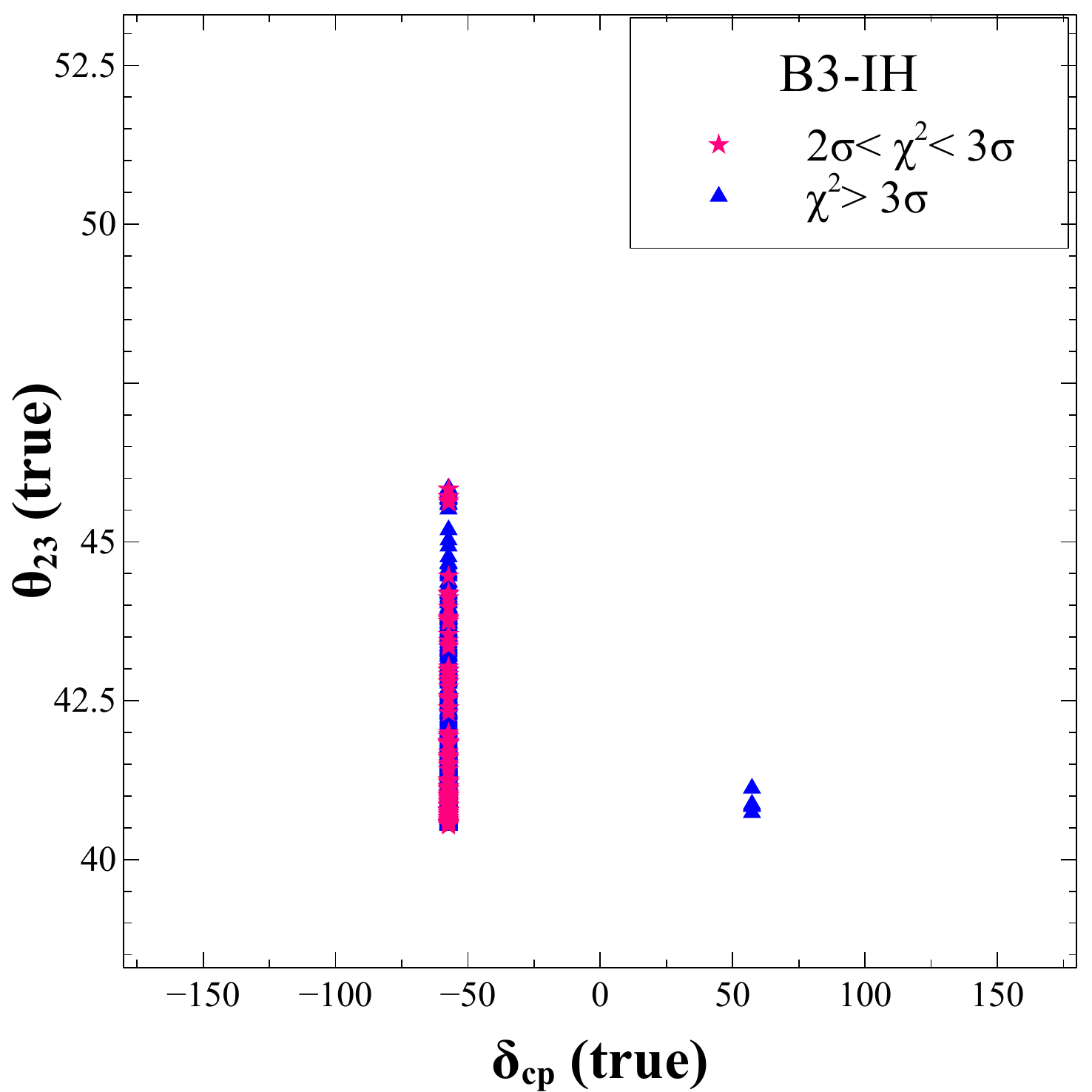}
\caption{Region of the $\theta_{23}-\delta_{cp}$ parameter space for which DUNE can establish two-zero texture against the present oscillation scenario for assumed true IH}
\label{fig94}
\end{center}
\end{figure*}

DUNE is an upcoming experiment and hence it is very much important to ask the following question: if Nature chooses one of these mass texture i.e. the true values of $\tz$ and $\dcp$ lies in the range restricted by the texture, then how DUNE can exclude the present oscillation scenario? To answer that question, we choose the discrete data sets (from fig. \ref{fig3} to fig. \ref{fig2}) which are an outcome of a constraint equation from the textures, as the `data' or `true values' and in the `fit', we consider the standard oscillation scenario i.e. we vary $\dcp$ and $\tz$ in their 3$\s$ allowed range. Then $\chi^2$ is calculated between `data' and `fit'. We show the results (from fig. \ref{fig91} to fig. \ref{fig94}) as a function of true $\tz$ and true $\dcp$. If for a particular choice of true parameters, all test values give $\chi^2 >4$ ($\chi^2>9$), then we can say that the standard oscillation scenario can be excluded at 2$\s$ (3$\s$) C.L. In $\tz-\dcp$ $(\rm true)$ parameter space, a point has been marked by magenta star if for that point $\chi^2_{\rm min}$ is greater than 4 but less than 9. All blue triangles represent the points which give $\chi^2_{\rm min}> 9$. So for all coloured points, we can say that DUNE can exclude the standard scenario at 2$\s$ C.L. DUNE can exclude the standard scenario at 3$\s$ C.L. only for the blue points.

\section{Conclusions}
\label{sec:conclude}
To conclude, in this work, we performed a phenomenological study on one-zero and two-zero textures from DUNE prospective. We first constrain the different possible one-zero and two-zero texture mass matrices from the latest experimental data on oscillation parameters, sum of absolute neutrino mass as well as the lower bound on the half-life of neutrinoless double beta decay. Using the most recent bounds on these parameters, we find that some of the texture zero mass matrices previously shown to be allowed become disfavoured or only marginally allowed. The allowed textures in particular the two-zero texture mass matrices also predict the neutrino parameters like $\theta_{23}, \dcp$ etc. to lie in a very specific range that can be probed at long baseline neutrino experiments like the DUNE.

In the second part of our work, we have studied the capability of DUNE to probe different mass textures once it takes data. We have shown our results for both one-zero and two-zero textures for both the hierarchies. Depending on the values of $\tz$ and $\dcp$ that Nature chooses, how DUNE allows/excludes these textures is presented in this work.
 From the analyses, we have observed that the most of the $\tz-\dcp$ parameter space is allowed for the one-zero textures in both the hierarchies at 3$\s$. DUNE excludes G4 (G6) texture at 3$\s$ almost for all $\dcp$ if true $\tz > 47.5^o$ ($\tz < 44^o$) and assumed true hierarchy is normal. DUNE can exclude G3 at 3$\s$ for any $\tz$ for all the CP conserving values of $\dcp$ while G4, G5 and G6 are excluded for $\dcp = \pm \pi$ at 3$\s$ for assumed true IH. Except A1, all other two-zero textures are excluded for CP conserving values at 4$\s$, irrespective of $\tz$ and hierarchies.
We have observed that these two-zero textures are very interesting from DUNE prospective as most of the true $\tz-\dcp$ parameter space is excluded at DUNE. Specially, for any $\tz$ in the HO, DUNE can exclude B1-NH for all true values of $\dcp$ at 3$\s$. If $\tz$ is such that $\tz>51^o$ and $\tz<43.5^o$, DUNE then can rule out B2-NH for all true $\dcp$ at 3$\s$ C.L.. Similarly, if $\tz >48^o$ ($\tz<44^o$), DUNE excludes B3-NH (B4-NH) for all true $\dcp$ at 3$\s$. 

Again, for assumed true IH, we have found that B1 and B2 textures can be excluded at DUNE for all true $\dcp$ in the upper half plan (UHP, $\dcp$ from $0^o$ to $180^o$) for any values of $\tz$ at 5$\s$ C.L. If Nature chooses $\tz$ such that $\tz<43.5^o$, DUNE can exclude B3 at 3$\s$ for all true $\dcp$.

From this study, we have observed that, irrespective of neutrino mass ordering, DUNE can exclude most of the true $\tz-\dcp$ parameter space of B1 texture and hence restricts B1 more tightly.

Since the texture zero mass matrices can be generated by different flavour symmetry models considered in several earlier works \cite{texturesym,texturesym1,texturesym2,texturesym3,texturesym4,texturesym5,texturesym6,texturesym7,texturesym8,texturesym9}, possible discrimination of different texture zeros at DUNE studied in this work could also disfavour certain flavour symmetry models leading to some particular textures. Although the new physics sector is not directly affecting the neutrino oscillation probabilities analytically, its presence is indirectly tested by probing the specific neutrino parameters it predicts through a particular texture zero mass matrix. This study can also be extended to other texture zero models like the ones with non-diagonal charged lepton mass matrix, the ones with additional light sterile neutrinos. We are working on these possibilities which will be presented in an upcoming work \cite{bbdsterile}.

\begin{acknowledgments}
We acknowledge the use of HRI cluster facility to carry out the computations. DD thanks Prof. Raj Gandhi for his support as well as discussions regarding DUNE. He acknowledges the support from Gauhati University and the DAE Neutrino project at HRI to visit Gauhati University and IIT, Guwahati. He also thanks Suprabh Prakash for some useful discussions during his HRI visit.
\end{acknowledgments}

\appendix

\section{Light neutrino mass matrix elements}
\label{appen1}
{\small \begin{widetext}
\begin{equation}
M_{ee} = c^2_{12}c^2_{13} m_1+c^2_{13} s^2_{12} m_2 e^{i 2\alpha} + s^2_{13} m_3 e^{i2\beta}
\end{equation}
\begin{align}
M_{e \mu}=M_{\mu e} &= c_{13}\bigg(s_{13}s_{23}m_3 e^{i(\delta_{\text{cp}}+2\beta)}-c_{12}m_1(c_{23}s_{12}+c_{12}s_{13}s_{23}e^{i\delta_{\text{cp}}}) \nonumber \\
& +s_{12}m_2 e^{i2\alpha} (c_{12}c_{23}-s_{12}s_{13}s_{23}e^{i\delta_{\text{cp}}}) \bigg) 
\end{align}
\begin{align}
M_{e\tau}=M_{\tau e}  &= c_{13} \bigg ( c_{23}s_{13}m_3 e^{i(\delta_{\text{cp}}+2\beta)}-s_{12}m_2 e^{i2\alpha} (c_{23}s_{12}s_{13}e^{i\delta_{\text{cp}}} \nonumber \\
& +c_{12}s_{23})+c_{12}m_1(-c_{12}c_{23}s_{13}e^{i\delta_{\text{cp}}}+s_{12}s_{23})\bigg )
\end{align}
\begin{equation}
M_{\mu \mu} = c^2_{13} s^2_{23} m_3 e^{i2(\delta_{\text{cp}}+\beta)}+m_1 (c_{23}s_{12} +c_{12} s_{13} s_{23} e^{i\delta_{\text{cp}}})^2+m_2 e^{i2\alpha}(c_{12}c_{23}-s_{12}s_{13}s_{23}e^{i\delta_{\text{cp}}})^2
\end{equation}
\begin{align}
M_{\mu \tau} =M_{\tau \mu} &= c^2_{13}c_{23}s_{23}m_3 e^{i2(\delta_{\text{cp}}+\beta)}+m_1 (c_{12}c_{23}s_{13} e^{i\delta_{\text{cp}}}-s_{12}s_{23})(c_{23}s_{12}+c_{12}s_{13}s_{23}e^{i\delta_{\text{cp}}}) \nonumber \\
&-m_2e^{i2\alpha}(c_{23}s_{12}s_{13}e^{i\delta_{\text{cp}}}+c_{12}s_{23})(c_{12}c_{23}-s_{12}s_{13}s_{23}e^{i\delta_{\text{cp}}})
\end{align}
\begin{equation}
M_{\tau \tau} = c^2_{13}c^2_{23}m_3 e^{i2(\delta_{\text{cp}}+\beta)}+m_2e^{i 2\alpha} (c_{23}s_{12}s_{13}e^{i\delta_{\text{cp}}}+c_{12}s_{23})^2+m_1(c_{12}c_{23}s_{13}e^{i\delta_{\text{cp}}}-s_{12}s_{23})^2
\end{equation}

\end{widetext}}

\end{document}